\documentclass[twocolumn,showpacs,preprintnumbers,amsmath,amssymb]{revtex4}

\usepackage{graphicx}
\usepackage{dcolumn}
\usepackage{bm}
\usepackage{amssymb}

\newcommand\nuh{\nu_h \rightarrow \gamma  \nu} 

\newcommand\numub{\overline{\nu}_\mu}
\newcommand\pair{e^+ e^-}
\newcommand\mix{|U_{\mu h}|^2}

\def\address{\@ifstar{\address@star}%
  {\@ifnextchar[{\address@optarg}{\address@noptarg}}}

\begin{document}

\author{S.N.~Gninenko}
\affiliation{Institute for Nuclear Research, Moscow 117312}


\title{Resolution of puzzles from \\
the LSND, KARMEN, and MiniBooNE experiments}

\date{\today}

\begin{abstract}
 This work has attempted  to reconcile puzzling neutrino oscillation results from the LSND, KARMEN and MiniBooNE experiments.
 We show that the LSND evidence for $\overline{\nu}_\mu \to \overline{\nu}_e$ oscillations, its
long-standing disagreement with the results from  KARMEN, and  the 
 anomalous event excess observed by MiniBooNE in $\nu_\mu$ and $\overline{\nu}_\mu$ data could all be explained  by  
the existence of a  heavy sterile neutrino ($\nu_h$). 
All these results are found to be consistent with each other, assuming  
 that the $\nu_h$ is  created in  
$\nu_\mu$  neutral-current  interactions and  decays radiatively  into  a photon and a light neutrino.
Assuming the $\nu_h$ is produced through mixing with $\nu_\mu$, 
the combined analysis of the LSND and MiniBooNe excess events suggests that 
 the $\nu_h$  mass is in the range from 40 to  80   MeV,  
the mixing strength is  $|U_{\mu h}|^2 \simeq 10^{-3}-10^{-2}$, and
 the lifetime is $\tau_{\nu_h} \lesssim 10^{-9}$ s.
 Surprisingly, this LSND-MiniBooNE  parameter window  is found to be unconstrained 
by the results from the most sensitive 
 experiments. 
We set new limits on $|U_{\mu h}|^2$ 
for the favorable mass region from the precision measurements of the Michel spectrum by 
the TWIST experiment.   
The results obtained provide  a strong motivation for a 
sensitive search for the $\nu_h$ in a near future $ K$ decay or neutrino experiments, 
which fit well in the existing and planned experimental programs at CERN or FNAL. 
The question of whether the heavy neutrino is   
 a Dirac or Majorana particle  is briefly discussed.

\end{abstract}


\maketitle


\section{Introduction}
Over the past 10 years the LSND collaboration \cite{a} has observed an event excess
 with a significance of 3.8 $\sigma$  at LANSCE \cite{lsnd96, lsndfin}. 
This excess, originally interpreted as a signal from 
$\overline{\nu}_\mu \to \overline{\nu}_e$ 
 oscillations was not confirmed by further measurements by a  similar experiment 
KARMEN \cite{b},  which was running at the ISIS neutron spallation facility of the RAL 
\cite{karmen}.  
The MiniBooNE experiment at FNAL \cite{c}, designed to examine the LSND effect, 
  did not find evidence  for $\nu_\mu \to \nu_e$ oscillations.
However,  an anomalous  excess of low energy electron-like events
in charge-current quasielastic ($CCQE$) neutrino events 
 over the expected standard  
 neutrino interactions  has been observed \cite{minibnu1}. This MiniBooNE anomaly 
has been  confirmed by the finding of more excess events \cite{minibnu2}. 
 Recently, the MiniBooNE experiment has reported new  results from a search for
$\overline{\nu}_\mu \to \overline{\nu}_e$ oscillations \cite{minibnub}.
 An excess of events was  observed which have a small probability of being  identified as   
background-only events. The data are found to be  consistent with $\overline{\nu}_\mu \to \overline{\nu}_e$ oscillations in the 0.1 eV$^2$
range and with the evidence for antineutrino oscillations from the LSND experiment.

The new observations  bring more confusion than  clarity to the experimental situation. 
The inconsistency  between the results of the experiments, 
in particular between the LSND and KARMEN experiments,  is also confusing  in light of the apparent simplicity of the primary reaction,  $p(\overline{\nu}_e, e)n$, used by these experiments 
for the oscillation signal 
search,  and also in view of the fact that other results, e.g. the inclusive cross section for $^{12}C(\nu_e, e)^{12}N^*$  with an electron in the final state, measured by LSND \cite{lsndcn} and  KARMEN \cite{karmencn}, agree quite well with each other  and also with 
theoretical calculations. To  reconcile the LSND, KARMEN,  and MiniBooNE results   in 
terms of the, so-called  (3+1)-$\nu$  oscillations scheme  or (a yet unknown)  experimental background seems quite difficult \cite{tsch}. Therefore, it is obviously important to ask whether neutrino
oscillations are the only possible explanation for the observed anomalies.

This work has attempted to reconcile puzzling neutrino oscillation results
 from the LSND, KARMEN and MiniBooNE experiments.
 Our discussion is based on the fact  that 
signals produced by  electrons or  by  converted photons in these experiments  are 
indistinguishable. This hint suggests  that the excess events observed by LSND and MiniBooNE 
 could originate from  converted photons, and not from electrons. 
As an input, we use a natural extension of the model  developed in 
Ref.\cite{sng} for an explanation of the MiniBooNe anomaly observed in $\nu_\mu$
  data in terms of the radiative decays of a heavy  neutrino.
We show that   
  the LSND evidence for $\overline{\nu}_\mu \to \overline{\nu}_e$ oscillations, 
 its long-standing disagreement  with the results from KARMEN, and  the 
 anomalous event excess observed by MiniBooNE in $\nu_\mu$ and $\overline{\nu}_\mu$ data
   could all be explained  by  
the existence of a  heavy neutral lepton ($\nu_h$).    
All these observations are found to 
be consistent with each other, assuming that the $\nu_h$'s are produced  
 in  $\nu_\mu$ neutral-current interactions ($NC$) and that they  decay radiatively  into  a photon and a light neutrino $\nu$.
 The  $\nu_h$'s could be Dirac or Majorana type and could decay dominantly into $\gamma \nu$ if, e.g.,  there is a large enough  transition magnetic moment 
between the $\nu_h$ and $\nu$ mass states. 
 Discussions of other decay modes  suggested  for the explanation of the LSND signal can be found in Ref.\cite{lsnddec}. 

\begin{figure}[tbh!]
\begin{center}
    \resizebox{7cm}{!}{\includegraphics{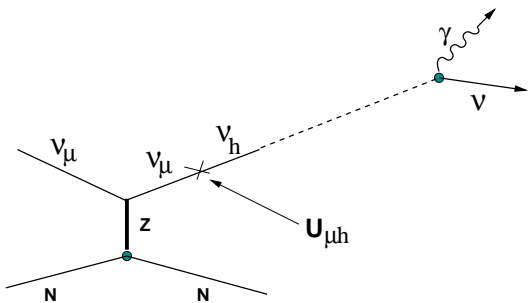}}
     \caption{ Schematic illustration of the $NCQE$ production and the decay of heavy 
neutrino. }
\label{diag}
\end{center}
\end{figure}

We may consider the $\nu_h$ as a very weakly interacting particle 
directly produced by the $\nu_\mu$ flavor eigenstate in neutrino-nucleus reactions. However,
it is known, that the neutrino weak flavor eigenstates ($\nu_e,~\nu_{\mu},~\nu_{
\tau},...$) 
can be different from the mass
eigenstates ($\nu_1,~\nu_2,~\nu_3,~\nu_4...$), but they  are related to them, in 
general,
through a unitary transformation.   A generalized mixing:
\begin{equation}
\nu_l= \sum_i U_{li} \nu_i;~~~l=e,\mu,\tau,...,~i=1,2,3,4,...
\label{mixing}
\end{equation}
results in neutrino oscillations when the mass differences are small,
and in neutrino decays when the mass differences are large. 
Hence, it would also be natural to assume 
that the $\nu_h$, if it exists, is a component of muon neutrinos which is  produced  
 in $\nu_\mu$ $NC$ interactions  by muonic mixing, as illustrated  in Fig.\ref{diag}. 
  This  assumption provides us with a useful framework for further discussions.
An immediate consequence  is that the $\nu_h$ can also be produced 
through $CC$ interactions 
in leptonic and semileptonic decays of sufficiently heavy mesons and
baryons  according to the proper mixing strength, as follows from Eq.~\eqref{mixing}, and
phase space and helicity  factors \cite{shrock1, shrock2} (see also \cite{sgdg}).
 Note that, although  $CC$ weak interactions of ordinary particles are 
$V-A$, one could 
 assume that the heavy neutrinos may dominantly be produced  by non-left-handed $V,A$ couplings; see e.g., the discussion in 
Ref.\cite{shrock1}. Therefore, it would be interesting and important to have a general analysis  
of the production of heavy neutrinos of Dirac or Majorana type, e.g. in $\nu_\mu NC$ interactions,
 for arbitrary weak couplings 
including the leptonic mixing and helicity effects. This is, however, beyond the scope of the present work.

The rest of the paper is organized as follows.
In Sec.II  we describe the formalism for the radiative neutrino decay, specifying the difference 
between the Dirac and Majorana decay modes.
The results from the LSND and KARMEN experiments are described in Sec.III. 
Here we  show how the suggested model explains those results.
In Sec.IV  we briefly describe the MiniBooNE experiment and 
give an explanation of the anomalous  excess of events  observed in  $\nu_\mu$ and $\overline{\nu}_\mu$ data. The final results from the 
combined analysis of the LSND and MiniBooNE data are reported in Sec.V.
The discussion and  review of the experimental and some cosmological and astrophysical  constraints on 
the mixing strength  $|U_{\mu h}|^2$ and neutrino magnetic moment are presented  in Sec.VI. 
We find that, quite surprisingly,  
the ($ m_{\nu_h};|U_{\mu h}|^2$) parameter space favorable 
for the explanation of the LSND and MiniBooNE results is  unconstrained by 
the results  
from the most sensitive experiments, e.g. searching for a  $\nu_h$ peak in $\pi_{\mu 2 }, K_{\mu 2 }$ decays.
 Moreover, we show that taking into account the dominance of the radiative $\nu_h$  decay and 
its  short lifetime, makes existing experimental  bounds weaker, 
allowing them to be extended to the higher mass region.
In Sec.VII, several proposed experiments to search for the  $\nu_h$  are described. 
We also show that, several tests can be 
applied to existing data.   Section VIII contains concluding remarks.

\section{Radiative neutrino decay}

Let us consider the decay of a heavy neutrino $\nu_h$ of mass $m_{\nu_h}$ and energy $E_{\nu_h}$
into a lighter neutrino $\nu$ and a photon:
\begin{equation}
\nu_h \to \nu + \gamma
\label{nurad}
\end{equation}
with the partial lifetime $\tau_{\nu_h}$.
The energy of the decay photon in the $\nu_h$ rest frame given by
\begin{equation}
E_\gamma^0=\frac{m_{\nu_h}}{2}(1-\frac{m_{\nu}^2}{m_{\nu_h}^2})
\end{equation}
is in the range $0 < E_\gamma < m_{\nu_h}/2$, depending 
on the mass of the $\nu$, which may be in the range $0 < m_\nu < m_{\nu_h}$.
 Furthermore, for simplicity we assume that the particle $\nu$ is almost massless, and the photon 
energy in the rest frame is $E_\gamma^0 = m_{\nu_h}/2$.  
 The energy of the decay photon in the laboratory frame depends on 
the $\nu_h$ initial energy  and on the center-of-mass angle $\Theta$ between the photon momentum 
and  the $\nu_h$ direction of flight:
\begin{equation}
E_\gamma=\frac{E_{\nu_h}}{2}(1+\frac{P_{\nu_h}}{E_{\nu_h}}cos\Theta)\simeq \frac{E_{\nu_h}}{2}(1+cos\Theta)
\end{equation}
 \begin{figure}[tbh!]
\begin{center}
    \resizebox{8cm}{!}{\includegraphics{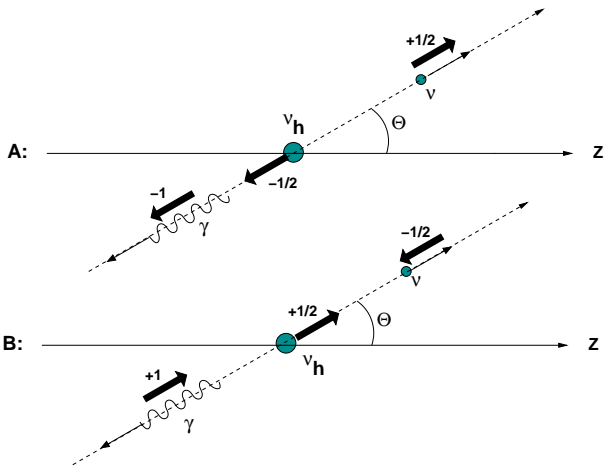}}
     \caption{ Two amplitudes, A and B, describing the decay $\nuh$ for different 
     $\nu_h$ helicities.  }
\label{decay}
\end{center}
\end{figure}
Hence, the energy distribution of photons in the laboratory system depends on their 
 angular distribution in the rest frame, which is not generally isotropic \cite{vogel}:
\begin{equation}
\frac{dN}{dcos\Theta} = \frac{1}{2}(1+a cos\Theta)
\label{angular}
\end{equation}
Here, the angle  $\Theta$ is defined as above 
 and  $a$ is the asymmetry parameter. It is also possible  to define $\Theta$ as the angle between 
 the direction of spin, the only direction available in the rest frame, and 
 the photon momentum. However, if we assume that the spin of $\nu_h$ is (anti)parallel to its momentum,
 both definitions are equivalent. 

The decay of a spin-$\frac{1}{2}$ neutrino into another spin-$\frac{1}{2}$ particle and a photon  can be generally described by two helicity amplitudes $A$ and $B$ corresponding 
to the final states shown in Fig. \ref{decay}. For the most general coupling given by \cite{wil, pal,  shrockdec}
\begin{equation}
\overline{\psi}(\nu) \sigma_{\mu \nu} (\alpha+\beta\gamma_5) \psi(\nu_h) \partial^\mu A^\nu 
\end{equation}
the amplitudes $A$ and $B$ are proportional, respectively to $(\alpha-\beta)$ and  $(\alpha +\beta)$.
If 
 $CP$ is conserved, the helicity  amplitudes  $|A|$ and  $|B|$ for the decay of Majorana neutrinos are equal. In this case the decay $\nuh$ would be isotropic and independent of the $\nu_h$ polarization, and hence  $a=0$ in Eq.(\ref{angular}). Indeed, suppose  the $\nuh$ decay  
is  anisotropic in the  center-of-mass system and photons are emitted preferably, say   
opposite to the $\nu_h$ spin direction.  Because of  $CP$ conservation,  
the $CP$-mirror image of this process should also exist, and it 
would correspond  to the $\nuh$ decay with photons emitted preferably along the $\nu_h$ spin direction. But, if the $\nu_h$ is 
  its own  antiparticle, the decay anisotropy must be the same for the $\nu_h$ and its 
 $CP$-mirror image. Hence, the decay must be isotropic. For the Dirac case, the $\nu_h$ and its $CP$-mirror image are  not identical
and the above arguments do not hold. For Dirac $\nu_h$   
the angular anisotropy is the result of parity nonconservation in the decay 
\eqref{nurad} and of nonvanishing polarization of the neutrinos. 
The decay  asymmetry parameter given by 
\begin{equation}
a= -2 \frac{Re(\alpha^* \beta)}{|\alpha|^2+|\beta|^2}
\end{equation}
is, in general, not constrained, and it may be in the range  $-1 < a < +1$ \cite{wil,pal}.
In the standard model $\beta/\alpha = (m_{\nu_h} - m_{\nu})/(m_{\nu_h} + m_{\nu})$, so that 
\begin{equation}
a = \frac{ m_{\nu}^2- m_{\nu_h}^2 }{m_{\nu_h}^2 + m_{\nu}^2}
\end{equation}
is equal to zero only when $m_{\nu_h}\simeq m_{\nu}$.
 For left-handed Dirac neutrinos and  $m_{\nu}\ll m_{\nu_h}$, one has $a=-1$, 
which means that  the decay photons are  emitted preferably backward \cite{wil, pal, shrockdec}, shifting 
 the energy spectrum in the  laboratory frame towards lower energies. 
For the right-handed Dirac neutrinos, 
one has $a=+1$, and the photons are  emitted preferably 
in the forward direction, making the energy  spectrum harder. 
Hence, the energy spectrum and angular distribution of the decay photons are
  sensitive to the type of $\nu_h$.
Note that if  $CP$ is conserved, the decay rate and the center-of-mass angular distributions for the Dirac case  
are the same for the $\nuh$ as for  the $\overline{\nu}_h \to \overline{\nu}  \gamma$ decay modes with respect to the 
beam direction.
Furthermore,  we assume that the decay $\nuh$ is generally $CP$ conserving (see also Sec.IV.B).

As mentioned above, the most natural way to allow  the  radiative  decay  of heavy neutrino 
is to introduce a nonzero   transition magnetic moment ($\mu_{tr}$)  between the $\nu_h$ and $\nu$ mass states;
see e.g. \cite{moh, bovo}. Such coupling of neutrinos with photons is a generic consequence 
of the finite neutrino mass. Observations of the neutrino magnetic moment could allow to distinguish if neutrinos  are of  Dirac or Majorana type since the Dirac neutrinos can only have flavor conserving 
transition magnetic moments while  the Majorana neutrinos can only have a changing one.
In addition, Dirac neutrinos can have diagonal magnetic moments while  Majorana neutrinos cannot.
The nonzero magnetic moment of the neutrino, although tiny, is predicted even in the standard model.
 The detailed calculations of the radiative neutrino decay rate in terms of 
the neutrino masses and mixings of Eq.(\ref{mixing}) were performed long ago,
see e.g. \cite{wil, pal, shrockdec}. 
The radiative decay mode could even be dominant, if  the $\mu_{tr}$ value is large enough; see  \cite{moh, bovo}. Originally, the idea of a large (Dirac) magnetic moment ($\gtrsim 10^{-11} \mu_B$, where
$\mu_B$ is the Bohr magneton) of the electron neutrino has been 
suggested in order to explain the solar neutrino flux variations \cite{vol}.  
Taking into account that in many extensions of the standard model the value of the $\mu_{tr}$
 is typically  proportional to the $\nu_h$ mass,  the intention  to make the radiative decay 
 of a $\nu_h \lesssim 100$ MeV 
 dominant by introducing a large transition magnetic moment (or through another mechanism) 
 is not particularly exotic from a theoretical viewpoint. 
Such  types of heavy neutrinos are present  in many interesting extensions of 
the standard model, such as GUT, superstring inspired models, left-right symmetric models and others, for a review; see e.g. Ref.\cite{moh}.

The total $\nu_h$ decay width can be defined as 
$\Gamma_{tot} = \Gamma(\nu_h \to \nu \gamma)+ \Sigma \Gamma_i$,
where $\Gamma(\nu_h \to \nu \gamma)$ is the $\nuh$ decay rate, and $\Sigma \Gamma_i$ is 
the sum over  decay modes whose decay rate is 
  proportional to the square of the mixing $|U_{\mu h}|^2$. For the $\nu_h$ with a mass $\lesssim 100$ MeV, 
the dominant contribution to  $\Sigma \Gamma_i$ comes from   
 $\nu_h \to\nu_\mu \pair$ and  $\nu_\mu \nu_l\overline{\nu}_l~(l=e,\mu,\tau)$  
decays,  for which the rate calculations can be found, e.g. in \cite{bolton, gs, atre}.
The $\nuh$ decay rate due to a transition moment $\mu_{tr}$ is given by \cite{marc}
\begin{eqnarray}
\Gamma_{\nu \gamma}=\frac{\mu_{tr}^2}{8\pi}m_{\nu_h}^3\bigl(1-\frac{m_{\nu}^2}{m_{\nu_h}^2}\bigr)^3 
\label{ratemagmom}
\end{eqnarray}
The decay rate $ \Gamma(\nu_h \to\nu_\mu ee)$  can be estimated as
\begin{equation}
\Gamma(\nu_h \to\nu_\mu \pair) \simeq \bigl(\frac{m_{\nu_h}}{10~MeV}\bigr)^{5} |U_{\mu h}|^{2}\cdot s^{-1}, 
\label{ratedec}
\end{equation}
and the sum rate $\Sigma \Gamma_i \simeq 9 \cdot \Gamma(\nu_h \to\nu_\mu \pair)$.
For $m_{\nu_h} \simeq 50$ MeV, $|U_{\mu h}|^2\lesssim 10^{-2}$ and $\mu_{tr} > 10^{-10}{\mu_B}$, 
we found that 
the radiative decay is dominant, as its branching fraction $Br(\nu_h \to \gamma \nu)=\frac{\Gamma(\nu_h \to \nu \gamma)}{\Gamma_{tot}}>0.99 $.
\begin{table*}
\caption{\label{tab:table3}Comparison of experimental parameters of the LSND and KARMEN experiments.}
\begin{ruledtabular}
\begin{tabular}{ccc}
 & LSND& KARMEN\\ \hline
$p$ beam kinetic energy, MeV & 800 & 800 \\
total number of POT's  &$1.8 \times 10^{23}$ &$5.9 \times 10^{22}$ \\
distance to target, m & 30 & 17\\
angle between the $\nu$  and $p$ beams   & 12$^o$ & 90$^o$\\
total $\overline{\nu}_\mu$ flux & $1.2\times 10^{22}$ &  $2.71\times 10^{21}$\\  
$\overline{\nu}_\mu, \nu_e / cm^2$ from $\mu^+$ DAR & 1.26$\times 10^{14}$& 8.86$\times 10^{13}$\\  
$\nu_\mu, \overline{\nu}_e / cm^2$ from $\mu^-$ DAR & 1.08$\times 10^{11}$&7.6$\times 10^{10}$\\
$\nu_\mu / cm^2$ from $\pi^+$ DIF & 2.2$\times 10^{12}$&$<10^{11}$ \\
$\overline{\nu}_e p \to e^+ n$ efficiency & 0.17 & 0.19\\
$e^+$ energy range , MeV  & 20 - 60  & 16 - 50 \\
observed events & 87 & 15\\
background & $53.8$& $15.8\pm 0.5$\\
event excess, $R_\gamma > 1$ & $87.9\pm 22.4\pm 6.0$ & $10\pm 32$\\
event excess, $R_\gamma > 10$& $32.2\pm 9.4$ & $< 5.1$(90\% C.L.)\\
$\overline{\nu}_\mu \to \overline{\nu}_e$ oscillation probability &$(2.64\pm0.67\pm0.45)\times 10^{-3}$& $<0.85\times 10^{-3}$(90\% C.L.)\\ 
\end{tabular}
\end{ruledtabular}
\end{table*}

\section{Interpretation of the LSND and KARMEN results}

The LSND  and KARMEN  experiments used neutrinos 
produced in the beam stop of a proton accelerator. LSND finished data taking at LANSCE at the end of 1998, while KARMEN finished data taking in 2001.
In these experiments, neutrinos were produced by the following decays of pions and muons occurring in the proton target: 

\begin{itemize}
\item $\pi^+\to \mu^+ \nu_\mu$ decays in flight (DIF) or at rest (DAR), 
\item $\mu^+ \to e^+ \nu_e \overline{\nu}_\mu$ DAR,
\item $\pi^- \to \mu^- \overline{\nu}_\mu$ DIF,
\item $\mu^- \to e^- \overline{\nu}_e \nu_\mu$ DAR.
\end{itemize}
The main detector properties and the neutrino fluxes 
in these experiments  are summarized in Table 1. 

\subsection{The LSND signal of $\overline{\nu}_{\mu} \to \overline{\nu}_e$ oscillations}

In 1996 the LSND experiment published  evidence for 
$\overline{\nu}_{\mu} \to \overline{\nu}_e$ oscillations, based on the  observation of 
an excess of $ \overline{\nu}_e$-like events \cite{lsnd96}.
Measurements performed from 1996-1998 with a different target configuration
confirmed the evidence and improved the significance of the 
observed excess. 
The LSND detector is described in detail in Ref. \cite{lsnd_det}.
It was located at a distance of 30 m downstream of 
the main LANSCE beam-stop A6 at a small angle of $\simeq 12^o$ relative to the 
primary proton beam. The detector was a cylindrical volume filled with 167 t 
of a dilute mineral oil (CH$_2$) based liquid scintillator viewed by  
photomultipliers (PMT) and surrounded by 
an active  4$\pi$ veto shield. The low light-yield of the scintillator allowed for
the detection of Cherenkov light generated by relativistic muons, electrons, and 
 converted photon tracks. This feature was of great 
importance for particle identification and reconstruction of its direction. 
The energy resolution of the detector was
about $\simeq 6\%$ at 50 MeV electron energy.  

The search for $\overline{\nu}_{\mu} \to \overline{\nu}_e$ oscillations was based on 
the appearance of $\overline{\nu}_e$  in the neutrino beam, detected through the reaction 
$\overline{\nu}_e p \to e^+ n$ resulting in a prompt relativistic $e^+$, followed 
by a 2.2 MeV gamma signal from the neutron capture $p(n,\gamma)d$.
The $e^+$ candidate events identification and separation from the background 
were based on the detection of the prompt and directional Cherenkov light, and scintillation 
light which is delayed and isotropic.
The 2.2 MeV signal from the reaction $p(n,\gamma)d$  is correlated in time 
with the positron one. It was identified and separated from accidental low-energy 
$\gamma$'s by means of a likelihood parameter $R_\gamma$, which is defined as the ratio
of the likelihood of a low-energy event being correlated or being accidental.
The parameter $R_\gamma$ was defined by three values: i) the PMT multiplicity, which 
is proportional to the $\gamma$ energy, ii) the radial distance between the 
reconstructed positions of the $e^+$ and $\gamma$, and iii) the time difference between   
 the $e^+$ and $\gamma$, which is defined by the capture time of 186 $\mu$s of neutrons 
 in mineral oil, while accidentals are distributed uniformly in time.
 A $\chi^2$ fit to the $R_\gamma$ distribution obtained from the 1993-1998
 measurements resulted, after subtraction of background from DAR and DIF neutrino events,  
 $(19.5 \pm 3.9)$ and $(10.5 \pm 4.6 )$, respectively, 
 in a beam on-off excess of $(87.9 \pm 22.4 \pm 6.0  )$ events.
 The neutrino background was carefully evaluated both from independent 
 measurements and calculations. This excess was attributed to the appearance of
  $\overline{\nu}_e$   from $\overline{\nu}_{\mu} \to \overline{\nu}_e$ oscillations
 and corresponds to the oscillation probability 
 $P(\overline{\nu}_{\mu} \to \overline{\nu}_e)= (2.64 \pm 0.67 \pm 0.45)\times 10^{-3}$.

\begin{figure}[tbh!]
\begin{center}
    \resizebox{9cm}{!}{\includegraphics{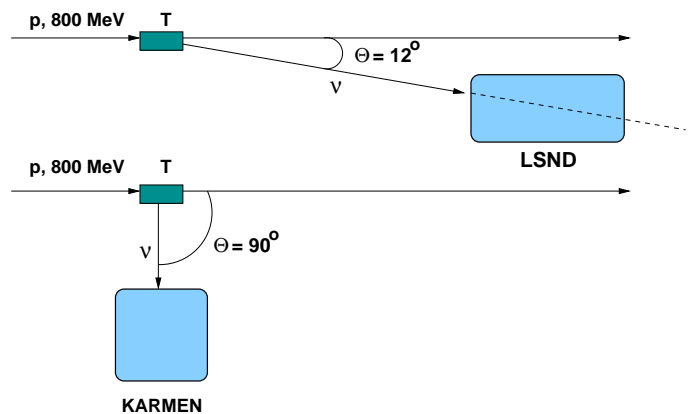}}
     \caption{ Schematic illustration of the location and orientation  of the LSND and KARMEN 
detectors relative to the 
     incident  proton beam direction. }
\label{setup}
\end{center}
\end{figure}

The KARMEN experiment  used a technique  similar to the LSND experiment
and observed no beam excess  \cite{karmen}. 
The signatures of the 15 candidate events were found to be 
in good agreement with those from the ($15.8\pm 0.5$) expected background 
events. 

\begin{figure}[h]
 \begin{center}
    \resizebox{9cm}{!}{\includegraphics{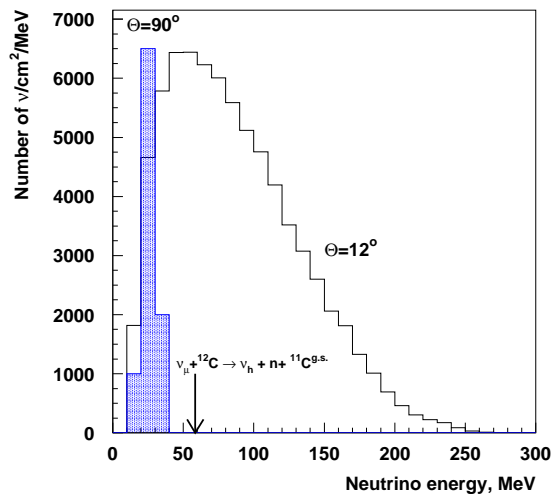}}
     \caption{ The shape of the energy distributions of the $\nu_\mu$'s from 
$\pi^+$ DIF  in the LSND ($\Theta=12^o$)   and KARMEN ($\Theta=90^o$, hatched) detectors. 
      The arrow shows the  production threshold of $E_{th} = 58.6$ MeV 
for the heavy neutrino with a mass of 40 MeV
 in the reaction $\nu_\mu + ^{12}C \to \nu_h +n+ ^{11}C_{g.s.}$, 
in which the $\nu_h$ production  is accompanied 
by the emission of a neutron and the isotope $^{11}$C in the ground state.
The distributions are normalized to a common maximum value.}
\label{numusp}
\end{center}
\end{figure}

Let us explain the discrepancy between the results of these two experiments 
in terms of the  production and radiative decay of a heavy neutrino, as  
illustrated in Fig.\ref{diag}. 
The location of the LSND and KARMEN detectors relative  to their  proton beam directions is 
schematically illustrated in Fig. \ref{setup}.
The energy distributions of the $\nu_\mu$'s from $\pi^+$ DIF  in the LSND (from Ref.\cite{lsndmu}) and KARMEN (simulated) detectors are shown in Fig.\ref{numusp}. The distributions are normalized to a common maximum value in order to place them 
on a similar scale.
One can see that the spectra are quite different. The LSND distribution is peaked at about 55 MeV; it has an average 
energy $\simeq$ 100 MeV, and a high energy tail up to $\sim$ 300 MeV, 
while the maximum of the energy spectrum in KARMEN, which is located at 90$^o$ with respect to the beam, 
is $\simeq 20$ MeV and the whole spectrum is well below 50 MeV.  
For a heavy neutrino with a  mass of $m_{\nu_h} = 40$ MeV, the production threshold in 
the reaction $\nu_\mu  ^{12}C \to \nu_h n ^{11}C_{g.s.}$ is  58.6 MeV, as shown in Fig.\ref {numusp}. 
Here we assume that the  $\nu_h$ production 
 is accompanied by the emission of a recoil neutron and  the 
isotope $^{11}$C in the ground state. 

Thus, our interpretation of the excess of events observed by LSND is the following. 
Positive pions  generated in proton collisions  produce the flux of $\nu_\mu$'s from the $\pi^+ \to \mu^+ \nu_\mu$ DIF  
in the target. The excess events are generated in the LSND detector 
by these $\nu_\mu$'s through the reaction
\begin{equation} 
\nu_\mu ^{12}C\to\nu_h   n  X \to \gamma \nu  n X,
\label{reaction}
\end{equation}
with the emission of a recoil neutron and a heavy neutrino, 
and not by $\overline{\nu}_\mu$'s from muon decays at rest via 
$\overline{\nu}_{\mu} \to \overline{\nu}_e$ 
oscillations, as was originally assumed \cite{lsndfin}.  
The $\nu_h$'s decay promptly into a photon and a light neutrino, with 
the subsequent
Compton scattering or $\pair$ pair  conversion of the decay
photon in the detector fiducial volume.
The former process dominates for photon energies  below the  critical 
energy of the LSND liquid of 85 MeV. 
In the laboratory system, the differential Compton scattering cross section 
has a sharp peak in the forward direction, and the vast majority of  events are 
in a narrow cone of $ \lesssim \sqrt{m_e/E_\gamma} \lesssim 100 $ mrad,
for $E_{\gamma} > 20 $ MeV. For the photon conversion into an  $\pair$ pair, its
opening angle is $\simeq m_e/E_{\gamma}< 25$ mrad for $E_{\gamma} > 20 $ MeV, which is 
too small to be resolved in LSND into two separate Cherenkov rings (here,
$m_e, E_{\gamma}$ are the electron mass and the photon energy, respectively).
Therefore, the  excess events are  originated from  photons of the reaction 
\eqref{reaction}  detected 
in coincidence with the associated 2.2 MeV $\gamma$ tag 
from the neutron capture and misidentified as  single electron events.
In the KARMEN experiment,  $\nu_\mu$'s  from $\pi$ decays in flight  
 cannot produce heavy neutrinos accompanied by the 
emission of a neutron because  
their  energy is below the $\nu_h$ production threshold, see  Fig. \ref{numusp}. 
Therefore, KARMEN should  observe no excess of $\overline{\nu}_{\mu} \to \overline{\nu}_e$-like events.
Note that  
the maximum energy of $\overline{\nu}_\mu$'s from muon DAR is about 50 MeV 
and is also less than the energy threshold of 58.6 MeV for the production of the 40 MeV $\nu_h$ 
and a recoil  neutron in collisions with  the carbon nucleus. 

To make quantitative estimates, 
we performed simplified simulations of the $\nu_h$ production in the inclusive reaction \eqref{reaction}
with the emission of a recoil neutron and followed by the decay $\nuh$, as shown in Fig. \ref{diag}, in the LSND detector.
In these simulations  we used the integral $\nu_\mu$ DIF energy spectrum, shown in Fig. \ref{numusp},
which was calculated in \cite{lsndmu}. There is also a contribution from $\overline{\nu}_\mu$ DIF events, which, however, 
is small and is neglected at the level of accuracy of our analysis. 
The energy of most of the $\nu_\mu$'s   
is well above the threshold for the production of 40-80 MeV $\nu_h$'s in the LSND detector.
Once produced, the $\nu_h$'s  decay at an average distance 
$\simeq c \tau_h E_{\nu_h}/m_{\nu_h}$  from the primary vertex. 
Since in the LSND  experiment the average $\nu_h$ kinetic energy is 
$E_{\nu_h}\simeq 50$ MeV and $\nu_h$'s 
would decay over the  average 
distance of $\lesssim$ 5 m from the primary vertex, the sensitivity is 
restricted to the  $\nu_h$ lifetimes  $\tau_{\nu_h} \lesssim 10^{-8}$ s
for the $\nu_{h}$ masses $m_{\nu_h} \gtrsim 40$ MeV. 
The decay
photon absorption occurs at a distance of the 
order of the Compton scattering length ($\simeq 40 $ cm ) of the LSND liquid from the $\nu_h$ decay point which is much less than the detector size. 
\begin{figure}[h]
\begin{center}
    \resizebox{9cm}{!}{\includegraphics{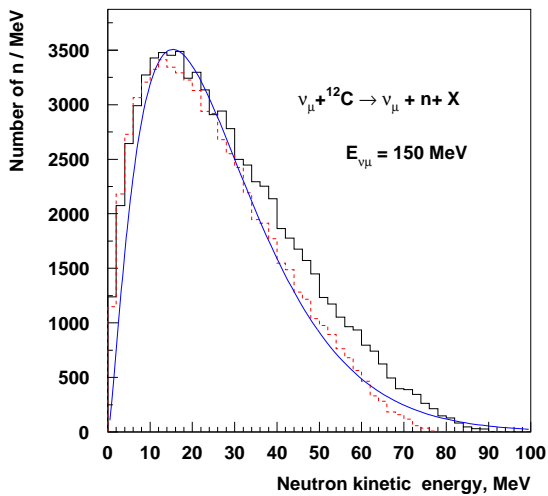}}
     \caption{ The shape of the distributions of the recoil neutron kinetic energy 
in the reaction $\nu_\mu  ^{12}C \to \nu_\mu   n X $ obtained in the 
present work (solid histogram ) and calculated  in \cite {hor} (solid curve )
for the $\nu_\mu$  energy $E_\nu$=150 MeV. The distribution of the kinetic energy of 
neutrons ejected in the reaction $\nu_\mu  ^{12}C \to \nu_h   n X $ for a heavy neutrino mass of 60 MeV 
is also shown for comparison (dashed histogram). The binding energy corrections are not applied. The 
 distributions are normalized to a common maximum value.}
\label{neutrcomp}
\end{center}
\end{figure}
\begin{figure}[h]
\begin{center}
    \resizebox{9cm}{!}{\includegraphics{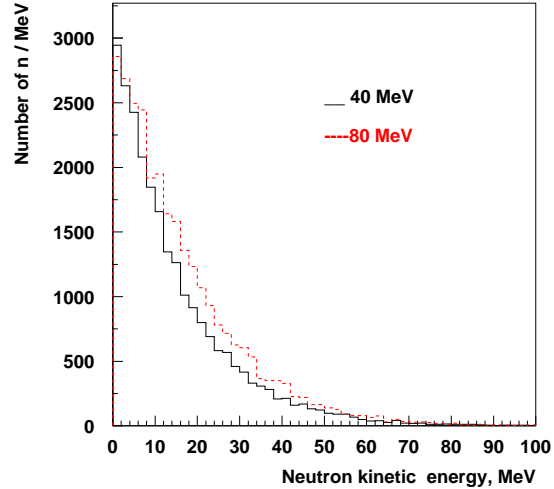}}
     \caption{ The $\nu_\mu$ flux-averaged distributions of the recoil neutron kinetic energy 
from the reaction $\nu_\mu ^{12}C \to \nu_\mu  n  X $
calculated for  $\nu_h$ masses of 40 and 80 MeV with the binding energy 
correction included. The histograms are normalized as in Fig.\ref{neutrcomp}.}
\label{neutron}
\end{center}
\end{figure}

The total cross section 
of the reaction  $\nu_\mu  ^{12}C \to  \nu_h  n  X$ for 100\% mixing is estimated  by extrapolating  the available  
cross section for the reaction $\nu_\mu  ^{12}C \to  \nu_\mu  n  X$ ($\simeq 24\times 10^{-40}$ cm$^2$)
calculated for the incident neutrino energy of 150  MeV
 \cite{hor,mart} to the neutrino 
 energies in the range  50-250 MeV (see Fig. \ref{numusp}), and by taking into account the corresponding  phase space 
factor.
 Note that  the average $\nu_\mu$ energy for the spectrum above 
the production threshold of the 40 MeV $\nu_h$ is 110 MeV.
The $\nu_\mu$ flux-averaged cross section is found to be 
$\sigma (\nu_\mu  ^{12}C \to  \nu_h  n  X) \simeq (16 \pm 6.5) \times 10^{-40}$ cm$^2$
for the production of the 40 MeV $\nu_h$  with the  uncertainty taken to be 40\%
due to accuracy of the extrapolation procedure.

 The crucial test of the  
$\overline{\nu}_{\mu} \to \overline{\nu}_e$  oscillation hypothesis in the 
LSND experiment was to check whether there is an excess of events with more 
than one correlated 2.2 MeV $\gamma$. If the excess of events is indeed due to the 
reaction  $\overline{\nu}_e p \to e^+ n$, then there should be
no excess with more than one correlated $\gamma$ because the recoil neutron 
is too low in energy  ( $< 5$ MeV) to knock out additional neutrons.
If, on the other hand, the excess involves  higher energy neutrons, 
which can break the $^{12}$C nucleus and produce another neutron(s),  
then one would expect an excess of events with $> 1$ correlated 
$\gamma$. As the LSND did not observe many of such (latter)  events \cite{lsndfin},
 the  energy spectrum of the ejected neutrons  is an important
characteristic of the  reaction \eqref{reaction}, as it affects  the likelihood ratio  
$R_\gamma$ and the number of correlated $\gamma$'s. The $\nu_\mu$ 
 flux-averaged cross sections and particle emission spectra in the 
LSND detector
are predicted  quit well for the charge current reactions $\nu_\mu ^{12}C \to \mu X$
from calculations performed in different theoretical frameworks;
see e.g. Ref.\cite{kolbe} and references therein. 
However,  much less is known for the  $\nu_\mu$ induced neutral-current 
reactions in the  detector. 
Therefore, we performed simulations of the recoil neutron kinetic energy  distributions of  in the reaction \eqref{reaction} by using the  Fermi gas nuclei model, but without taking 
into account nuclear effects, such  
as  the neutron re-scattering in nuclear matter and
 the carbon nucleus level structure.  
The Fermi momentum and the neutron binding energy for the $^{12} C$ nucleus are taken to be $ 200$ MeV and  
 18 MeV, respectively.
\begin{figure*}
 \resizebox{22cm}{!}{\includegraphics{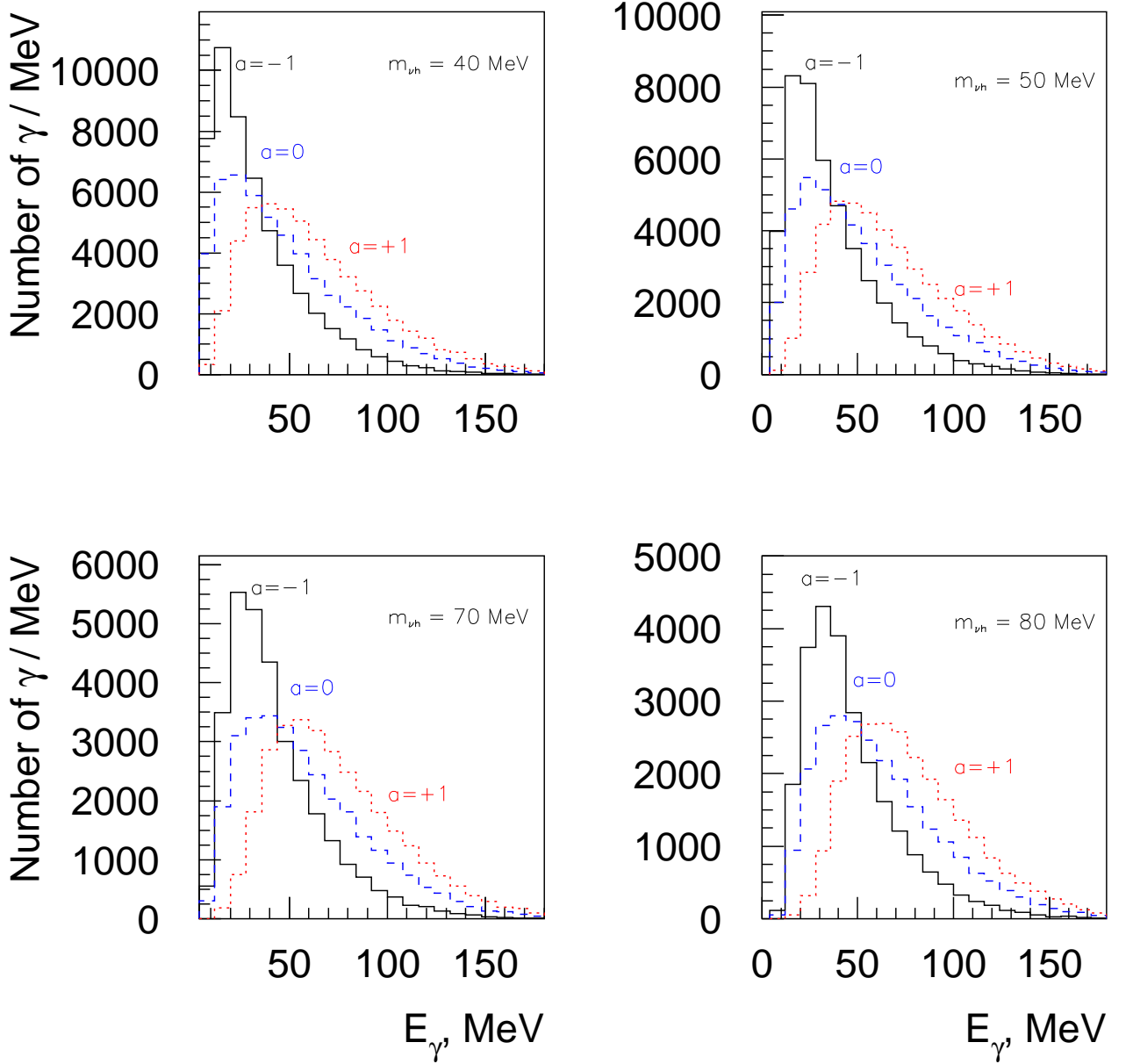}}
\caption{ The $\nu_\mu$ flux-averaged distributions of the energy of photons from the radiative decay of heavy 
neutrinos produced in the reaction $\nu_\mu ^{12}C \to \nu_\mu  n  X $
calculated for 100\% mixing strength and $\nu_h$ masses from  40 to 80 MeV with no photon detection efficiency and 
the neutron binding energy 
correction included . The spectra are  calculated for the $a=-1$ (solid line), 
$a=+1$ (dotted line), 
 and $a=0$ (dashed line) cases for the same $\nu_\mu$ flux.}
\label{spectra}
\end{figure*}

To evaluate uncertainties of our calculations we have compared our results 
with others which take into account 
nuclear effects \cite{hor}-\cite{ventel}.
 Figure\ref{neutrcomp} shows the distribution of the kinetic energy of neutrons
ejected in the reaction  $\nu_\mu ^{12}C \to \nu_\mu n X$ and the 
 analogous spectrum from Ref.\cite{hor}, both 
calculated for the massless case for the incident neutrino energy of 150 MeV 
without nucleon binding energy corrections. 
 One can see that our simulations reproduce the more precise results 
quite reasonably. The comparison of the calculations results in an 
uncertainty of about 20\%-30\%. Figure\ref{neutrcomp} also 
shows the neutron energy distribution 
calculated for the reaction \eqref{reaction} for the $\nu_h$  mass of 60 MeV. 
It is seen that in this case the neutron energy   spectrum is shifted towards lower energies.   

\begin{figure*}
 \resizebox{17cm}{7cm}{\includegraphics{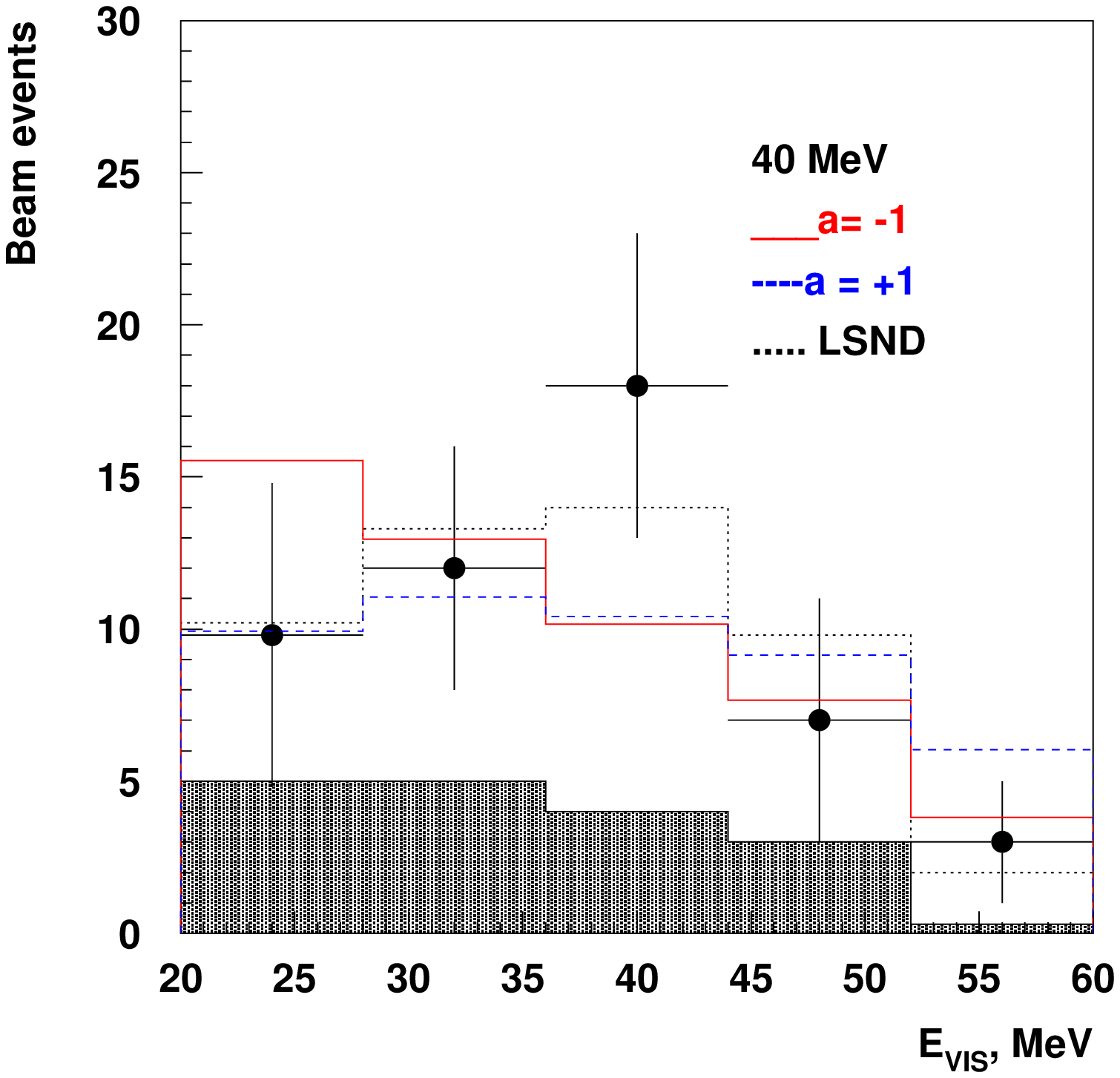}{\hspace{-3cm}\includegraphics{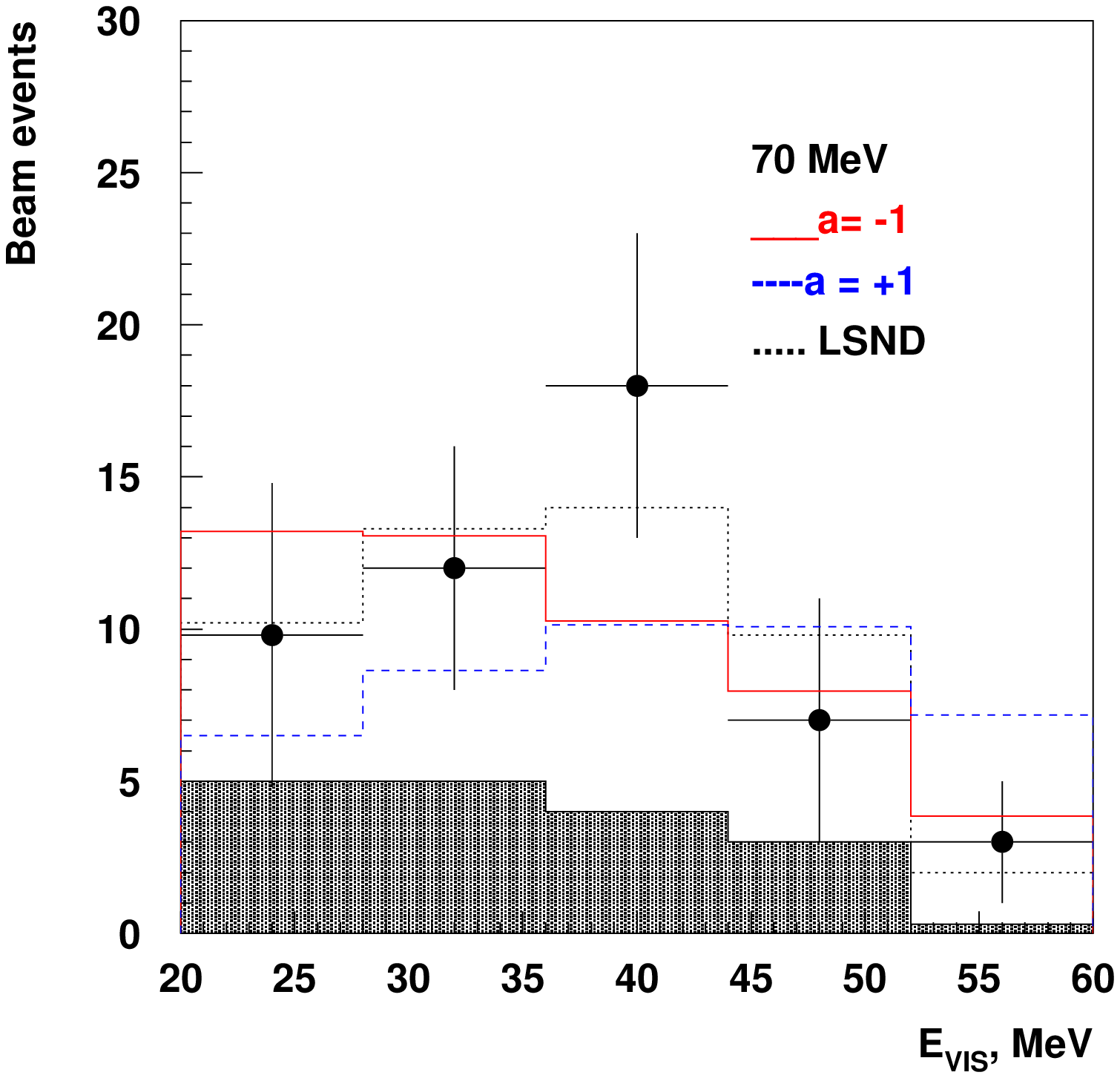}}}
\caption{
Distributions of the excess events reconstructed as $\overline{\nu}_e \rm{CC}$ events  in the LSND detector
as a function of visible energy $E_{vis}$ for $R_\gamma> 10$
from the 1993-1998 data sample (dots), and from a combination of 
the  $\nuh$ decay plus expected neutrino background  calculated for $a= -1$ (solid line) and $a= +1$ (dashed line),  
 $\nu_h$ masses of 40 and 70 MeV  shown in the plots,
the mixing strength $|U_{\mu h}|^2 =3\times 10^{-3}$, and the $\nu_h$ lifetime $\tau_{\nu_h}= 10^{-9}$ s. A
 combination of neutrino background plus neutrino oscillations at low $\Delta m^2$ (dotted line) and
the expected distribution from neutrino background 
(shaded, from Ref. \cite{lsndfin}) are also shown.}
\label{evisdmn}
\end{figure*}

\begin{figure*}
 \resizebox{17cm}{7cm}{\includegraphics{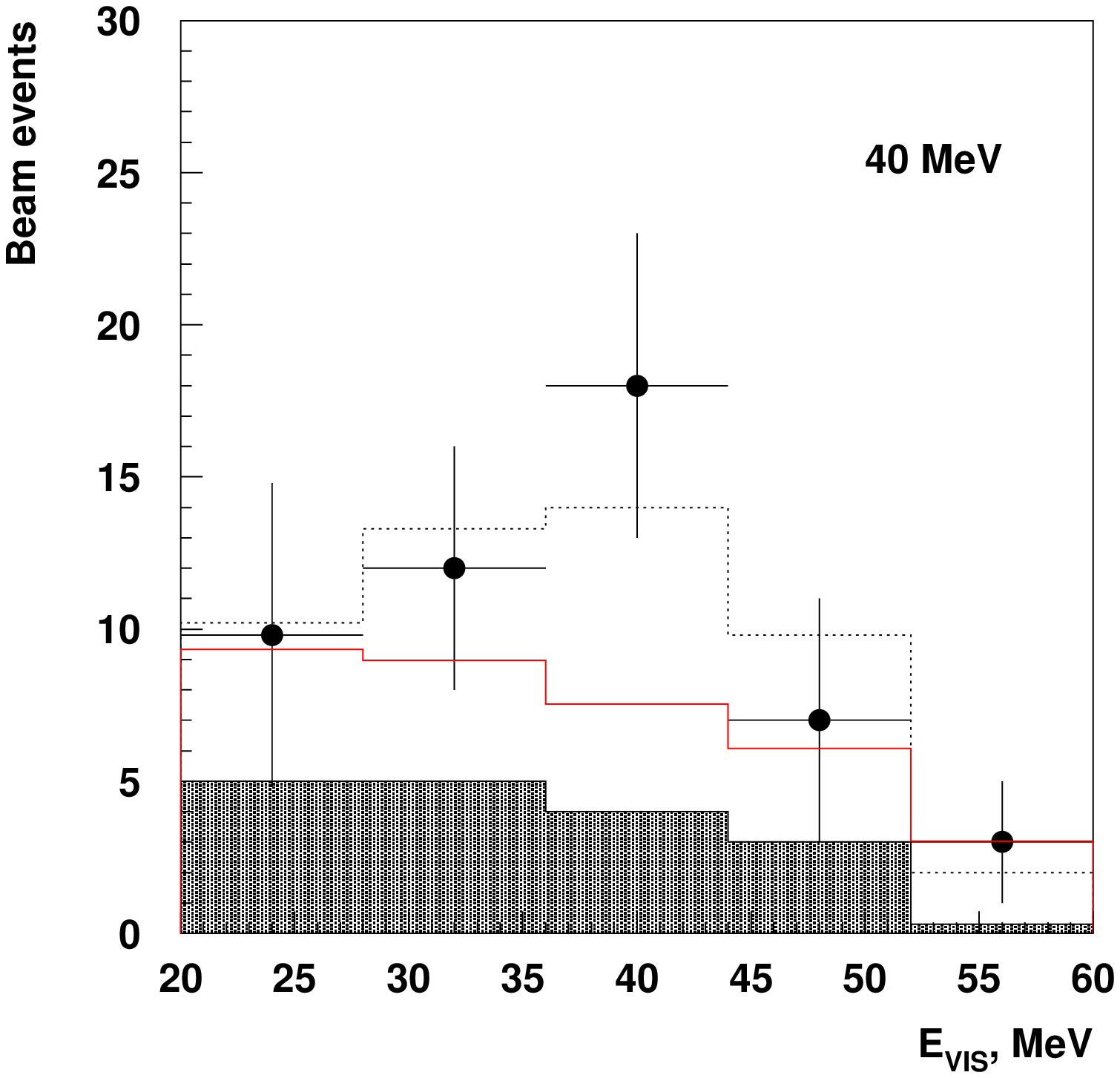}{\hspace{-3cm}\includegraphics{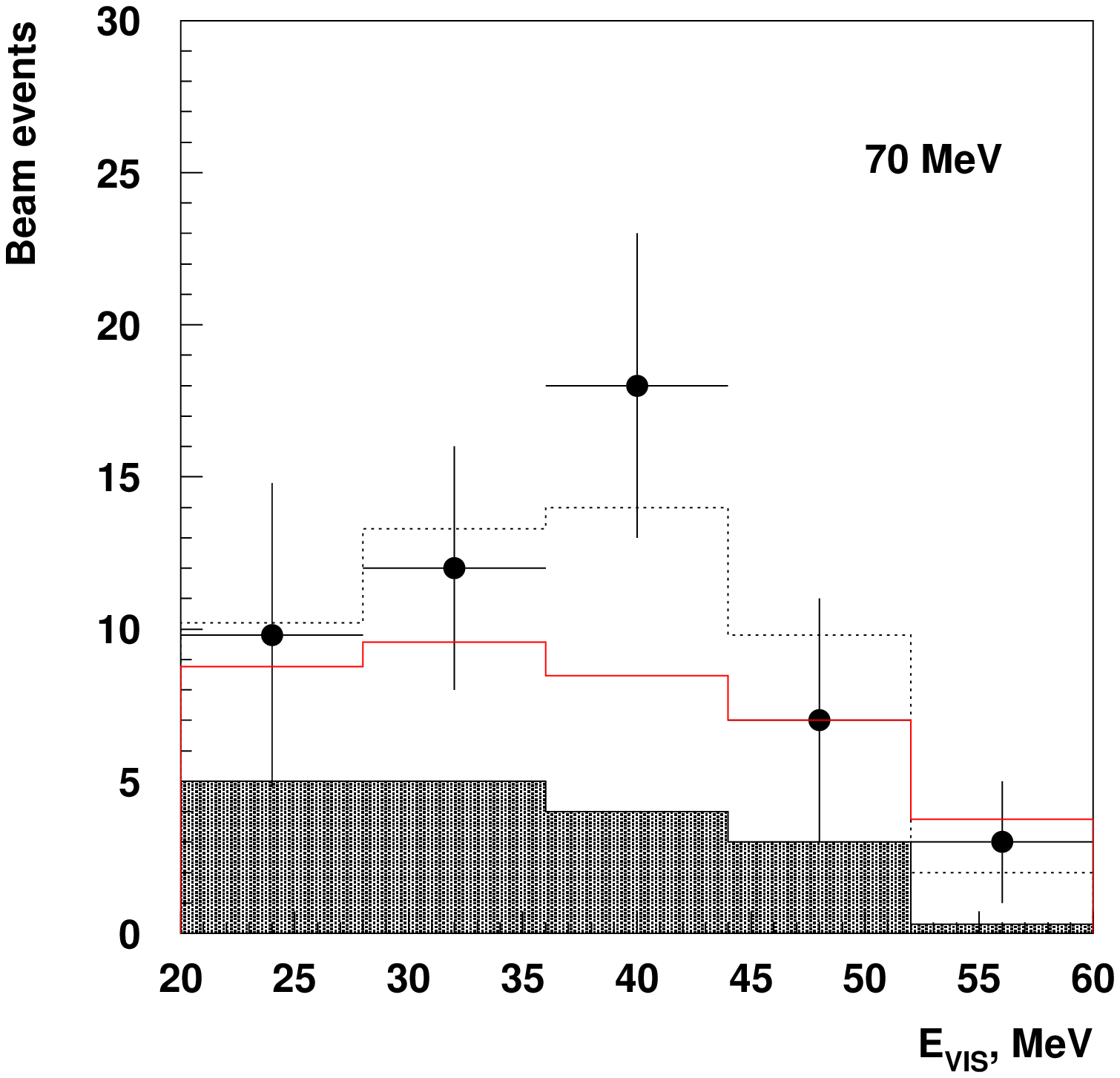}}}
\caption{ The same as Fig. \ref{evisdmn} for the case $a=0$.}
\label{evismn}
\end{figure*}

Having this reasonable agreement in mind, we have performed calculations of  
the LSND $\nu_\mu$ flux-averaged distributions of the kinetic energy of knockout neutrons 
produced in the reaction \eqref{reaction}. The results are shown   
in Fig.\ref{neutron} for the $\nu_h$ masses of 40 and 80 MeV.
The average energies of the recoil neutrons are,  respectively,  14 and 16 MeV.
To decrease this energy to the typical  energy of neutrons from the reaction  
$\overline{\nu}_e p \to e^+ n$  ($< 5$ MeV)
takes about $n_{coll} \simeq 6$ neutron collisions in mineral oil. Taking into account that the average collision length is $L_{coll} \lesssim  10$ cm results in 
a displacement of the neutron from the primary vertex of the order 
$\Delta r \simeq L_{coll} \sqrt{n_{coll}}\simeq 25$ cm, 
which is significantly less 
than the value of $\simeq 70$ cm for the neutrons from the reaction
$\overline{\nu}_e p \to e^+ n$, defined mainly  by  the 
reconstruction accuracy \cite{lsndfin}. Thus, one would expect no significant 
contribution to  the likelihood ratio $R_{\gamma}$ due to  this effect.
The energy decreasing time $\Delta t \simeq \Delta r/\beta c \simeq 5$ ns
 is also much less than the neutron capture time of 186 $\mu$s. 

As discussed above, the neutrons from the higher energy tail of the  
distribution shown in Fig. \ref{neutron} can knock out an 
additional neutron(s), resulting in the observation in LSND of a  number of 
events with more than one   correlated capture $\gamma$'s. 
To estimate this number, we use the results of the measurement of the 
neutron yield from the 70 MeV proton beam collisions with a thick
graphite target \cite{graphite}, assuming that this yield 
is approximatelly the same for the neutron induced reaction of the 
same energy. The measured number of neutrons per 
proton is found to be $\simeq  0.06$. The neutron 
 energy threshold to produce 
a secondary neutron in collisions with a $^{12}C$ nucleus is about 20 MeV.
 The fraction of
neutrons with energy greater than 20 MeV in the distribution shown in Fig.\ref{neutron}, is  
$ \simeq 20-25 \%$ depending on the value of the 
$\nu_h$ mass. Taking this into account, we find that the total 
 fraction of events with more than one correlated gamma from the 
flux-averaged reaction \eqref{reaction} is $\lesssim 2\%$, 
which can be neglected. 

  Note that in our calculations we  overestimate the fraction of high energy neutrons.
The calculations of the reaction $\nu_\mu ^{12}C \to \mu_\nu n ^{11}C$  performed in 
Ref.\cite{garvey} show that  the cross section and recoil neutron energy spectrum 
are essentially dependent on the details of the $ ^{12}C$ level structure for neutron energies below 30 MeV.
The consideration of such an effect,  including the rescattering of outgoing neutrons,
is quite important for the emission spectrum, as it 
will shift  the spectrum to lower neutron energies. 
Therefore, 
we may assume, that the fraction of events with $>1$ correlated $\gamma$ 
is even  less than 2\%, or 0.6 events.
This number should be compared with the background of $\lesssim 5$ such events 
expected  in the LSND experiment  at the 2 $\sigma$ level  \cite{lsndfin}.
 Thus, our estimate is compatible with the (approximately zero) 
 number of events with $>1$ correlated $\gamma$'s observed by LSND for the full $20 < E_{vis} < 60 $ MeV energy region.

\begin{figure*}
\resizebox{17cm}{7cm}{\hspace{-2cm}\includegraphics{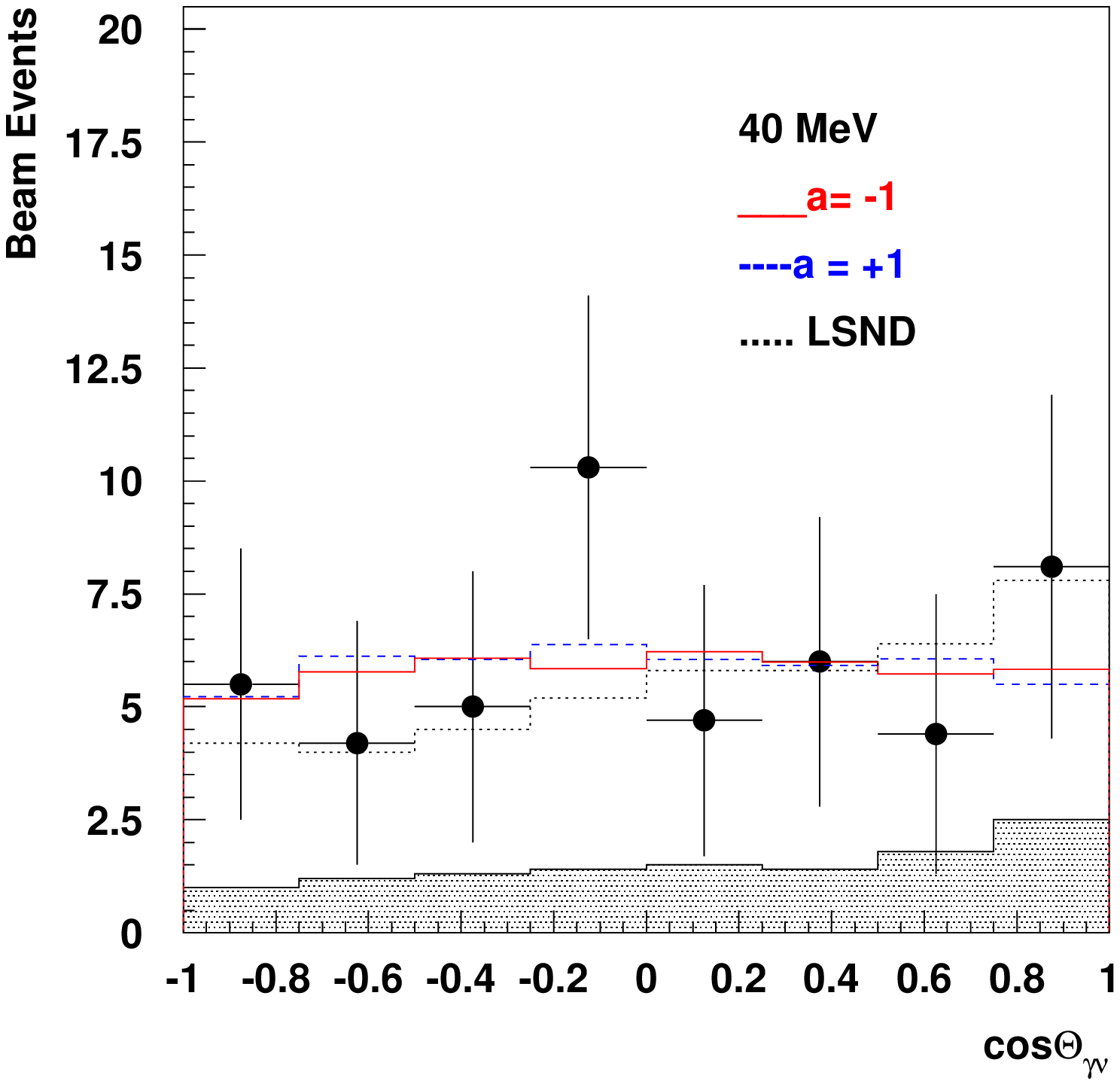}{\hspace{-3cm}\includegraphics{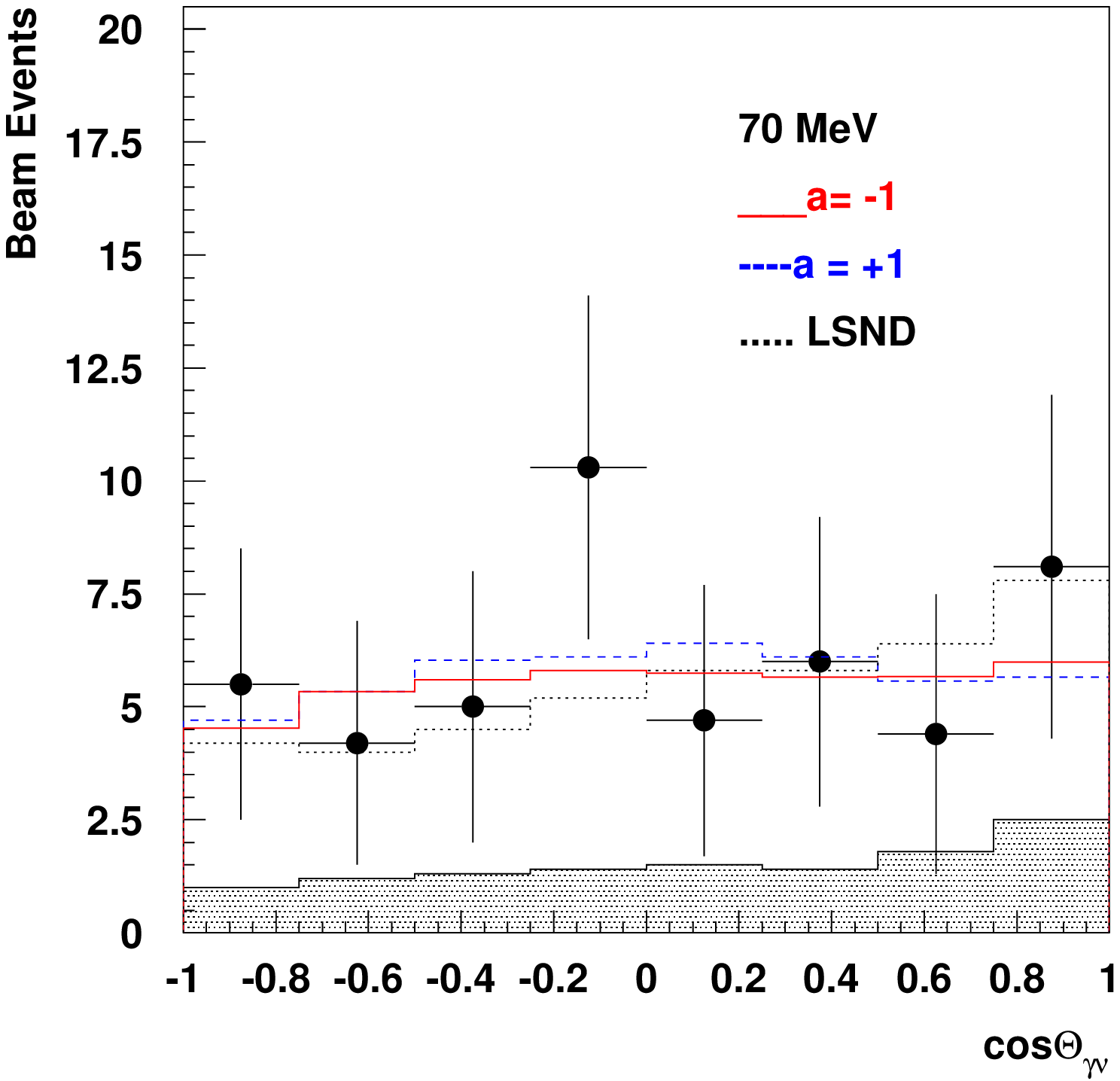}}}
\caption{
Distributions of the excess events reconstructed as $\overline{\nu}_e \rm{CC}$ events with $36< E_{vis} < 60$ MeV in the LSND detector as a function of $cos\Theta$   from the 1993-1998 data sample (dots), and from
 a combination of the 
 the  $\nu_h \to \gamma \nu $ decay plus expected neutrino background  calculated for $a= -1$ (solid line) and $a= +1$ (dashed line),  
 $\nu_h$ masses of 40 and 70 MeV  shown in the plots,
the mixing strength $|U_{\mu h}|^2 =3\times 10^{-3}$, and the $\nu_h$ lifetime $\tau_{\nu_h}= 10^{-9}$ s. A
 combination of neutrino background plus neutrino oscillations at low $\Delta m^2$ (dotted line) and
the expected distribution from neutrino background 
(shaded, from Ref. \cite{lsndfin}) are also shown.}
\label{cosdmn}
\end{figure*}
\begin{figure*}
 \resizebox{17cm}{7cm}{\hspace{-2cm}\includegraphics{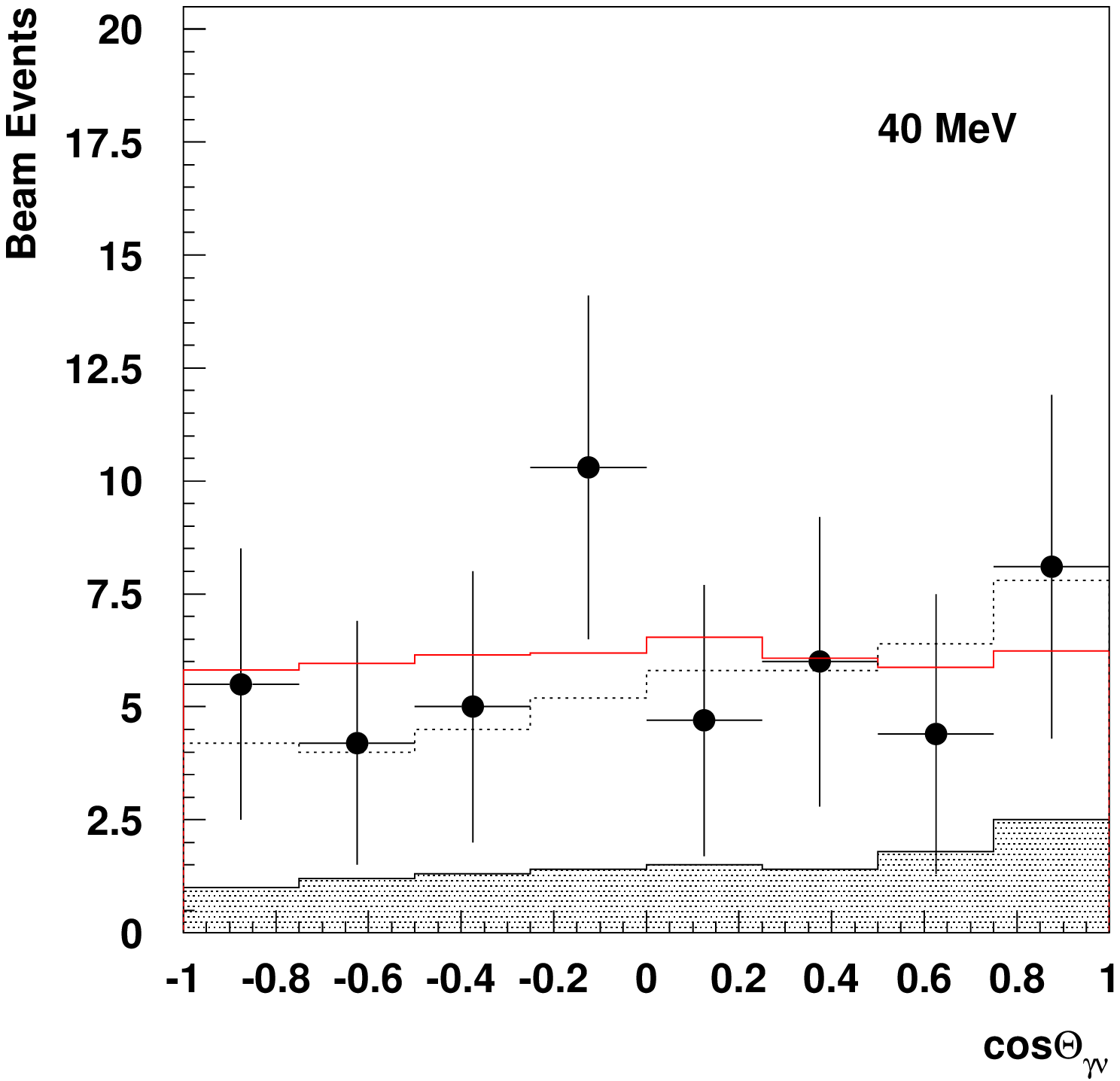}{\hspace{-3cm}\includegraphics{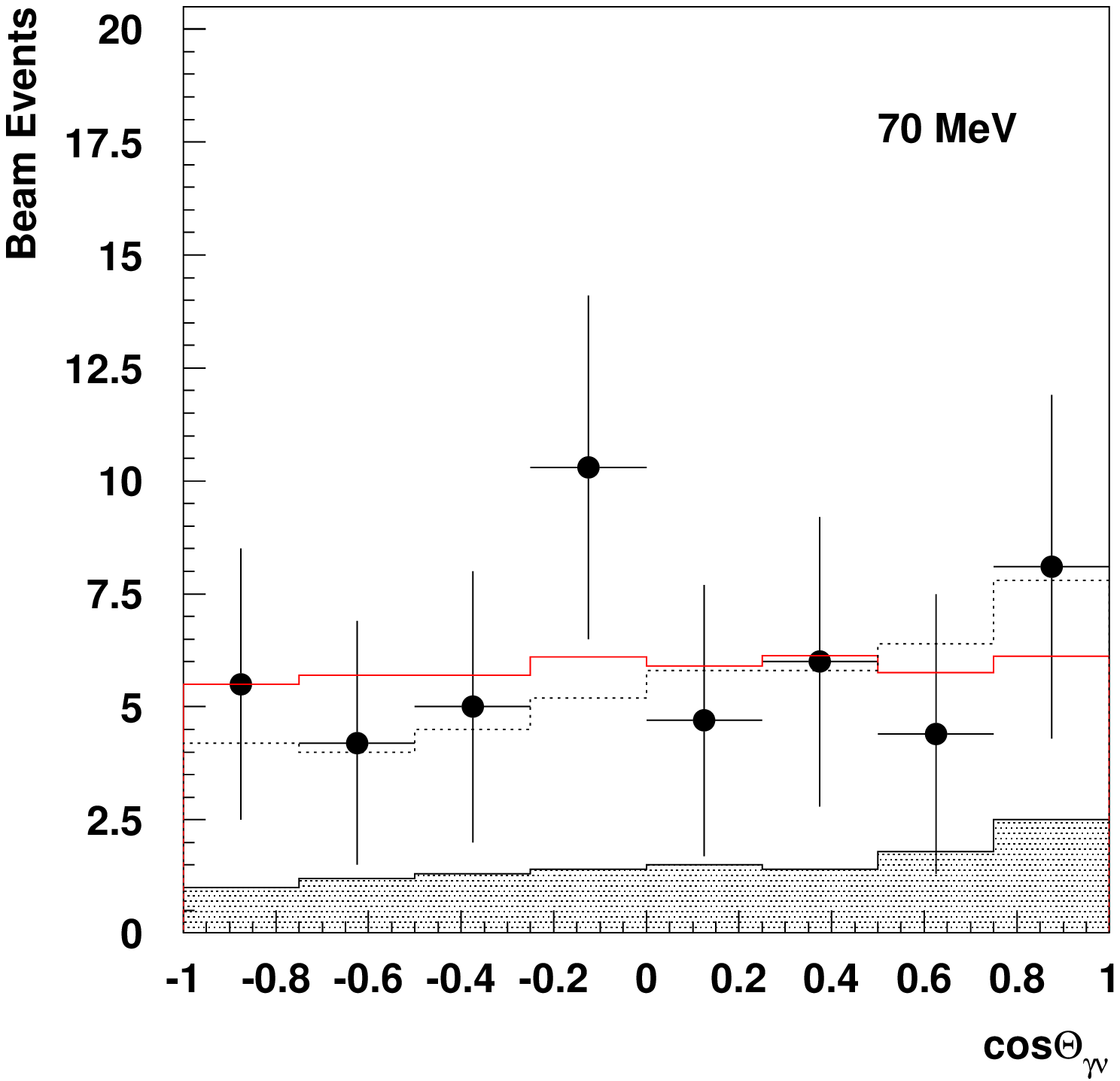}}}
\caption{ The same as in Fig. \ref{cosdmn} for the case  $a=0$.}
\label{cosmn}
\end{figure*}

A cross-check  for the fraction of neutron emitted in the reaction 
\eqref{reaction}  can be obtained from the comparison of results obtained  
for the inclusive reaction
$^{12}C(\overline{\nu}_\mu,\mu^+ n)X$ in this work, in Ref.\cite{kolbe}, and
 by the LSND Collaboration \cite{lsndmu}.
In this reaction the presence of a muon and a neutron is established by detection of the Cherenkov ring and of  the
$\gamma$ ray from the neutron's capture  as  described in detail in Ref. \cite{lsndgam}. 
 We found that  the fraction of events accompanied by the emission of a 
recoil neutron  is $\simeq$ 81\%.
This number has to be compared with the fractions of 79 \% predicted  from the calculations in  \cite{kolbe},  and $(79.6\pm 40.0)$\% obtained by  LSND \cite{lsndmu}. The agreement is quite good. 

In Fig.\ref{spectra} the energy distributions of photons from the radiative decay of heavy 
neutrinos produced in the reaction $\nu_\mu ^{12}C \to \nu_\mu  n  X $ 
calculated for  several $\nu_h$ masses from  40 to 80 MeV and for three of the most interesting 
cases   of the decay asymmetry parameter,
$a=\pm 1$  and  $a= 0$,  are shown. The calculations are performed  with no photon efficiency and no binding  energy 
corrections included. 
 One can see that for the  Dirac case with $a=-1$, 
 the simulated  events are mainly distributed in the narrow region $0\lesssim E_{\gamma} \lesssim 60$ MeV.
 The fraction of photons in this  region varies from 0.86 to 0.77 for a  $\nu_h$ mass from 40 to 80 MeV, respectively.
 The remaining events are distributed over the region  $60\lesssim E_{\gamma} \lesssim 150$ 
MeV,  where they can  be hidden by the low statistics.   
 For the $a=0$ and $a= +1$ cases, the fraction of photon events in the region 
 $0\lesssim E_{\gamma} \lesssim 60$ MeV varies from 0.71 to 0.57; and from 0.54 to 0.46, 
 respectively. 
Further, we  will discuss  mainly the cases $a=-1$ and $a=0$,
because  for  the case $a=+1$ the  fraction of events above 60 MeV  
is too high compared to the LSND observations. 

The  visible energy distributions expected from a  combination of
the  $\nuh$ events plus  neutrino background in the LSND
detector \cite{lsndfin} are shown in Figs. \ref{evisdmn},\ref{evismn}
for the energy range  $20\lesssim E_{vis} \lesssim 60$ MeV.
 The spectra are calculated for $m_{\nu_h}=$ 40 and 70 MeV, 
 $|U_{\mu h}|^2 = 3\times 10^{-3}$, and the $\nu_h$ lifetime $\tau_{\nu_h}= 10^{-9}$ s
  by taking into account the decay 
 photon efficiency and the neutron binding energy corrections. The photon efficiency has been estimated 
by a simple Monte Carlo calculation as a fraction of 
photons with  energies $20 < E_{\gamma} < 60$ MeV in the detector fiducial volume times  the  detection efficiency,
which  is taken to be  essentially constant, $\epsilon_\gamma \simeq 0.4$, for this energy range. 
The distribution expected from a 
 combination of  neutrino background plus neutrino oscillations at low $\Delta m^2$  \cite{lsndfin} is also  shown for comparison.
The  clean experimental sample of the oscillation candidate events shown in 
 Fig.'s \ref{evisdmn},\ref{evismn} is obtained by enforcing 
strongly correlated gammas with the cut $R_\gamma > 10$. In this 
case the beam on-off excess is $49.1\pm 9.4$ events while the estimated 
neutrino background is only $16.9 \pm 2.3$  resulting in 
a total excess of $32.2 \pm 9.4 \pm 2.3$ events   \cite{lsndfin}.
The analogous distributions for $cos\Theta_{\gamma \nu}$, the cosine of the angle between the incident 
neutrino beam and decay photon momenta, for events with $36\lesssim E_{vis} \lesssim 60$ MeV are shown in 
Figs. \ref{cosdmn}, \ref{cosmn}.
The distributions are obtained assuming  that the energy deposited by the
decay  photon  is misreconstructed  as the energy from a
 single electron track.
 
Simulations are in reasonable agreement with the experimental distributions.  
For instance, for the distribution
 shown in Fig.\ref{evisdmn} for $m_{\nu_h} =40$ MeV, the comparison with 
the LSND data yields a $\chi^2$ of 3.6 for 3 DF, corresponding to 34\%  C.L. 
The best fit results suggest that the $\nu_h$ mass is in the region $ 10 \lesssim m_{\nu_h} \lesssim 90$ MeV
and the lifetime is $\tau_{\nu_h} \lesssim 10^{-8}$ s.  However, to avoid the production of 
$\nu_h$'s in the KARMEN experiment, the low mass limit is set to 40 MeV. The 
mass upper bound is set to 80 MeV 
because for higher  masses the production of $\nu_h$ in LSND is suppressed by the 
phase space factor.
The simulations showed that the shape of the $ E_{vis}$  
distribution  is sensitive to the choice of the $\nu_h$ mass : the   
higher the mass, the harder the visible energy spectrum.  

Before the calculation of the required mixing strength $|U_{\mu h}|^2$, 
let us  estimate, for a  cross-check,
the number of events expected for $\overline{\nu}_\mu \to \overline{\nu}_e$ oscillations followed by 
$\overline{\nu}_e p \to e^+ n$ scattering in the LSND detector. This number could be estimated as 
\begin{equation} 
\Delta N_{\overline{\nu}_\mu \to \overline{\nu}_e} \simeq A \Phi_{\overline{\nu}_\mu} P_{osc} \sigma_{\overline{\nu}_e} f_e \epsilon_e 
\label{numbere}
\end{equation}
where $A = 7.4\times 10^{30}$ is the number of free protons in the LSND fiducial volume,
$\Phi_{\overline{\nu}_\mu}$ is the neutrino flux $1.26 \times 10^{14}~ \nu / cm^2$ (see Table 1), 
 $P_{osc}$ is the $\overline{\nu}_\mu \to \overline{\nu}_e$ oscillation probability averaged over the incident 
neutrino energy,
 $\sigma_{\overline{\nu}_e}= 0.95 \times 10^{-40}~šcm^2$ is the cross section averaged  over the entire energy range,
$f_e \simeq 0.89$ is the fraction of events in the energy range $20 < E < 60$ MeV, and $\epsilon = 0.42 $ is  the average  positron reconstruction efficiency \cite{lsndfin}.
Using the above values, we found that the LSND  experiment should detect an excess of 
$\Delta N_{\overline{\nu}_\mu \to \overline{\nu}_e} \simeq 70$ events if the 
oscillation probability is $P_{osc} \simeq 2.6\times 10^{-3}$. This value is in 
a good agreement with the number of events $87.9 \pm 22.4 \pm 6.0$ quoted in Ref.\cite{lsndfin}.

Consider now the case of heavy neutrino. 
 The estimate of the mixing parameter $|U_{\mu h}|^2$ was 
performed by using the following relations.
For a given flux of muon neutrinos, $\Phi_{\nu_\mu}$,  the expected number  
 of the $\nuh$ decays events  in the LSND  detector is given by
\begin{equation} 
\Delta N_{\nuh} \simeq  A \int\Phi_{\nu_\mu} \sigma_{\nu_\mu} |U_{\mu h}|^2 f_\gamma f_n f_{phs}P_{dec} P_{abs}\epsilon_\gamma dE  
\label{numberg}
\end{equation}
where  $\Delta N_{\nuh}= 32.2\pm 9.7$ is the number of excess events 
 observed in the 1993-1998 data sample (with errors combined in quadrature),
$A=3.7\times 10^{30}$ is the number of carbon nuclear in the LSND fiducial volume, 
$\sigma_{\nu_\mu n}$ is the 
 cross-section for the reaction $^{12}C(\nu_\mu, \nu_\mu n)X$ 
 with the emission of a recoil neutron for the massless case, $f_\gamma \simeq 0.5-0.76$ is the fraction of events in the energy range $20 < E < 60$ MeV, 
$f_n \simeq 0.4-0.8$ is the 
fraction of events with the emission of a recoil neutron in reaction 
\eqref{reaction},   
  $P_{dec}$ is the probability for the $\nuh$ decay within the detector 
fiducial volume,  $P_{abs}$ is the probability of the decay photon absorption 
in the detector, and $\epsilon_\gamma \simeq 0.4$ is the overall efficiency 
for decay photon detection.   
In Eq.(\ref{numberg}) the number of heavy neutrinos produced is proportional to  the product of the 
$^{12}C(\nu_\mu, \nu_\mu n)X$ cross section, the mixing 
$|U_{\mu h}|^2$, and the phase space factor $f_{phs}$, which  takes into account the threshold effect due to the heavy 
neutrino mass. 
The value of the total $\nu_\mu$ DIF flux ( see Table 1), the number of 
reconstructed $\overline{\nu}_\mu \to \overline{\nu}_e$-like \cite{lsndfin} and   $\nu_\mu CC$ events \cite{lsndmu} in the detector were
 used for cross checks and normalization.
The $\nu_\mu$ flux averaged $^{12}C(\nu_\mu, \nu_\mu n)X$ cross section is
$<\sigma(E)_{\nu_\mu n}> \simeq 16\times 10^{-40}$ cm$^2$. 
The probability  of the heavy neutrino to decay radiatively 
in the fiducial volume at a distance $r$ from the primary vertex is given by 
\begin{equation}
P_{dec}=[1-\exp (\frac{-rm_{\nu_h}}{p_{\nu_h} \tau_{\nu_h}})]\frac{ \Gamma (\nuh)}{\Gamma_{tot}} 
\label{probdec}
\end{equation}
where the last term is the branching fraction $Br(\nuh) \simeq 1$. 
Assuming that almost all $\nuh$ decays occur inside 
the fiducial volume of the detector, we estimate the $\mix$  
to be in the range 
\begin{equation}
|U_{\mu h}|^2 \simeq (3-9)\times 10^{-3}.
\label{lsndmix}
\end{equation}
This result is mainly defined by the uncertainty on  
the number of excess events and  is valid 
for  the mass region 
\begin{equation}
40 \lesssim m_{\nu_h} \lesssim  80~MeV
\label{lsndmass}
\end{equation}
and the $\nu_h$ lifetime
\begin{equation}
\tau_{\nu_h} \lesssim 10^{-8} s
\label{lsndtau}
\end{equation}

\subsection{The LSND signal of $\nu_{\mu} \to \nu_e$ oscillations}

During the first three years of LSND data taking, the target area of the LANSCE 
accelerator consisted of a 30 cm long water target located $\simeq$1 m upstream of the 
beam stop. This configuration enhanced the probability of pion decay in flight, 
allowing LSND to search for $\nu_\mu - \nu_e$  oscillations using $\nu_\mu$ with energy above 60 
MeV. In this case, one expects to observe an excess of events from the reaction 
$\nu_e ^{12}C\to e^- X$ above the expected backgrounds. 
This reaction has only one signature (a prompt 
signal),  but the higher energy and  the longer track of the events 
allow good electron identification and measuring its direction.  
In this search  LSND has observed 40 events to be compared with 
$12.3 \pm 0.9$ events from cosmic ray background and $9.6 \pm 1.9$ events from machine-related 
(neutrino-induced) processes  \cite{lsndmue}. 
The excess of  $(18.1 \pm 6.6)$ events 
corresponds to a $\nu_\mu - \nu_e$ oscillation probability of $(2.6 \pm 1.0)\times 10^{-3}$ 
, consistent with the value found from the study of the $\overline{\nu}_e p \to  e^+ n$ reaction below 60 MeV. 

The number of the $\nuh$ events that would be observed by LSND after 
applying the high energy cut $E> 60$ MeV, is about   $3-10$ 
events depending 
on the $\nu_h$ mass and mixing obtained from the combined analysis, as shown below in Sec.V.
For example, out of 10 events,  $\simeq$ 5 ($\simeq$ 2) events are from the 
reaction $\nu_\mu ^{12}\rm{C}\to  \nu_h X$
occurring on protons (neutrons) of the $^{12}\rm{C}$ nucleus,  which is not expected to produce free neutrons, and 
$\simeq$  3  events are from the reaction \eqref{reaction}
with a recoil neutron production, which is identified by the presence of the 2.2 MeV 
photon from the capture reaction.   
Thus,  the ratio of the number of excess events 
with and without photon tag is $\simeq$  3 : 7, which is in 
 agreement within errors with the observed numbers $\simeq (4\pm 2.5):(15\pm 5)$ of events 
in the LSND experiment \cite{lsndmue}.   

\section{The  MiniBooNe anomalies}

The MiniBooNE detector is described in detail in Ref. \cite{mbdet}. 
It uses an almost pure  $\nu_\mu$ beam originated from
the  $\pi^{+}$ decays in flight, which are  generated 
  by 8 GeV protons from the FNAL booster.
The detector consists of a target, which is  a
12.2 m diameter sphere filled with 800 t of mineral oil, 
surrounded by an outer  veto region. 
The Cherenkov light rings generated by  muon, electron and converted photon 
tracks are used for the reconstruction of the events. The resolutions 
reached on the vertex position, the outgoing particle 
direction and the visible energy are 20 cm, 4$^o$,  and 
12\%, respectively for $CCQE$ electrons \cite{mbreco}.
The $\nu_\mu$ beam is peaked  
around $\sim 600$ MeV, has a mean energy of  
 $\sim 800$ MeV and a high energy tail up to $\sim$ 3 GeV \cite{mbbeam}.

Below, we consider  the MiniBooNE anomalous event excess observed in $\nu_\mu$ and 
$\numub$ data and the interpretation  of these results in terms of the
heavy neutrino decay.  
\begin{figure}[h]
\begin{center}
    \resizebox{9cm}{!}{\includegraphics{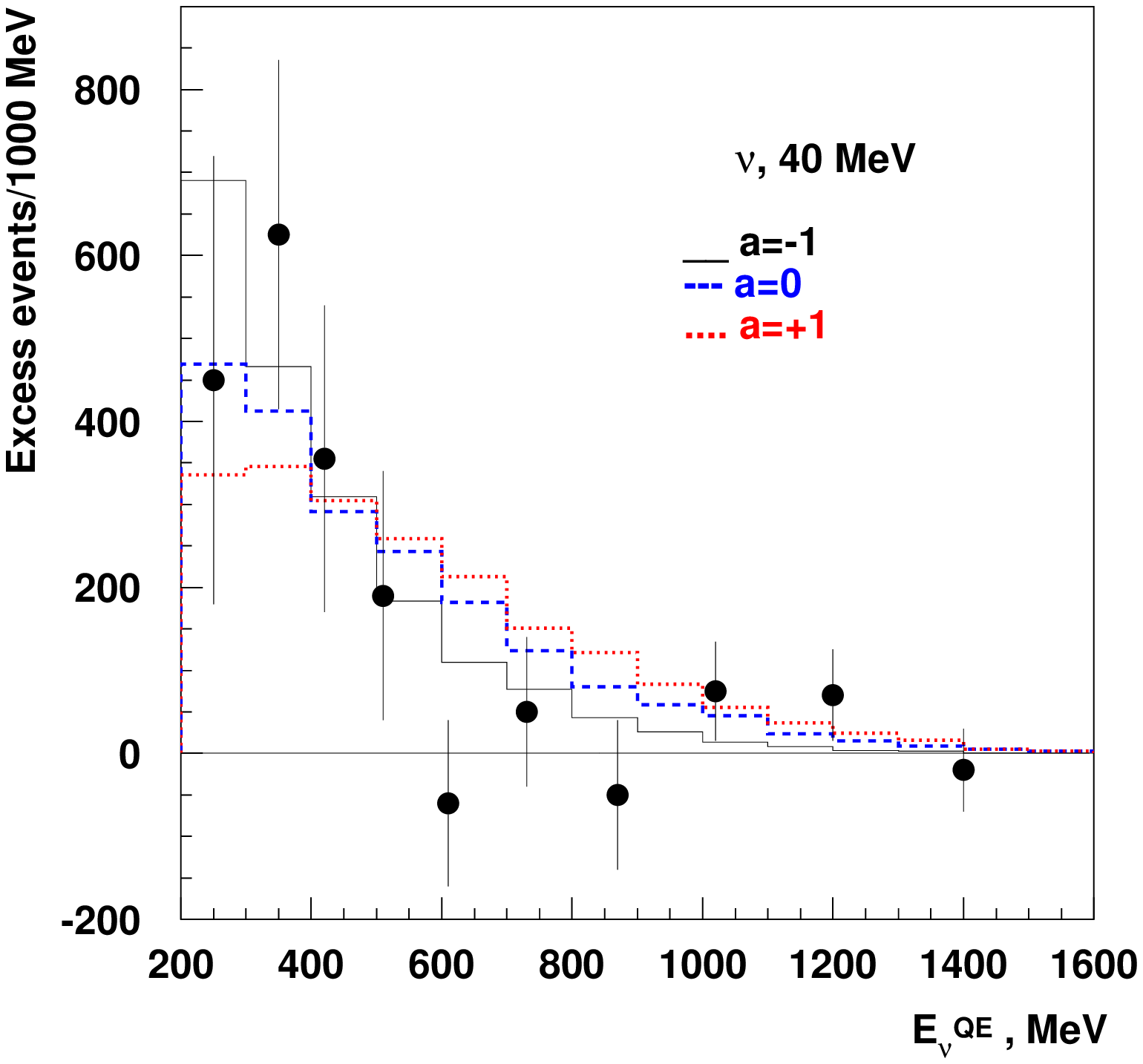}}
     \caption{ Distributions of the excess events in the MiniBooNE detector from the $\nuh$ decay
 reconstructed as $\nu_e CC$ events as a function of $E^{QE}_\nu$ for 
$|U_{\mu h}|^2= 3 \times 10^{-3}$,  $m_{\nu_h} = 40$ MeV,
and $\tau_{\nu_h}= 10^{-9}$ s, and  for different values of the asymmetry parameter $a$.
The dots are experimental points
for the excess events in the MiniBooNE detector. Error bars 
include both statistical and systematic errors \cite{minibnu2}. 
The  comparison of the distributions with the experimental data
yields a $\chi^2$ of 7.1 ($a=-1$), 9.3 ($a=0$), and 10.1 ($a=+1$) for 8 DF.} 
\label{eque40nu}
\end{center}
\end{figure}
\begin{figure}[h]
\begin{center}
    \resizebox{9cm}{!}{\includegraphics{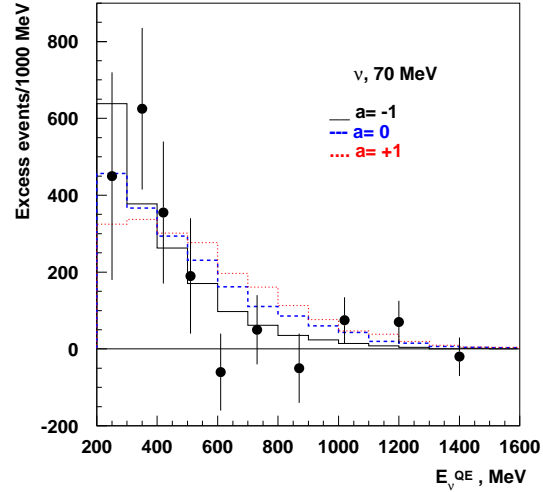}}
     \caption{Same as Fig.\ref{eque40nu} for the 70 MeV $\nu_h$.
The  comparison of the distributions with the experimental data yields a $\chi^2$ of 7.5 ($a=-1$),  9.2 ($a=0$), and 10.3 ($a=+1$) for 8 DF.}
\label{eque70nu}
\end{center}
\end{figure}

\subsection{Excess of events in $\nu_{\mu}$ data}

An excess of $\Delta N=$128.8$\pm 20.4 \pm 38.3$ electronlike events 
 has been observed in the data   
accumulated with $6.64\times 10^{20}$ protons on target.
For the following discussion several distinctive features of the excess events
are of importance \cite{minibnu2}: a) the excess is  observed for single track events, 
originating either from an 
electron, or from  a photon converted into a $\pair$ pair with a typical 
opening angle $\simeq m_e/E_{\pair}< 1^o$ (for $E_{\pair} > 200$ MeV), which is 
too small to be resolved into two separate Cherenkov rings (here,
$m_e, E_{\pair}$ are the electron mass and the $\pair$ pair energy);
b) the reconstructed neutrino 
energy is in the range $200 < E^{QE}_\nu < 475$ MeV, while 
there is no significant excess for the region $E^{QE}_\nu > 475$ MeV
(the variable 
 $E^{QE}_\nu$  is calculated under 
the assumption that the observed  electron track originates from $\nu_e QE$
interaction);  c) the visible 
energy $E_{vis}$   is in the  narrow  region 
$200\lesssim  E_{vis} \lesssim 400$ MeV for events with $E^{QE}_\nu > 200$ MeV; and
d) the angular distribution of 
the excess events with respect to the incident neutrino direction 
is wide and consistent with the shape expected 
from $\nu_e CC$ interactions. 
\begin{figure}[h]
\begin{center}
    \resizebox{9cm}{!}{\includegraphics{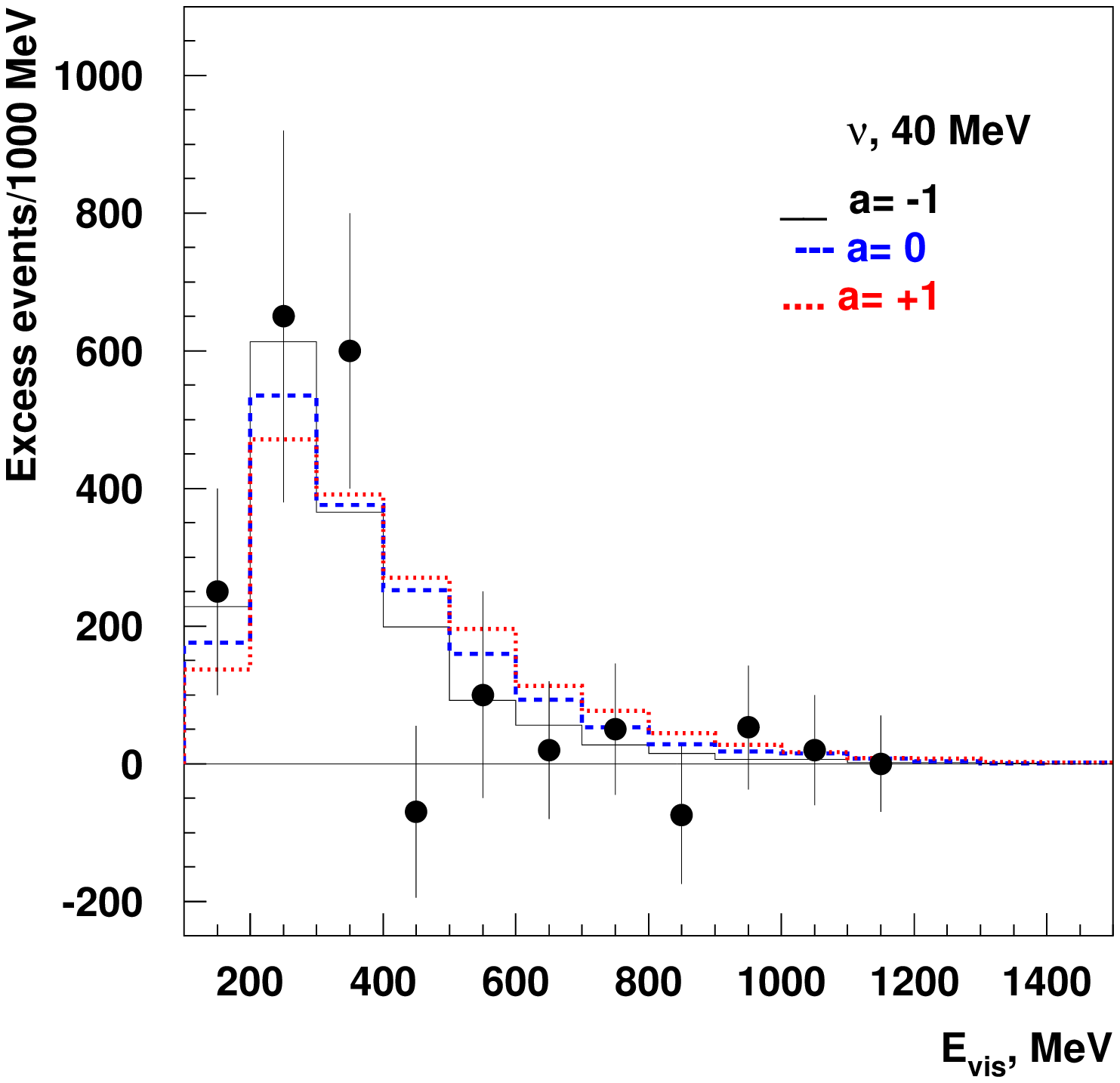}}
     \caption{ Distributions of the excess events in the MiniBooNE detector from the $\nuh$ decay
  reconstructed as $\nu_e CC$ events as a function of $E_{vis}$ for 
$E^{QE}_\nu > 200$ MeV, 
$|U_{\mu h}|^2= 3 \times 10^{-3}$,  $m_{\nu_h} = 40$ MeV
and $\tau_{\nu_h}= 10^{-9}$ s, and for  different values of the asymmetry parameter $a$.
The dots are experimental points
for the excess events in the MiniBooNE detector. Error bars 
include both statistical and systematic errors \cite{minibnu2}. 
The  comparison of the distributions with the experimental data
yields a $\chi^2$ of 7.2 ($a=-1$), 9.4 ($a=0$), and 10.3 ($a=+1$) for 8 DF.} 
\label{evis40nu}
\end{center}
\end{figure}
\begin{figure}[h]
\begin{center}
    \resizebox{9cm}{!}{\includegraphics{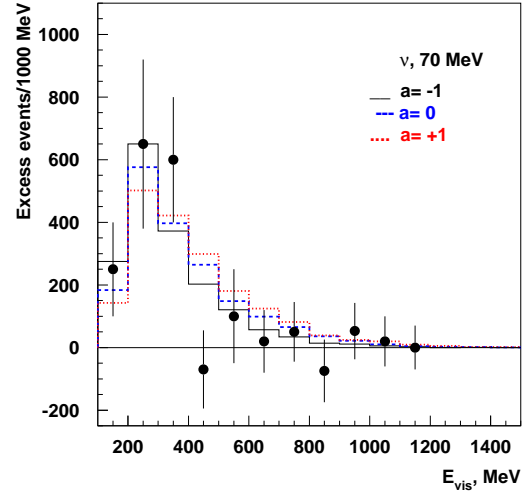}}
     \caption{Same as Fig.\ref{evis40nu} for the 70 MeV $\nu_h$.
The  comparison of the distributions with the experimental data
yields a $\chi^2$ of 6.8 ($a=-1$),  8.8 ($a=0$), and 9.7 ($a=+1$) for 8 DF.}
\label{evis70nu}
\end{center}
\end{figure}
To satisfy the criteria a)-d), we propose that  the excess events are  
originated from the decay of the heavy neutrino $\nu_h$ considered in  Sec.III. 
The $\nu_h$'s  are produced by mixing in  
$\nu_\mu$ neutral-current $QE$ interactions and  
depositing their  energy via the 
visible decay mode  $\nu_h \to \nu \gamma$,  as  
shown in Fig.\ref{diag}, with the subsequent conversion of the decay
photon into the $\pair$ pair in the MiniBooNE target. To make a quantitative estimate, 
we performed simplified simulations of the production 
and decay processes shown in Fig.\ref{diag}.
In these simulations  we used a $\nu_\mu$ energy spectrum parametrized from the  
reconstructed $\nu_\mu CCQE$ events \cite{mbbeam}. 
Since in the MiniBooNE experiment the $\nu_h$'s have higher energies and  
decay over an average distance of 
$\lesssim$ 5 m from the production vertex, the sensitivity in the LSND $\nu_h$  mass range of Eq.(\ref{lsndmass}) is 
restricted to  the   $\nu_h$ lifetimes  
\begin{equation}
\tau_{\nu_h}\lesssim  10^{-9} s,
\label{minibtau}
\end{equation}
 to be compared  with \eqref{lsndtau}.

\begin{figure}[h]
\begin{center}
    \resizebox{9cm}{!}{\includegraphics{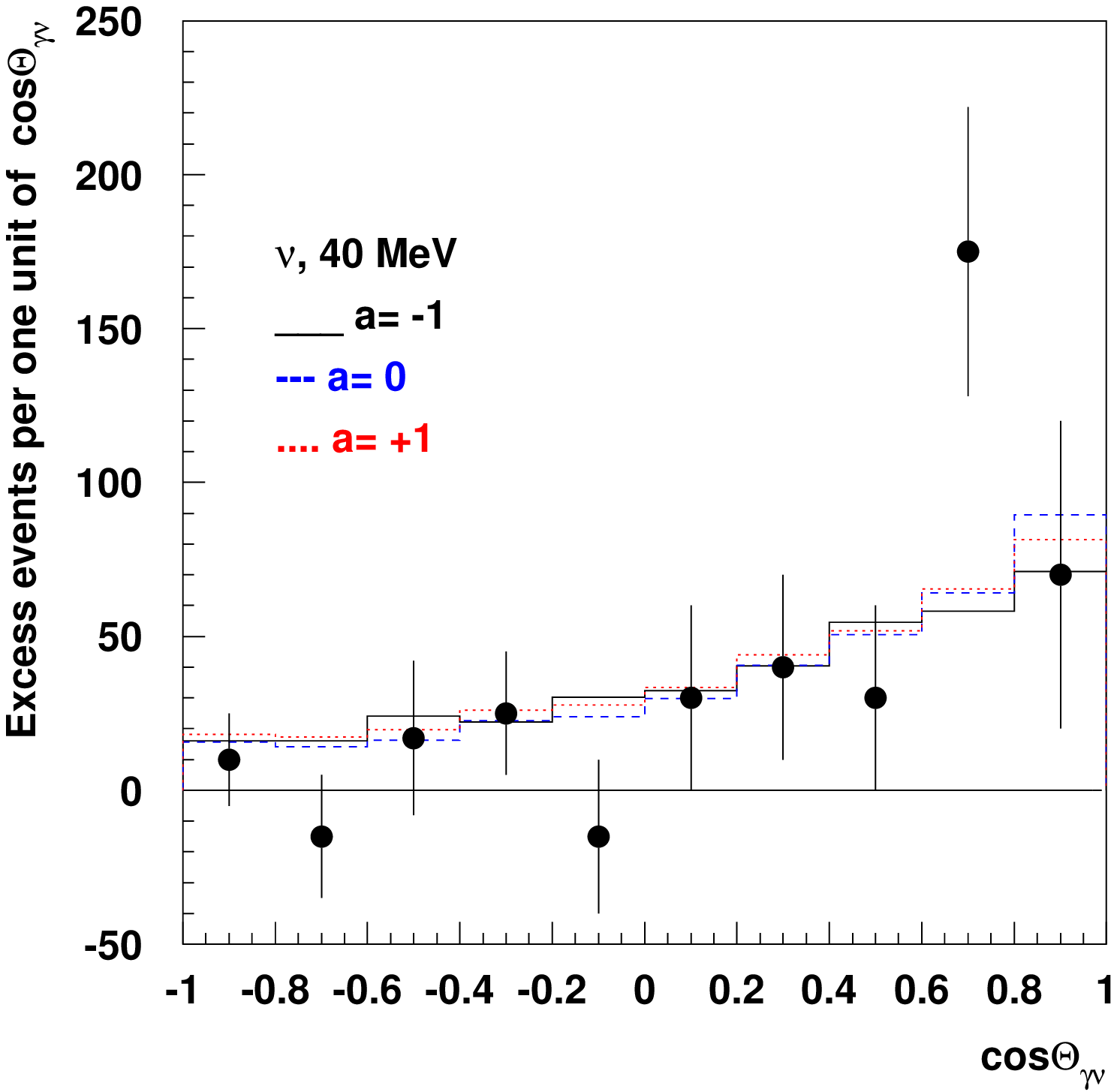}}
     \caption{ Distribution of the excess events in the MiniBooNE detector from the $\nu_h$ decay
 reconstructed as $\nu_e CC$ events 
as a function of  $\cos \Theta_{\gamma \nu}$ for 
$300< E^{QE}_\nu < 400$ MeV, $|U_{\mu h}|^2= 3 \times 10^{-3}$,  $m_{\nu_h} = 40$ MeV,
and $\tau_{\nu_h}= 10^{-9}$ s, and  for different values of the asymmetry parameter $a$. The dots are experimental points
for the excess events in the MiniBooNE detector. Error bars 
include both statistical and systematic errors \cite{minibnu2}. 
The  comparison of the distributions with the experimental data
yields a $\chi^2$ of 10.1 ($a=-1$), 9.8 ($a=0$), and 9.7 ($a=+1$) for 8 DF.}
\label{cos40nu}
\end{center}
\end{figure}

\begin{figure}[h]
\begin{center}
    \resizebox{9cm}{!}{\includegraphics{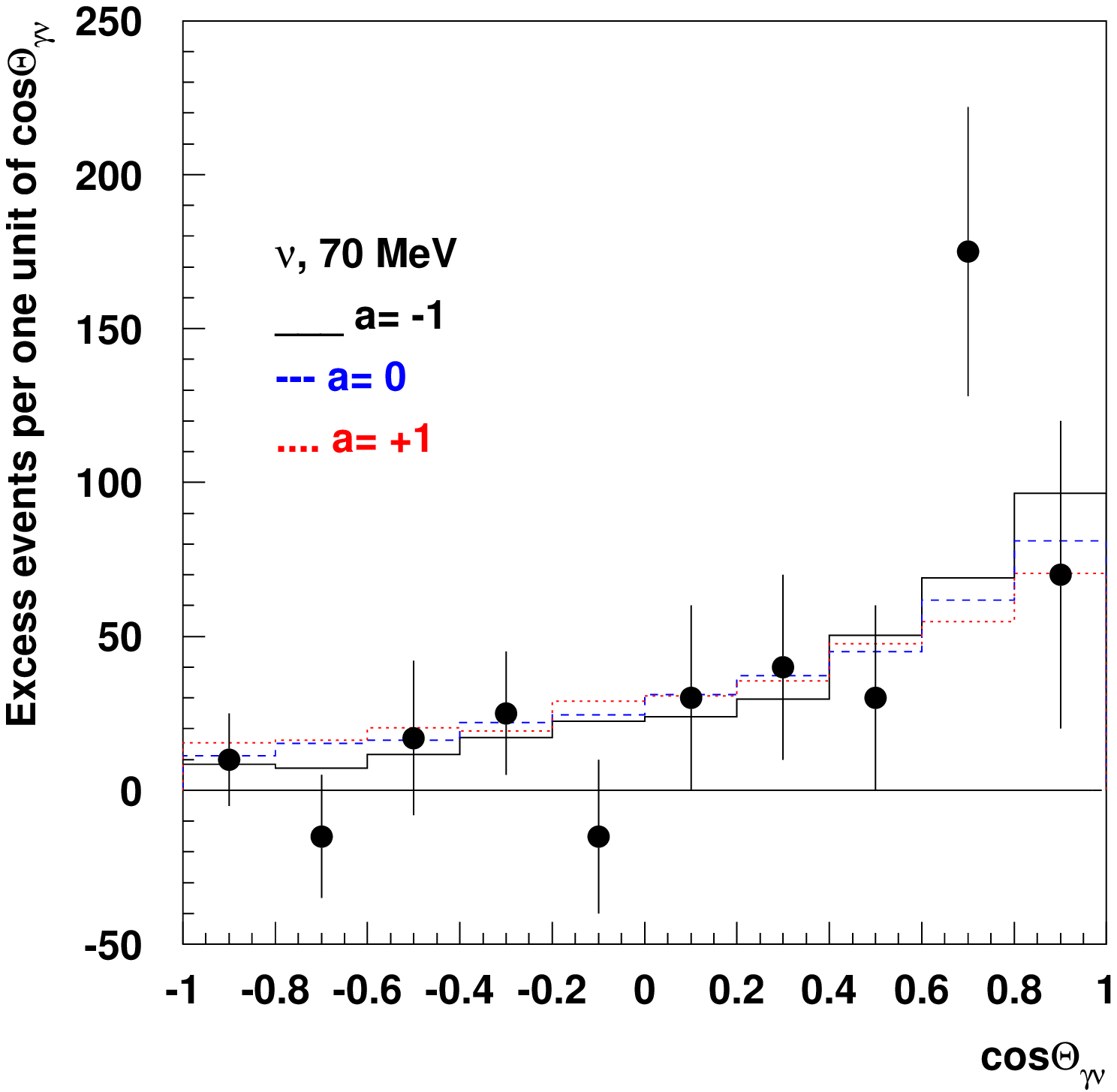}}
     \caption{ Same as Fig.\ref{cos40nu} for the 70 MeV $\nu_h$.
The  comparison of the distributions with the experimental data
yields a $\chi^2$ of 9.3 ($a=-1$),  10.1 ($a=0$), and 10.0 ($a=+1$) for 8 DF.} 
\label{cos70nu}
\end{center}
\end{figure}

In Figs.\ref{eque40nu}-\ref{cos70nu} the distributions of 
the kinematic variables $E^{QE}_\nu, E_{vis}$ and $\cos \Theta_{\gamma \nu}$ 
for the $\nuh$ events are shown 
for  $m_{\nu_h}=40$ and 70 MeV and 
$\tau_{\nu_h}= 10^{-9}$s.  
These distributions were obtained assuming 
 that the  $\pair$ pair from the converted photon 
 is misreconstructed  as a  single track from the $\nu_e QE$ reaction.
Simulations are in reasonable agreement with the experimental distributions.  
For instance, for the distributions
 shown in Figs.\ref{eque40nu},\ref{eque70nu}, for the case $a= -1$, the
comparison with MiniBooNE  data
yields a $\chi^2$ of 7.1 (7.5) for 8 DF corresponding to 53\% ($\simeq 47\%$) C.L. 
for $m_{\nu_h} =40~(70)$ MeV and $\tau_{\nu_h}= 10^{-9}$ s. 
The simulated excess events, shown in Figs.\ref{evis40nu},\ref{evis70nu},
are  mainly distributed  in the
narrow region $ 200 \lesssim E_{vis} \lesssim 400$ MeV.
Their fraction in this region is $\sim 70\%$. 
The remaining  events are distributed over the region 
$400\lesssim E_{vis} \lesssim 1200$ MeV, where they can be  hidden by the low 
statistics. 

The simulations show that the shape of the $E^{QE}_\nu$ and $ E_{vis}$  
distributions 
is sensitive to the choice of the $\nu_h$ mass: the heavier the $\nu_h$, the harder the visible energy 
spectrum.  The best 
fit results suggest  that the $\nu_h$ mass 
 is in the region $20 \lesssim m_{\nu_h}\lesssim 600$ MeV and the 
lifetime is in the range $\tau_{\nu_h} \lesssim 10^{-9}-10^{-7}$ s, 
respectively; see also \cite{sng}.

The estimate of the mixing parameter $|U_{\mu h}|^2$ was 
performed by using a relation similar to Eq.(\ref{numberg}). 
The flux  $\Phi(\nu_h)$ was estimated from the expected 
number of the  $\nu_\mu NC$  events  times the mixing 
$|U_{\mu h}|^2$, taking into account the threshold effect due to the heavy 
neutrino mass. The total number of 
reconstructed $\nu_\mu CC$ events in the detector \cite{mbbeam} was used for 
normalization.
The probability  of the heavy neutrino to decay radiatively 
in the fiducial volume at a distance $r$ from the primary vertex is given by Eq.(\ref{probdec}),
assuming  the branching fraction $Br(\nuh) \simeq 1$.
Taking into account  the 
ratio $\nu_\mu NCQE / \nu_\mu CCQE \sim $ 0.43, and 
the number of  $\nu_\mu CCQE$ events observed 
\cite{minibnu1,minibnu2} and assuming that almost all $\nuh$ decays occur inside 
the fiducial volume of the detector, we estimate the $\mix$ to be in the range 
\begin{equation}
|U_{\mu h}|^2 \simeq (1-4)\times 10^{-3}.
\label{minibmix}
\end{equation}
This result is mainly defined by the uncertainty on  
the number of excess events. Equation(\ref{minibmix}) is valid for  the mass region 
$40 \lesssim m_{\nu_h} \lesssim  80$ MeV
favored by the LSND data.

\subsection{Excess of events in $\overline{\nu}_\mu$ data}
\begin{figure}[h]
\begin{center}
    \resizebox{9cm}{!}{\includegraphics{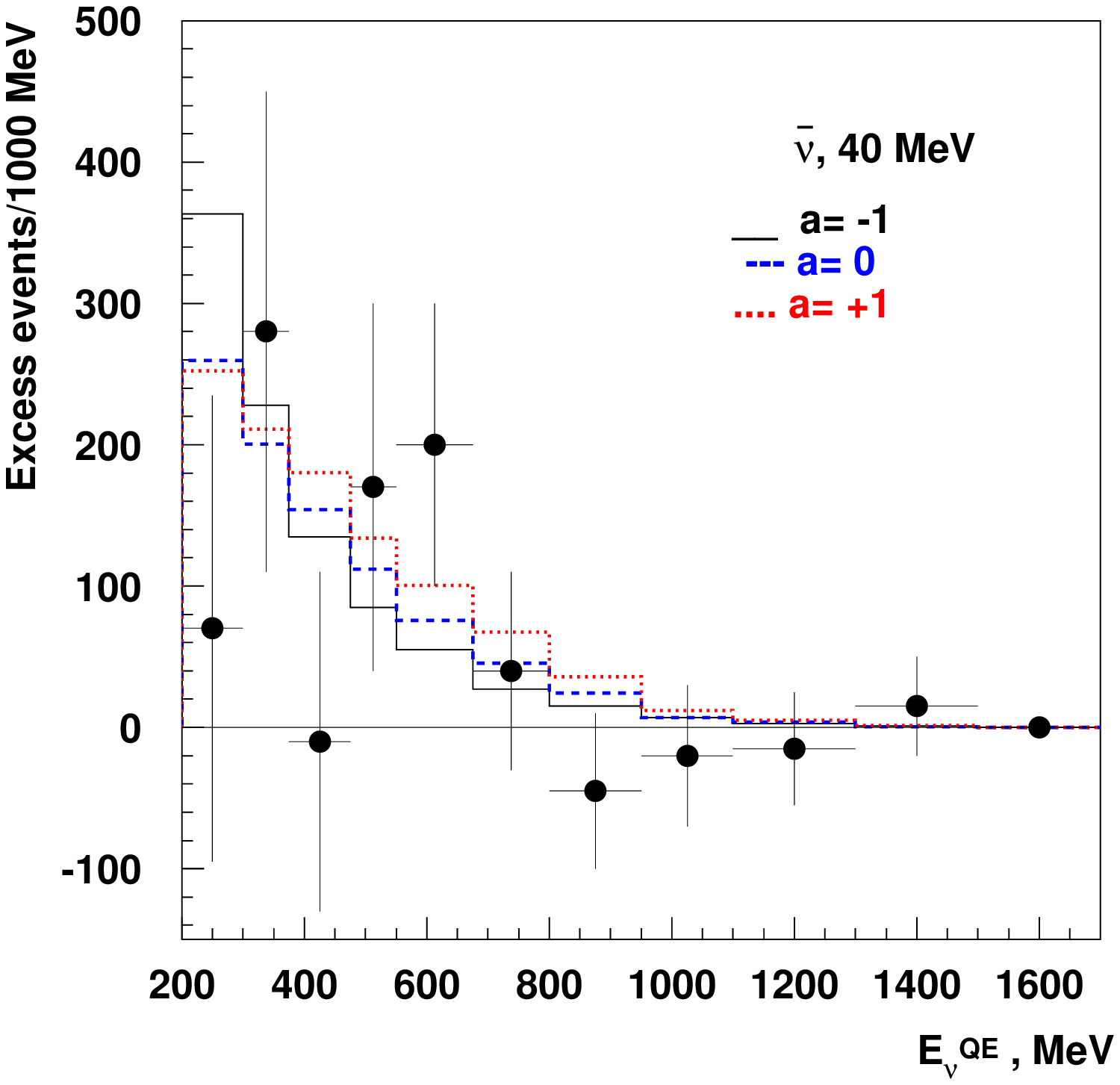}}
     \caption{Distributions of the excess events in the MiniBooNE detector from the $\nuh$ decay
 reconstructed as $\overline{\nu}_e CC$ events as a function of $E^{QE}_\nu$ for 
$|U_{\mu h}|^2= 3 \times 10^{-3}$,  $m_{\nu_h} = 40$ MeV,
and $\tau_{\nu_h}= 10^{-9}$ s, and  for different values of the asymmetry parameter $a$.
The dots are experimental points
for the excess events in the MiniBooNE detector. Error bars 
include both statistical and systematic errors \cite{minibnub}. 
The  comparison of the distributions with the experimental data
yields a $\chi^2$ of 8.2 (a=-1), 7.1 (a=0), and  6.7 (a=1) for 
9 DF.}
\label{eque40nuball}
\end{center}
\end{figure}
\begin{figure}[h]
\begin{center}
    \resizebox{9cm}{!}{\includegraphics{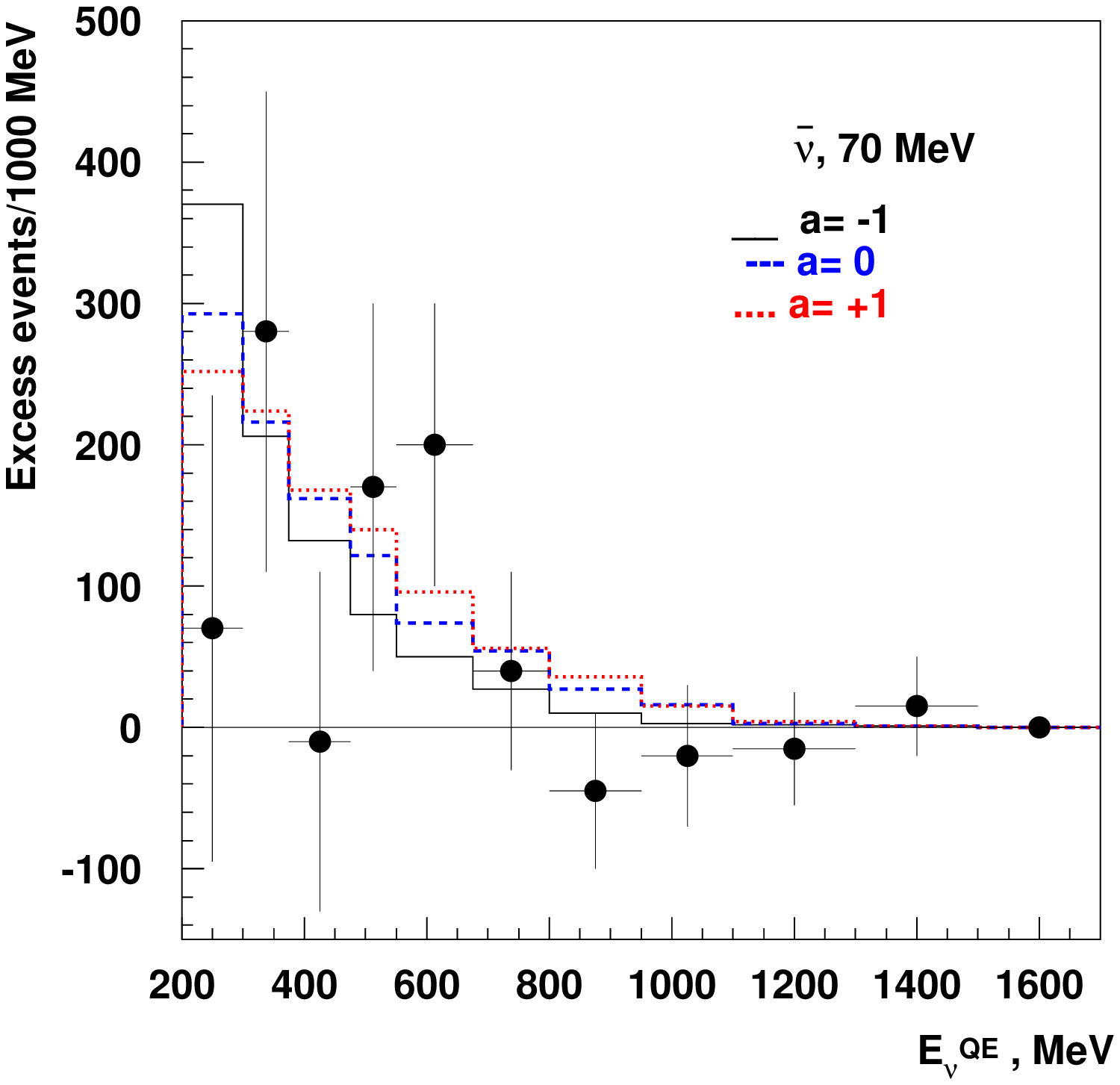}}
     \caption{Same as Fig.\ref{eque40nuball} for the 70 MeV $\nu_h$.
The  comparison of the distributions with the experimental data
yields a $\chi^2$ of 8.3 (a=-1), 7.5 (a=0), and 6.3 (a=1) for 
9 DF.}
\label{eque70nuball}
\end{center}
\end{figure}
Recently, the MiniBooNE experiment has reported  results from a search for
$\overline{\nu}_\mu \to \overline{\nu}_e$ oscillations 
using a data sample corresponding to $5.66\times 10^{20}$ protons on target \cite{minibnub}.
 An excess of $\Delta N=$43.2$\pm 22.5$ electronlike events is observed which, 
 when constrained by the observed  $\overline{\nu}_\mu$  events, has a probability for consistency with the 
background-only hypothesis of 0.5\% in the oscillation-sensitive energy range of 
$475<E<1250$ 
MeV. The data are consistent with $\overline{\nu}_\mu \to \overline{\nu}_e$ oscillations in the 0.1 eV 
range and with the 
evidence for antineutrino oscillations from the LSND. Note, that 
the low statistics antineutrino  data collected by the MiniBooNe experiment 
seem to show no low-energy excess  \cite{antinu}. 

Similar to the neutrino data \cite{minibnu2}: 
a) the excess is  observed for single track events, 
originating either from an 
electron, or from  a photon converted into an $\pair$ pair ;
b) the reconstructed neutrino 
energy is in the wider range $200 < E^{QE}_\nu < 700$ MeV, and
there is also an  
excess for the region $E^{QE}_\nu > 475$ MeV. 
The variable 
 $E^{QE}_\nu$  is calculated under 
the assumption that the observed  electron track originates from  a $\overline{\nu}_e$ interaction;
 c) compare to the $\nu_\mu$ data the visible 
energy $E_{vis}$   is in the wider range  
$200\lesssim  E_{vis} \lesssim 700$ MeV for events with $E^{QE}_\nu > 200$ MeV; and
d) the angular distribution of 
the excess events with respect to the incident neutrino direction 
is wide and consistent with the shape expected 
from $\overline{\nu}_e CC$ interactions. 
To satisfy the criteria a)-d), we propose again that  the $\overline{\nu}_\mu$ excess events are  
originated from the decay of the  heavy neutrino  considered in the previous sections. 
The $\nu_h$'s  are produced by mixing in  
$\overline{\nu}_\mu$ $NCQE$ interactions and  
deposit their  energy via the radiative  
decay mode,  as  
shown in Fig.\ref{diag}, with the subsequent conversion of the decay
photon into an $\pair$ pair in the MiniBooNE target. 

\begin{figure}[h]
\begin{center}
    \resizebox{9cm}{!}{\includegraphics{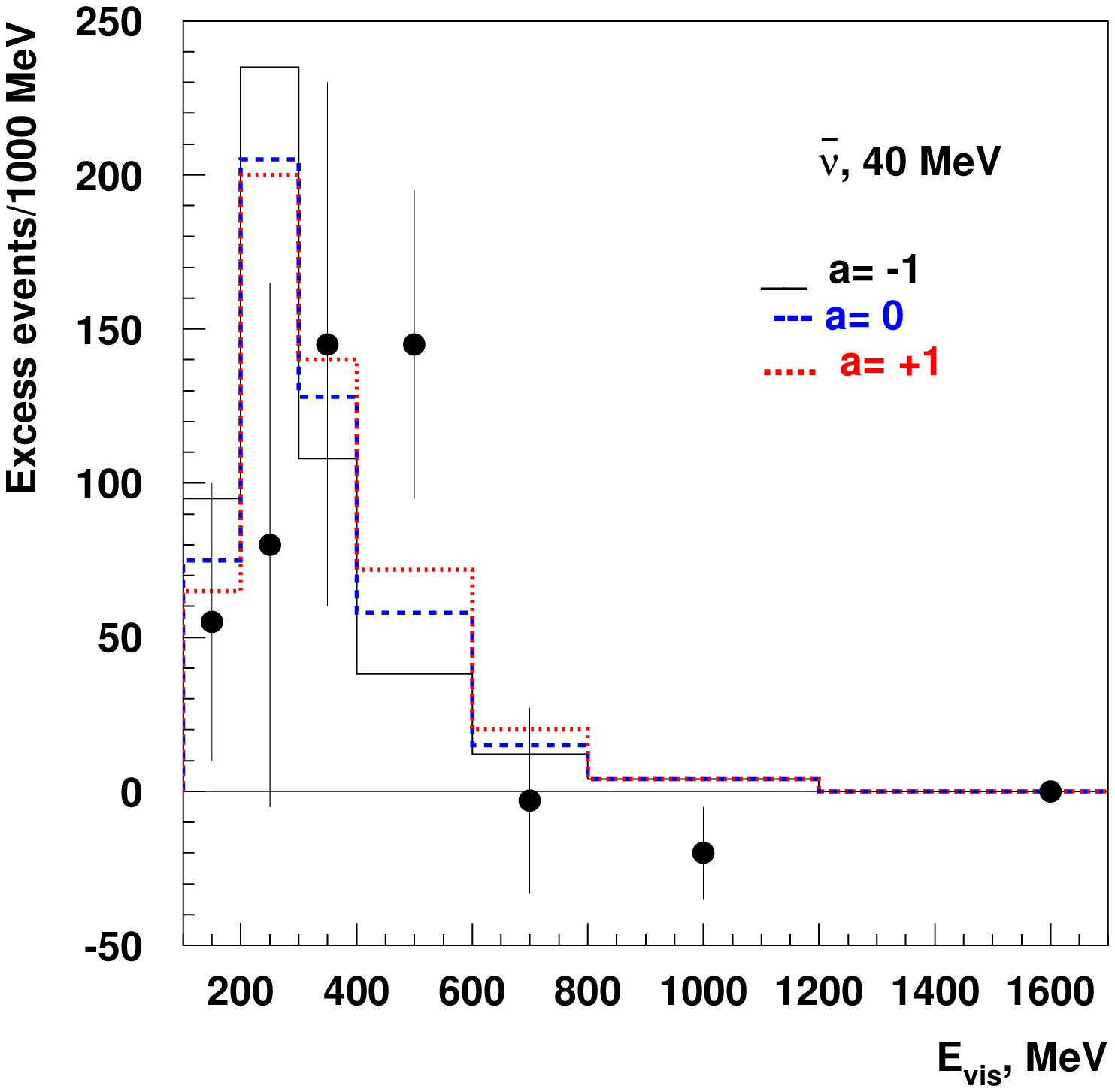}}
     \caption{ Distribution of the excess events in the MiniBooNE from the $\nu_h$ decay
 reconstructed as $\overline{\nu}_e CC$ events 
as a function of  $E_{vis}$ for 
$ E^{QE}_\nu > 200$ MeV, $|U_{\mu h}|^2= 3 \times 10^{-3}$,  $m_{\nu_h} = 40$ MeV,
and $\tau_{\nu_h}= 10^{-9}$ s, and  for different values of the asymmetry parameter $a$. 
The dots are experimental points
for the excess events in the MiniBooNE detector. Error bars 
include both statistical and systematic errors \cite{minibnu2}. 
The  comparison of the distributions with the experimental data
yields a $\chi^2$ of 9.5 (a=-1), 7.5 (a=0), and 6.2 (a=1) for 5 DF.}
\label{evis40nuball}
\end{center}
\end{figure}
\begin{figure}[h]
\begin{center}
    \resizebox{9cm}{!}{\includegraphics{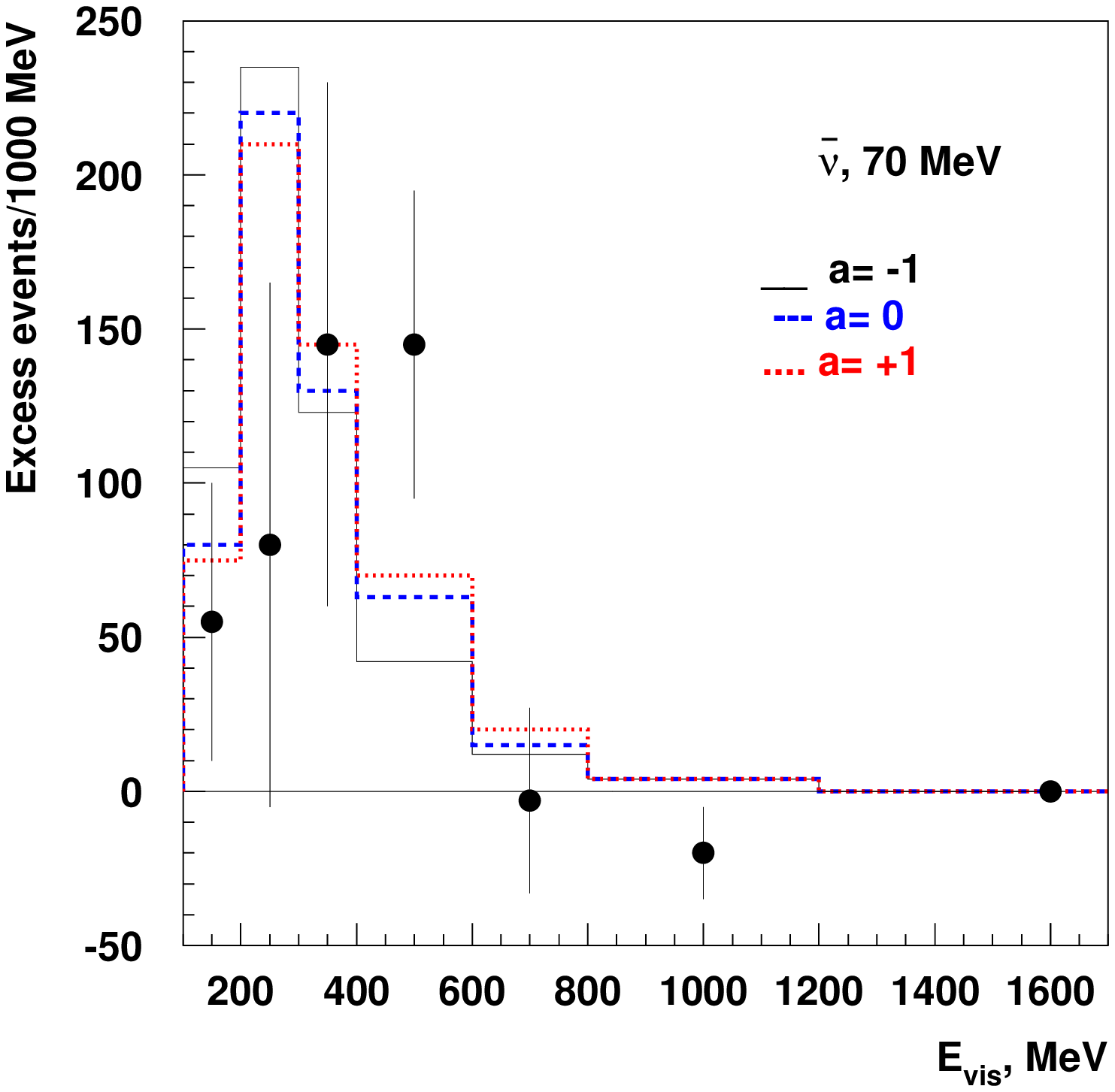}}
     \caption{ Same as  Fig.\ref{evis40nuball} for the 70 MeV $\nu_h$.
The  comparison of the distributions with the experimental data
yields a $\chi^2$ of 8.5 (a=-1), 7.2(a=0), and 6.4 (a=1) for 
5 DF.}
\label{evis70nuball}
\end{center}
\end{figure}

For simulations 
 we used a $\overline{\nu}_\mu$ energy spectrum parametrized from the  
reconstructed $\overline{\nu}_\mu CCQE$ events \cite{mbbeam}. 
Note that the antineutrino energy distribution has a maximum at $\simeq 400$ MeV and
an average energy of about 600 MeV.  

\begin{figure}
\begin{center}
    \resizebox{9cm}{!}{\includegraphics{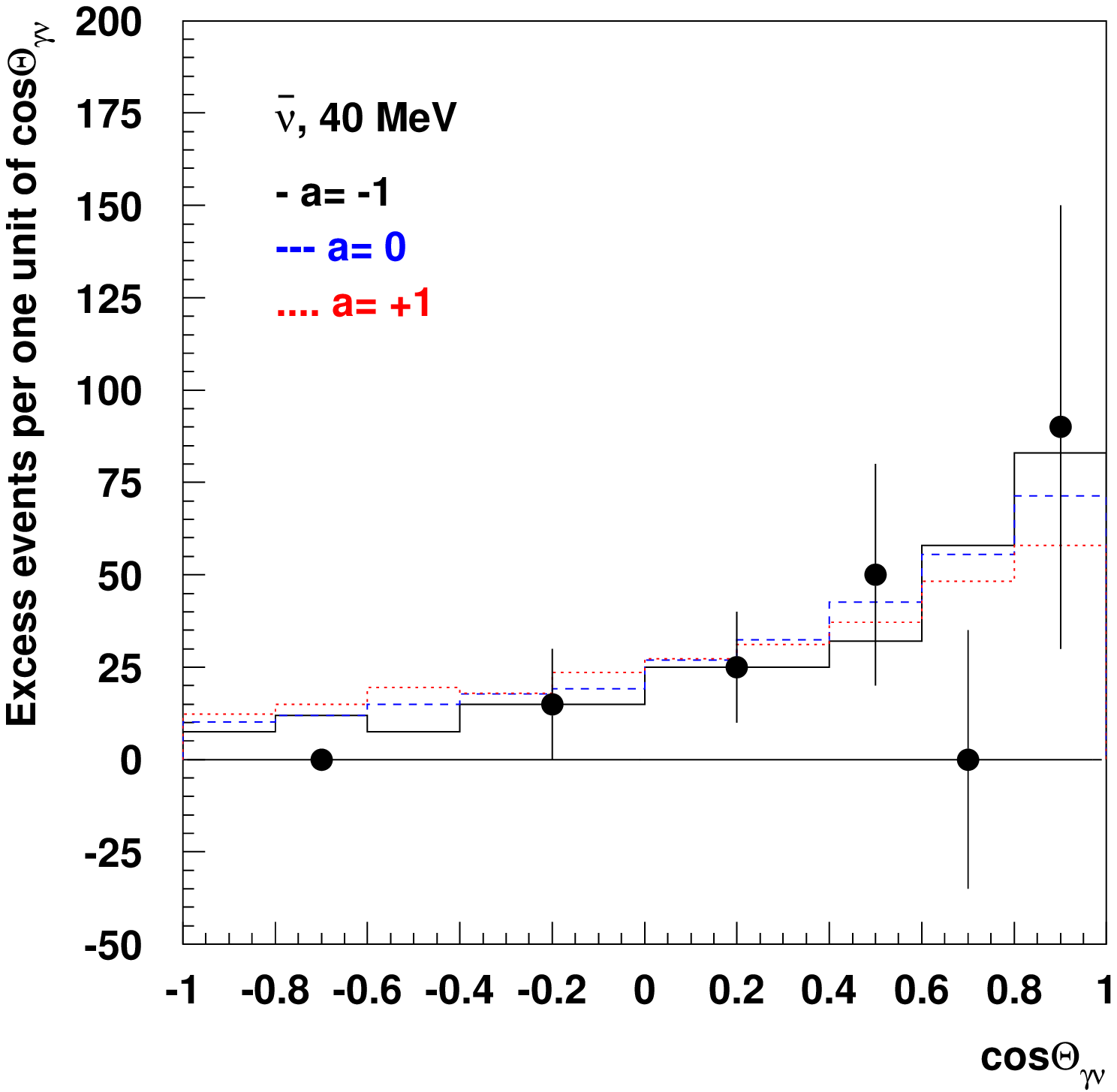}}
     \caption{Distribution of the excess events in the MiniBooNE detector from the $\nu_h$ decay
 reconstructed as $\overline{\nu}_e CC$ events 
as a function of  $\cos \Theta_{\gamma \nu}$ for 
$E^{QE}_\nu > 200$ MeV, $|U_{\mu h}|^2= 3 \times 10^{-3}$,  $m_{\nu_h} = 40$ MeV,
and $\tau_{\nu_h}= 10^{-9}$ s, and  for different values of the asymmetry parameter $a$.
 The dots are experimental points
for the excess events in the MiniBooNE detector. Error bars 
include both statistical and systematic errors \cite{minibnu2}. 
The  comparison of the distributions with the experimental data
yields a $\chi^2$ of 3.1 (a=-1), 2.7 (a=0), and 3.3 (a=1) for 4 DF.}
\label{cos40nuball}
\end{center}
\end{figure}

\begin{figure}[h]
\begin{center}
    \resizebox{9cm}{!}{\includegraphics{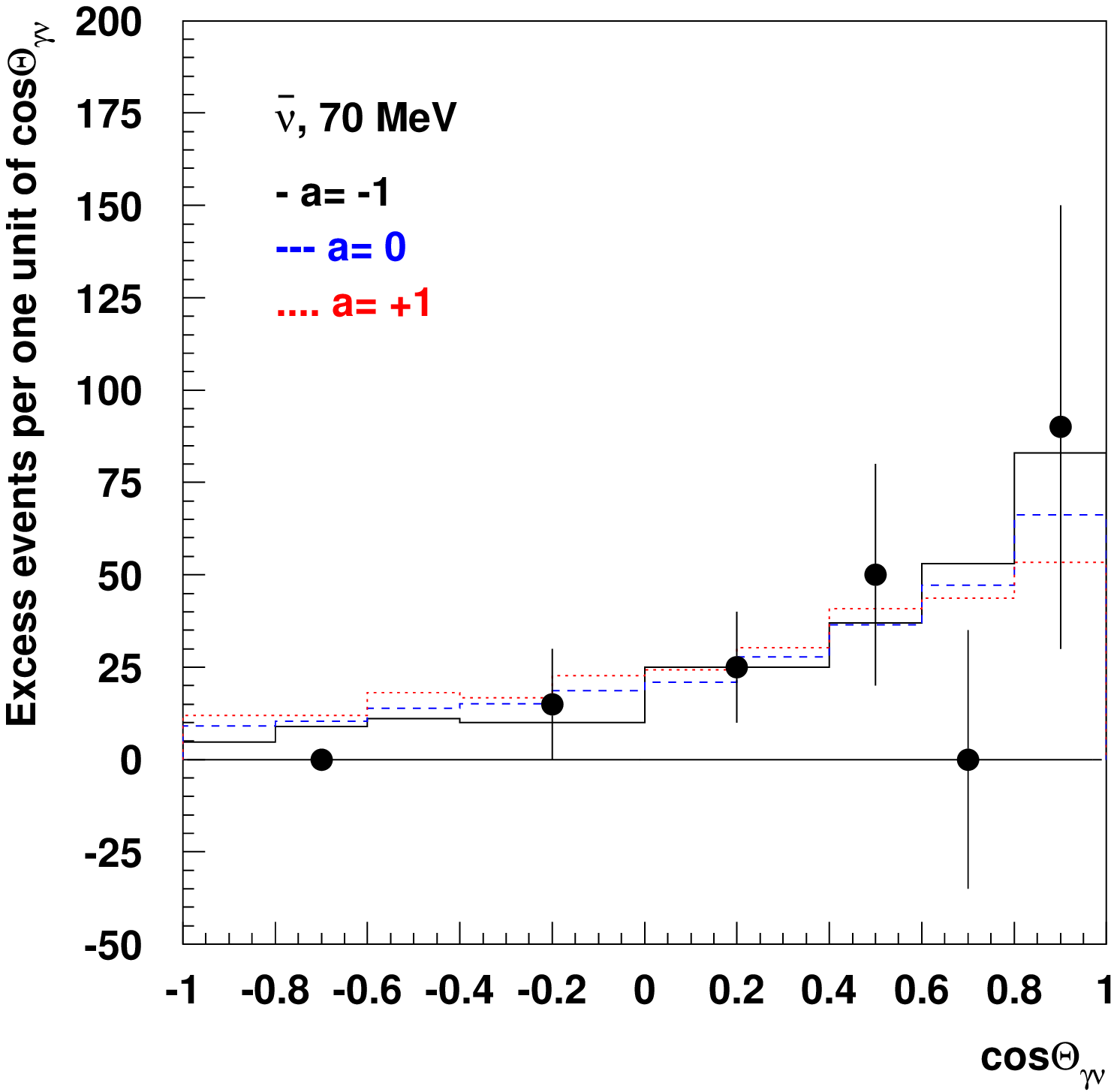}}
     \caption{ The same as in Fig.\ref{cos40nuball} for the 70 MeV $\nu_h$.
     The  comparison of the distributions with the experimental data
yields a $\chi^2$ of 2.5 (a=-1), 2.7 (a=0), and 2.6 (a=1) for 4 DF.}
\label{cos70nuball}
\end{center}
\end{figure}

In Figs.\ref{eque40nuball}-\ref{cos70nuball} the distributions of 
the kinematic variables $E^{QE}_\nu, E_{vis}$ and $\cos \Theta_{\gamma \nu}$ 
for the $\nuh$ events are shown 
for  $m_{\nu_h}=40$ and 70 MeV and 
$\tau_{\nu_h}= 10^{-9}$s.  
These distributions were obtained assuming 
 that the  $\pair$ pair from the converted photon 
 is misreconstructed  as a  single track from the $\overline{\nu}_e QE$ reaction.
In this calculation we assume that the angular distribution of photons in the $\nu_h$ rest frame 
has the same  asymmetry as for the $\nu_\mu$ case due to $CP$ conservation;
see Sec.II. 

Simulations are in reasonable agreement with the experimental distributions.  
For instance, for the $E_\nu^{QE}$-distributions
 shown in Figs.\ref{eque40nuball},\ref{eque70nuball} for the Dirac case with $a=-1$, the
comparison with MiniBooNE  data
yields a $\chi^2$ of 8.2 (8.3) for 9 DF corresponding to 47\% ($\simeq 45\%$) C.L. 
for $m_{\nu_h} =40 (70) $ MeV and $\tau_{\nu_h}= 10^{-9}$ s. 
The $E_{vis}$ distributions shown in Figs. \ref{evis40nuball},\ref{evis70nuball}
are also in  reasonable agreement with the experiment.
 For the case $a=-1$, the comparison with data yields a 
$\chi^2$ of 9.5 (8.5) for 5 DF corresponding to 10\% ($\simeq 14\%$) C.L. 
for $m_{\nu_h} =40 (70) $ MeV and $\tau_{\nu_h}= 10^{-9}$ s. The events 
are  mainly distributed  in the
 region $ 200 \lesssim E_{vis} \lesssim 600$ MeV, where  their  fraction is $\sim 90\%$. 
The remaining  events are distributed over the region 
$600\lesssim E_{vis} \lesssim 1200$ MeV,
 where the observed number of events is found to be consistent with  
the expected one. 

 The analysis of these data within the framework discussed above 
suggests that a smaller excess of events is observed 
mainly due to the lower $\overline{\nu}_\mu$
 energy and  $NC$ cross section. 
The estimate of the mixing parameter $|U_{\mu h}|^2$ was 
performed by using relation similar to Eq.(\ref{numberg}),
assuming  the branching fraction $Br(\nuh) \simeq 1$.
The flux  $\Phi(\nu_h)$ was estimated from the expected 
number of   $\overline{\nu}_\mu NC$  events  times the mixing 
$|U_{\mu h}|^2$, taking into account the phase space factor due to the heavy 
neutrino mass. The total number of 27,771 
reconstructed $\overline{\nu}_\mu CCQE$ events in the detector
 \cite{minibnub} was used for 
normalization. Taking into account  the 
ratio $\overline{\nu}_\mu NCQE /\overline{\nu}_\mu CCQE \sim $ 0.41 and
the total number of  $\overline{\nu}_\mu CCQE$ events observed, 
 and assuming that almost all $\nuh$ decays occur inside 
the fiducial volume of the detector, we estimate the $\mix$ to be in the range 
\begin{equation}
|U_{\mu h}|^2 \simeq (0-8)\times 10^{-3}.
\label{minibamix}
\end{equation}
which is consistent with the mixing from Eq.(\ref{lsndmix}).
This result is mainly defined by the  uncertainty in  
the number of excess events. Equation(\ref{minibamix}) is valid 
for  the mass region 
$40 \lesssim m_{\nu_h} \lesssim 600$ MeV, which includes the region 
favored by  the LSND data.

\section{Results from the LSND and MiniBooNE data and properties of the heavy neutrino}

To obtain the combined regions
 in the $(m_{\nu_h}; |U_{\mu h}|^2)$ parameter space, we 
have analyzed  both, the LSND  and MiniBooNe  data simultaneously. 
The  $E_{vis}$ and $cos \Theta_{\gamma \nu}$ distributions from LSND, and the $E^{QE}_\nu$, $E_{vis}$, and $cos \Theta_{\gamma \nu}$
distributions from MiniBooNE were
 used for comparison with the corresponding simulated  distributions 
from the $\nuh$ decay 
to constrain the mixing strength and heavy neutrino mass. The shape of the simulated distributions is defined 
by the mass ( and type) of the $\nu_h$, while the mixing strength  is  defined mainly by the 
overall normalization of distributions 
to the number of excess events observed in the experiments.  
The analysis  includes the following  constraints: i) the number of excess events can vary within the 2$\sigma$ range,  ii) the $\nu_h$ lifetime has to be less than 
$2\times 10^{-9}$ s, iii) the $\nu_h$ should be heavier than 40 MeV to avoid its production in the 
KARMEN experiment; and  iv) the number of excess events in the LSND detector 
in the  
 energy  region  $ 60<  E_{vis} < 150$ MeV with a recoil neutron 
should be  $\lesssim $10 events. 
The latter constrain came from the upper limit on the number of 
events with $>0$ recoil neutrons observed by LSND in this energy region \cite{lsndmue}.

\begin{figure}[h]
\begin{center}
    \resizebox{10cm}{!}{\includegraphics{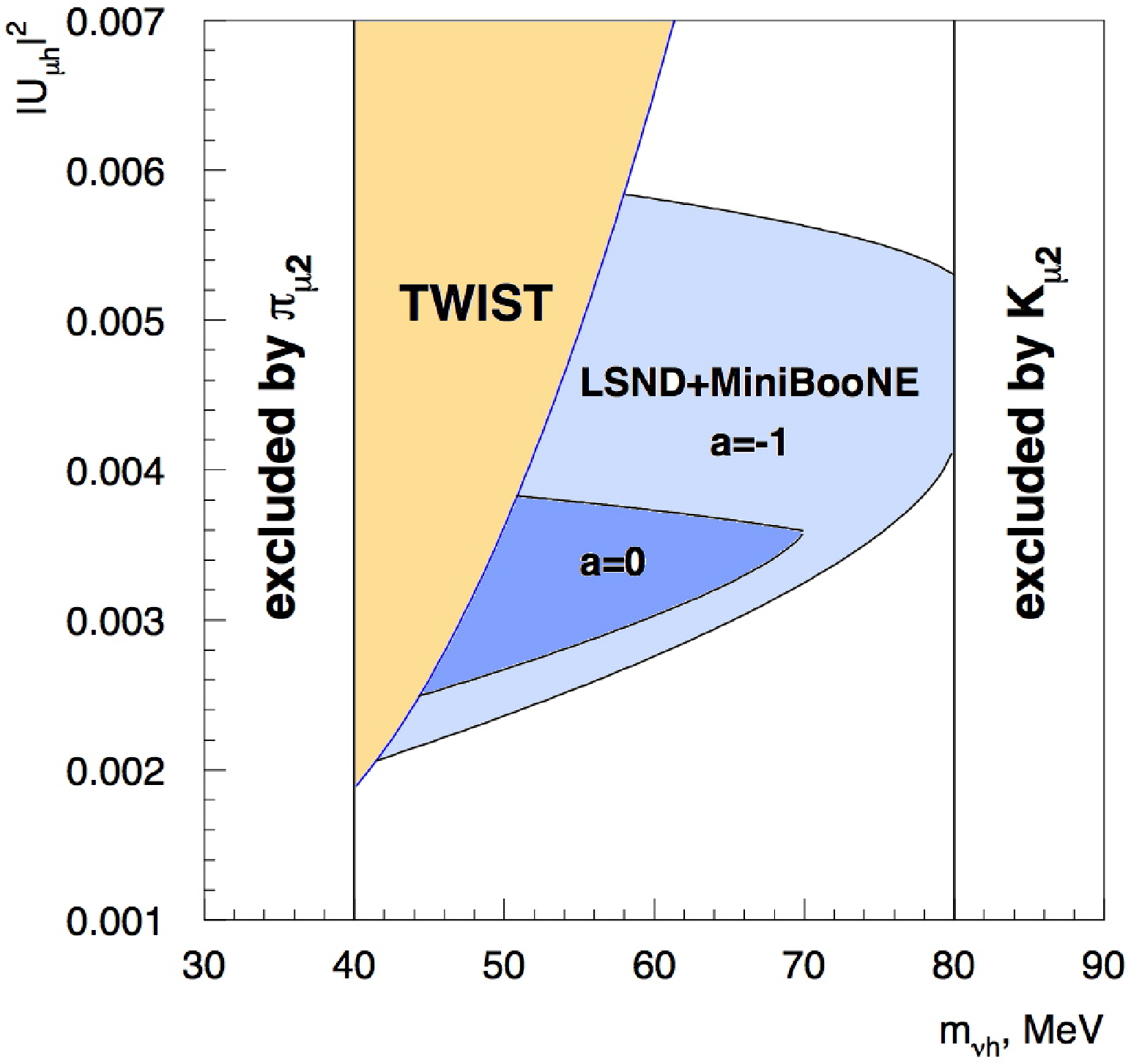}}
     \caption{ The 2$\sigma$ allowed  region (dark areas) in the $(m_{\nu_h}; |U_{\mu h}|^2)$ 
parameter space obtained for different values of the asymmetry parameter $a$
 from the combined analysis of LSND and MiniBooNE $\nu_\mu$ and 
$\overline{\nu}_\mu$ data.  The areas excluded by the $\pi_{\mu 2}$ and 
$K_{\mu 2}$ decay experiments \cite{pdg}, and the exclusion region obtained in the present work 
 from the results of precision measurements of the muon decay parameters  by the TWIST experiment 
\cite{twist} are also shown; see Sec. VI. } 
\label{plot}
\end{center}
\end{figure}

The summary results are shown in Fig.\ref{plot} together with the constrains 
from the $\pi_{\mu 2}, K_{\mu 2}$ experiments and the limit 
obtained 
from the recent results of the TWIST 
experiment on precision measurements of the Michel spectrum in muon decay, see Sec.VI.   
The  fit results suggest  that the $\nu_h$ mass 
is in the region $40 \lesssim m_{\nu_h}\lesssim 80$ MeV
for the Dirac $\nu_h$ with an asymmetry parameter $a= -1$, and in the region $40 \lesssim m_{\nu_h}\lesssim 70$ MeV
 for the case $a=0$. 
 The $\chi^2$ contribution from
the MiniBooNE $\nu_\mu$ data is  smallest for the whole mass range. As expected, for both cases 
the main contributions to $\chi^2$ are from the MiniBooNE $\overline{\nu}_\mu$ data. For higher $\nu_h$ masses, preferred by the MiniBooNE $\numub$ data,
  the region of allowed $|U_{\mu h}|^2$  moves towards smaller values , 
while it is cut by the LSND constrain iv).

The LSND results strongly restrict the allowed $\nu_h$ mass region  
and exclude  solutions  with $m_{\nu_h} > 80$ MeV, which are favored 
by the MiniBooNE $\numub$ data.  
The analysis gives a 14\% $\chi^2$  probability for 
compatibility between the LSND  and MiniBooNE data and  the $\nuh$ 
interpretation, demonstrating a reasonable level of agreement.

As already mentioned in Sec. II, the angular and energy distributions 
of decay photons are sensitive to the type of the heavy neutrino.
The analysis shows that better fit results can be obtained provided that
the $\nu_h$'s produced in the LSND and MiniBooNe experiments   
by the muon neutrino  decay radiatively as a left-handed Dirac neutrino 
with the asymmetry parameter $a=-1$,
 while the $\nu_h$'s produced by the muon antineutrinos   decay 
 as a right-handed Dirac neutrino with the asymmetry parameter $a=+1$.  
The positive sign of the asymmetry coefficient is 
preferred by the analysis  
of MiniBooNE $\numub$  data, while the negative sign provides a  better fit to the 
distributions  from LSND and MiniBooNE $\nu_\mu$ data.
 If the $\nu_h$'s with such exotic properties exist, that would 
  mean that the $\nuh$ decay is not $CP(CPT)$ conserving \cite{kayser,nieves}. 

\section{Limits  on $|U_{\mu h}|^2$ and $\mu_{tr}$}

One might reasonably ask if the mixing strength as large as the one  
shown in Fig.\ref{plot} is consistent with the results of 
previous searches for the MeV $\nu_h$'s \cite{pdg}. Below we discuss the most stringent 
limits which came from the $K$ meson \cite{kek1} - \cite{kek2} and muon decays \cite{twist}-\cite{derenzo}, neutrino scattering 
experiments \cite{ps191}, searches at the CERN LEP \cite{delphi, aleph}, and also from cosmology, astrophysics and 
atmospheric neutrino experiments \cite{dolgov1}-\cite{gr}. Finally, the limits on the transition magnetic moment
are also discussed.
 \subsection{ Limits from $K$ decays}
\begin{figure}[h]
    \resizebox{10cm}{!}{\includegraphics{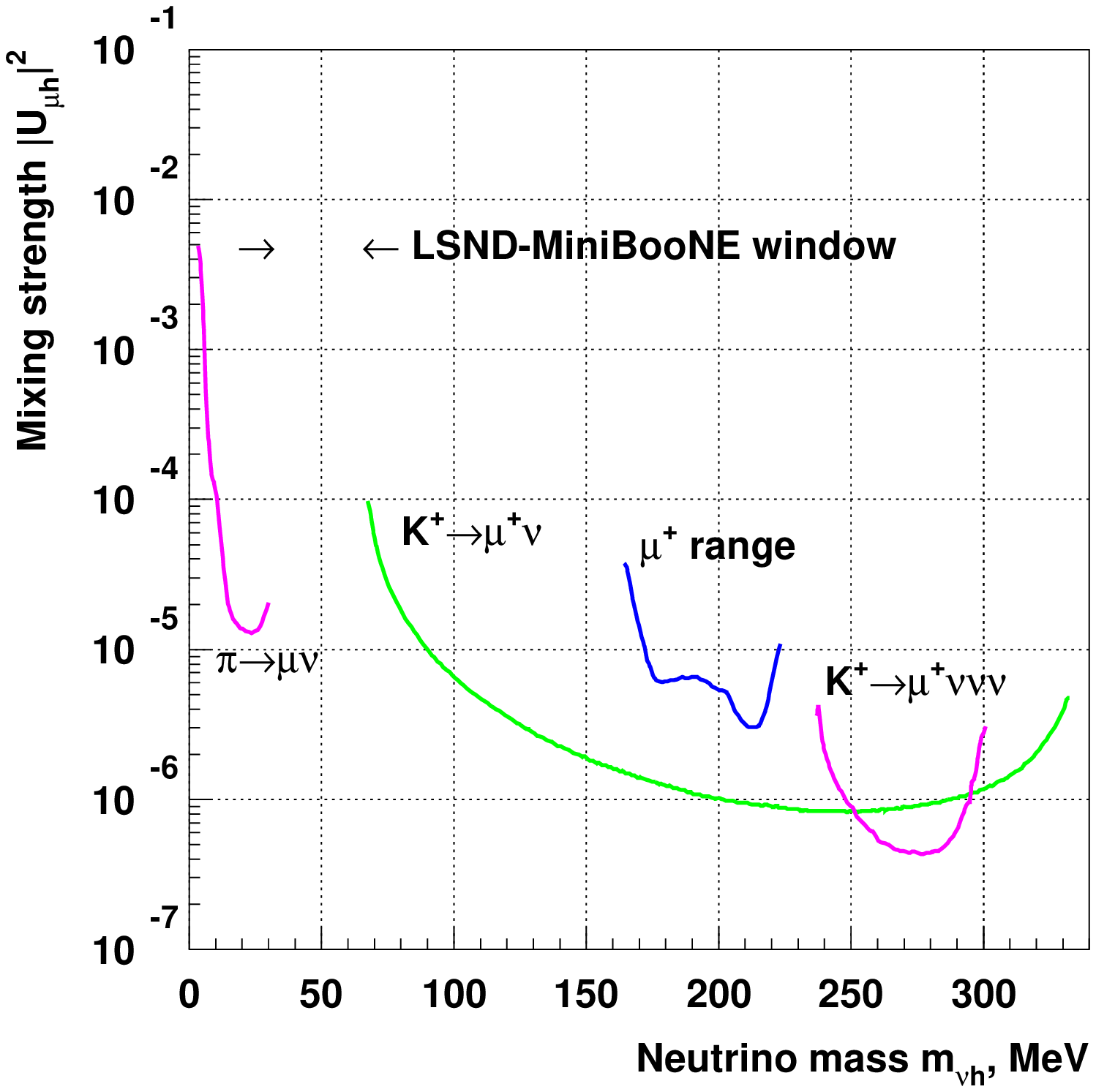}}
     \caption{Bounds on the muonic mixing strength $|U_{\mu h}|^2$ of the heavy neutrino vs its mass
     from $K_{\mu 2}$ range measurements \cite{kek1}, the $K^+\to \mu^+ \nu \nu \nu$ decay search 
     experiment \cite{lbl}, and from heavy neutrino searches in $\pi \to \mu \nu$ \cite{psi} and 
     $K \to  \mu \nu$  \cite{kek2} decays. The arrows show the unconstrained LSND-MiniBooNE mass window.} 
\label{limit}
\end{figure}

It is well known that heavy neutrino in the mass range  $\lesssim$ 400 MeV can be 
effectively probed through the two body decays of charged kaons \cite{shrock1}.
The $K$ meson, which normally decays into a 
$\mu$  and a $\nu_\mu$, might instead decay into a $\mu$ and a  $\nu_h$. 
The experimental signature of the presence of the decay $K^+ \to \mu \nu_h$ is a peak in the 
muon energy  distribution
 below the normal one from the ordinary $K_{\mu 2}$ decay 
 at the energy 
\begin{equation}
E_\mu = \frac{M_{K}^2 + m_\mu^2- m_{\nu_h}^2}{2 M_{K}}
\label{peak}
\end{equation}
The most stringent current experimental limits on  $|U_{\mu h}|^2$ 
for the $\nu_h$ mass region below 400 MeV, 
are summarized in Fig. \ref{limit} \cite{pdg}; for a recent review see \cite{gs, atre}. 
One can see, that  the limit  for the mass region around 100 MeV, derived from a search for the $\nu_h$   at KEK \cite{kek2}, is  $|U_{\mu h}|^2 \lesssim 10^{-5}$.
Surprisingly, the neighboring ($m_{\nu_h};|U_{\mu h}|^2$) region of parameters \eqref{lsndmix},\eqref{lsndmass} 
favorable for the explanation of the LSND and MiniBooNE results remains 
  unconstrained. The reason for that is because 
the $\nu_h$ in the mass range of $m_{\nu_h}\simeq 40 - 70$ MeV is 
 outside of the kinematical limits for $\pi_{\mu 2}$ decays and is
 not accessible in 
$K_{\mu 2}$ decay experiments due to experimental resolutions and a high background
level. For example, to resolve the  muon peak of 234 MeV/c from 
the 40 MeV $\nu_h$ and the main peak of 236 MeV/c,  a
 muon momentum resolution  better than 1\% is required. Another reason is
 related to  the  
$K^\pm \to \mu \nu \gamma$ and  $K^+\to \mu^+ \pi^0 \nu$ decays
which produce  a continuous background to the muon 
momentum distribution below the  main peak and 
essentially constrains 
the sensitivity of the search for the $\nu_h$ mass range  $\lesssim 100$ MeV 
again due to the requirement of a very high experimental resolution. 
Let us consider this in detail.
In the  most sensitive experiment performed at KEK \cite{kek2}, degraded $K^+$'s were stopped and 
decayed in the scintillator target. Charged particles from $K^+$ decays were momentum 
analyzed by a magnetic spectrometer. To achieve high sensitivity to small signals, the main background decay modes,
$K^+\to \mu^+ \pi^0 \nu$ and $K^+\to \mu^+ \nu \gamma $, were vetoed by using an almost  hermetic (92\% of 4$\pi$) low-energy threshold ($\simeq 1 $ MeV) NaI calorimeter, surrounding the kaon decay target. The veto efficiency for the $K^+\to \mu^+ \pi^0 \nu$ decay mode was quite high, better than 99\%, thanks to the emission of two photons. The decay $K^+\to \mu^+ \nu \gamma $ was difficult to suppress, and about 30\% of photons from this decay 
mode were undetected. The reasons for this are the following:
i) the low-energy photons are preferably emitted along  the decay muon momentum direction, so they escape undetected; ii) the photo-nuclear absorption cross section is high for photon energies $\simeq$ 50 MeV; and iii) there is an  
 absorption of decay photons due to the
presence of a dead material in the vicinity of the target.
These effects results in a  high background level which significantly decreases
the  sensitivity for the $\nu_h$ masses below $\simeq$ 80 MeV, 
as one can see from Fig. \ref{limit}.
Let us now  show that taking into account the dominance of the $\nuh$ decay and the 
relatively short $\nu_h$ lifetime makes existing experimental  bounds even 
weaker. Indeed,  in these searches  it was typically 
 assumed that the $\nu_h$ is a relatively long-lived particle, i.e.
 $\frac{L m_{\nu_h}}{p_{\nu_h} \tau_{\nu_h}} \ll 1 $, where $L \simeq 50$ cm 
is the typical size  of  the target region.
However, if the decay  $\nuh$ is dominant and the heavy neutrino is a short-lived particle,
the $\nu_h$ would decay presumably in the vicinity of the target.
In this case, the decay photon would be  vetoed by the  calorimeter, and the event  
would be rejected. An estimate shows that for a $\nu_h$ lifetime, as short as 
$\tau_{\nu_h} \lesssim 10^{-9}$ s more than  95\% of heavy neutrinos 
would decay in the  target region or inside the calorimeter of the experiment 
\cite{kek2}, producing a photon with an energy well above the veto energy 
threshold. Because of this self-veto effect, 
the  limit  $|U_{\mu h}|^2 \lesssim (2-4)\times 10^{-5}$ 
for the $\nu_h$ mass around $\simeq$ 80 MeV, 
 could  be worsened  by more than an order of magnitude,
and thus, would be in the region  $\simeq 10^{-3}$ close 
to values from \eqref{lsndmix}.
Thus, it would be important  to  perform 
an ''open mind'' search for heavy neutrino 
in a wider mass range, including the region around 80 MeV.  

\subsection{Muon decay constraints} 

\begin{figure}[h]
\begin{center}
    \resizebox{11cm}{!}{\includegraphics{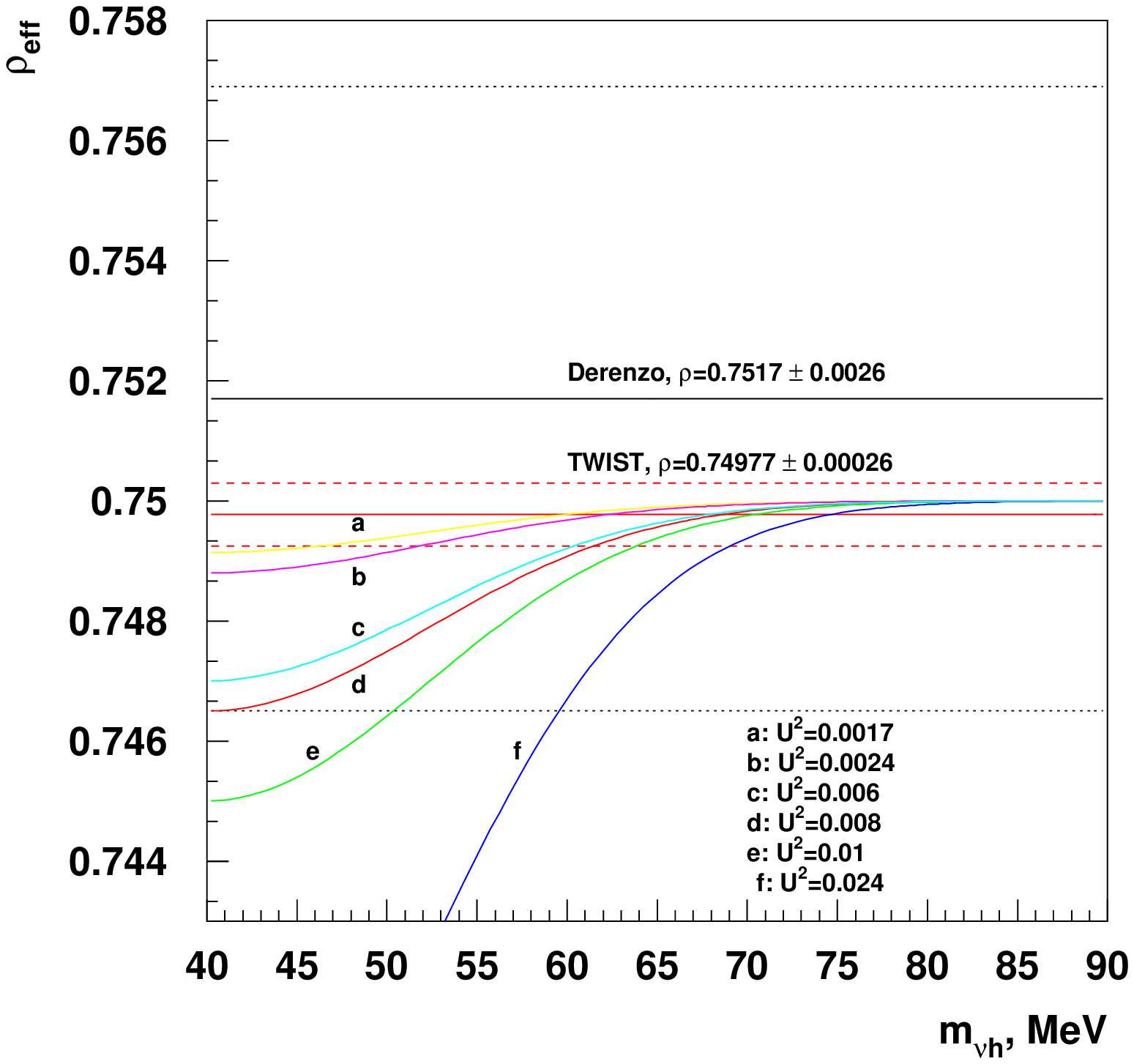}}
     \caption{ Dependence of the parameter $\rho_{eff}$ as a function of the $m_{\nu_h}$ mass 
     determined for the decay $\mu \to e \nu_e \nu_h $ for several values of  $|U_{\mu h}|^2$ shown in the plot. The 2 $\sigma$ bands around the central values of the $\rho$ parameter 
     measured by Derenzo \cite{derenzo} (dotted lines), and by the TWIST experiment \cite{twist} (dashed lines)
     are also shown.} 
\label{shrock}
\end{center}
\end{figure}

If a heavy neutrino with mixing into $\nu_\mu$ 
in the range  of Eq.(\ref{lsndmix}) exists, it would  notably  
 change  the shape of the Michel spectrum of the ordinary muon decay, which is well predicted in the standard model.
 This gives the possibility of using    
the results of precision measurements of the Michel spectrum 
in the ordinary muon decay in order to probe the possible existence of
a heavy neutrino \cite{shrock2}.
The relatively free of theoretical uncertainties limit 
on  mixing $|U_{\mu h}|^2$ 
for $\nu_h$ masses in the range from 30 to 70 MeV was 
originally set in [13, 49] by using results of the $\rho$ parameter measurement by Derenzo \cite{derenzo}. 
Following Shrock \cite{shrock2}, we  use the idea 
 that an admixture of a heavy neutrino to the Michel spectrum 
 could significantly alter the  $\rho$ parameter, 
resulting in an effective $\rho$ parameter 
$\rho_{eff}$ that is different from the canonical value $\rho=0.75$.  
Hence, one can extract  limits on the mixing  $|U_{\mu h}|^2$ from comparison 
of the measured  ( $\rho_{exp}$) and effective $\rho_{eff}$ values  
 by  requiring $|\rho_{eff} - \rho_{exp}| \lesssim \sigma_{exp}$, 
where $\sigma_{exp}$ is the error of the measurements. 
 Figure \ref{shrock} shows the dependency  of 
the effective $\rho_{eff}$ parameter as a function of $m_{\nu_h}$ for the region of interest from 
40 to 80 MeV, for several values of $|U_{\mu h}|^2$ 
 obtained by the fit of the Michel spectrum.
  The two sigma bands from the measurement of the $\rho$ parameter by
Derenzo  \cite{derenzo},  $\rho = 0.7518 \pm 0.0026$, and from the recently reported precision 
measurements by the TWIST Collaboration \cite{twist},  $\rho = 0.74977 \pm 0.00012 (stat)\pm 0.00023(syst)$, or $\rho = 0.74977 \pm 0.00026$ with all errors combined in quadrature,  are also  shown. 
 The theoretical expressions 
for the mixing of heavy Dirac neutrinos in muon decay  can be found in \cite{shrock2}, and, with 
radiative corrections included,  in \cite{kn}.
Our new $2\sigma$ limit shown  in Fig. \ref{plot} is
  derived  for the $\nu_h$ masses in the range from 40 to 80 MeV 
 by using the results of the TWIST experiment.
 For  very large masses 
the limit is less restrictive because the $\nu_h$ contribution is highly  suppressed. Comparison of
the bounds  obtained in \cite{shrock2, dixit} and in the present work by using the results of 
the measurements of the $\rho$ parameter by Derenzo \cite{derenzo}  shows good agreement. 

\subsection{Bounds from neutrino scattering experiments}

Next we consider the  neutrino experiments that searched for
$\nu_h$ decays. The direct searches for the radiative $\nuh$ decay were 
performed for the mass region $< 1$ MeV \cite{pdg}.
Others  experiments searched for 
heavy neutrino decaying into charged particles in the final 
state, such as e.g. $\nu_h \to ee\nu, \mu e\nu, \mu \pi \nu,..$. 
None of  these experiments has reported  a bound on the mixing 
strength $|U_{\mu h}|^2$, or on the combination $|U_{\mu h}|^2 \mu_{tr}$,
 for the radiative $\nuh$ heavy neutrino decay.
The experimental signature for the $\nu_h$ 
decaying into charged particles is quite clean.  
The selection of two tracks 
originating from a common vertex with nonzero invariant mass makes these 
searches almost background-free. In contrast, to 
search for an excess of a single converted photons  
from the radiative neutrino decay is  
more difficult. At high energies  the  
background level from the $\pi^0$ decays and bremsstrahlung photons 
is high. The uniqueness of the LSND and MiniBooNE experiments is that they run
at low energies when the production of the $\nu_h$'s is still possible 
and the background level is relatively small due to the high fraction of 
 $\nu_\mu NC QE$ events used for the production of $\nu_h$'s.

The best limit $|U_{\mu h}|^2 \lesssim 10^{-5}-10^{-6}$ 
for the mass region $m_{\nu_h}\simeq 40-80$ MeV was derived from a search for  $\nu_h \to \pair \nu$ decays  in 
the PS191 beam dump experiment at CERN \cite{ps191}.  
It was assumed that this decay mode is dominant and
 that the $\nu_h$ is a relatively long-lived particle,
i.e $\frac{L m_{\nu_h}}{p_{\nu_h} \tau_{\nu_h}} \ll 1 $, where $L \simeq 1.4\times 10^2$ m is the distance between the target and the detector.
Other decay modes with  charged particles in the final state, such e.g. as 
$\nu_h \to \mu \pi, \mu \mu \nu , \mu e \nu$ are forbidden by the energy 
conservation.  
The PS191 detector consist of 
a 12 m  long decay volume, eight chambers located 
inside the volume to detect charged tracks and  followed by a calorimeter.
 The decay volume  was
essentially  an  empty region   
filled with helium to reduced the number of ordinary neutrino 
interactions down to $\simeq 100$ events, with a total amount of dead material around 3.6  g/cm$^2$.
The events searched for in the experiment were requested to consist of two 
tracks originating from a common vertex in the ''vacuum'' part of the $\nu_h$ decay volume and giving 
rise to at least one shower in the calorimeter.

Consider now our case, e.g. with
 $|U_{\mu h}|^2 = 3\times 10^{-3}$, $m_{\nu_h} = 40$ MeV and the $\nu_h$
lifetime $\tau_{\nu_h}= 10^{-9}$ s. Because of  the
larger mixing
the $\nu_h$ flux from the $K$ decays in flight  would increase
by a factor $\simeq  10^3-10^4$. However, several new suppression factors have to be
taken into account. First,   the $\nu_h$ flux would decrease by a factor $\simeq 30$ 
due to the more  rapid decay of the $\nu_h$'s. Second, 
the experimental signature for the $\nuh$ decay would be an $\pair$ pair 
from the conversion of the decay photon in the decay region. However, 
the opening angle of the $\pair$ pair is $\simeq m_e/E_{\pair} \lesssim 10^{-3}$ rad, which 
is too small to be resolved in the detector, 
and thus the event 
 would be misidentified as a single track event. Such event would be rejected.
The rejection factor is estimated to be $\gtrsim 10^{-2}$.
 Third,  
the average probability of the photon conversion with the vertex 
located in the  low density decay region ( not in a chamber)
 is as small as $\lesssim 10^{-2}$.
Finally,  the total
number of signal events in PS191 would  decrease
 by a factor of more than $\simeq 10^2$  compared to the number of events expected
for a long-lived $\nu_h$'s produced and decaying through the mixing
$|U_{\mu h}|^2 = 10^{-5}$. 
Note, that for the above values of  $|U_{\mu h}|^2$, $m_{\nu_h}$  and $\tau_{\nu_h}$, the branching fraction of the direct $\nu_h$ decay into the $\pair \nu$ final state is  found to be  
$Br(\nu_h \to \pair \nu) < 10^{-5}$ \cite{gs}, which is also small enough to produce a detectable 
excess of $\pair$ events in the PS191 experiment. 
In this estimate the average $\nu_h$ momentum is
$<p_{\nu_h}>\simeq 4$ GeV, the decay region length is $l=12$ m 
 and  the typical photon energy is $\simeq$ 2 GeV \cite{ps191}.

\subsection{LEP constraints}

Next we consider bounds from LEP experiments \cite{pdg}.
For the mass region around 50 MeV, 
the model-independent limit from  the  searches for 
the $Z \to \nu \nu_h$ decay is $|U_{\mu h}|^2 \lesssim 10^{-2}$, 
(see e.g. \cite{delphi}) which is compatible with Eq.(\ref{lsndmix}). 
Direct searches for radiative decays of an excited 
neutrino $\nu^* \to \gamma \nu$ produced in $Z\to \nu^* \nu$ decays have 
also been performed \cite{pdg}.  The best limit from ALEPH  is  \cite{aleph}
\begin{equation}
Br(Z\to \nu \nu ^*) Br(\nu* \to \gamma \nu) < 2.7\times 10^{-5}.
\label{aleph}
\end{equation}
As the  experimental signatures for the 
$\nu* \to \gamma \nu$ and $\nuh$ decays
are the same, we will use this bound for comparison.
 The number of expected $\nuh$
events in ALEPH is proportional to 
$Br(Z\to \nu \nu_h) Br(\nuh) [1-\exp (-\frac{l m_{\nu_h}}{p_{\nu_h} \tau_{\nu_h}})]$, with $l\simeq 1$ m and $p_{\nu_h}\simeq 45$ GeV. 
Taking into account  
$\frac{Br(Z\to \nu \nu_h)}{Br(Z \to \nu \nu )} \simeq  |U_{\mu h}|^2$
and  using \eqref{aleph}, we find
\begin{equation}
|U_{\mu h}|^2\times \frac{m_{\nu_h}[MeV]}{\tau_{\nu_h}[s]} < 4.8 \times 10^{9}. 
\end{equation}
For the mass range \eqref{lsndmass} using Eq.(\ref{lsndmix}) 
results  in 
\begin{equation}
\tau_{\nu_h} \gtrsim 10^{-11}-10^{-10}~ s,
\label{leptau}
\end{equation}
 which is consistent with \eqref{lsndtau}.

\subsection{Bounds from cosmology, astrophysics, and the Super-K experiment}
Although a detailed analysis of the cosmological and astrophysical constrains 
on the properties of heavy sterile neutrinos is beyond the scope of this work, let us briefly 
discuss some of them. The most stringent bounds  $\mix < 10^{-10} - 10^{-3}$ for the $\nu_h$'s
 in the MeV mass range were obtained from the primordial nucleosynthesis and
SN1987A  considerations \cite{dolgov1,dolgov2,dolgov3,kus}. These limits are typically valid under assumption that the $\nu_h$ 
is a relatively long-lived particle with the dominant decay mode $\nu_h \to \nu e^+ e^-$ into  an active neutrino 
($\nu_e, \nu_\mu, \nu_\tau$) and an $\pair$ pair. In this case, for the required mass range 40 - 80 MeV and mixing 
$|U_{\mu h}|^{2} < 10^{-2}$, the $\nu_h$ lifetime estimated from Eq.(\ref{ratedec}) is 
$\tau_{\nu_h} \gtrsim 10^{-2}$ s,   
which is about 7 orders of magnitude  longer compared to the one required by \eqref{minibtau}. 

Another independent constrain  on $\mix$ can be set based on the nonobservation of 
atmospheric sterile neutrino decays by the Super-K experiment \cite{kus}. In this work
it is assumed that heavy neutrinos could be 
copiously produced in the Earth's atmosphere and could decay inside the 
Super-K detector, generating an excess event rate \cite{kus}.
The requirement for this rate to not exceed the rate of events observed by the experiment results in  upper limits on mixing strength       
 $|U_{\mu h}|^2 <  10^{-5}-10^{-4}$ for  the mass region $40 \lesssim m_{\nu_h}\lesssim 80$ MeV (see Fig. 6 in Ref.\cite{kus}). 
 In these calculations it is assumed that the typical Lorentz $\gamma$ factor of heavy 
 neutrinos is $\lesssim 10$. Taking into account \eqref{minibtau} results in a $\nu_h$ decay length   of the order of $l\lesssim 10$ m.  Assuming an average distance 
 between the $\nu_h$ production region and the Super-K detector of $L\simeq  1$ km
  gives a very high  $\nu_h$ flux suppression factor of $exp(-L/l)\ll 10^{-5}$.  Thus, one can see that 
  the stringent  bounds from cosmology, astrophysics, and the Super-K experiment are 
  evaded due to the short lifetime of the $\nu_h$ in accordance with \eqref{minibtau}; for more detailed discussions of the bounds on heavy neutrinos from cosmology and  astrophysics, see e.g. \cite{dolgov1, gr}. 
  
\subsection{Limits on $\mu_{tr}$}
For the light neutrino mass $m_\nu << m_{\nu_h}$ using Eq.(\ref{ratemagmom}) the $\nu_h$ lifetime  due to a transition 
moment $\mu_{tr}$ is given by 
\begin{eqnarray}
\tau_{\nu \gamma}^{-1}=\frac{\alpha}{8}\bigl(\frac{\mu_{tr}}{\mu_B}\bigr)^2 \bigl(\frac{m_{\nu_h}}{m_e}\bigr)^2 m_{\nu_h}
\label{rate1}
\end{eqnarray}
The requirement \eqref{minibtau} for the $\nuh$ decays to occur mostly inside the MiniBooNE fiducial volume 
results in   
\begin{equation}
\mu_{tr} \gtrsim 3 \times 10^{-8}{\mu_B}.
\label{magmom}
\end{equation} 

For $m_{\nu_h} \simeq 40-80$ MeV and $\mu_{tr} > 10^{-8}{\mu_B}$ the radiative decay is dominant, $Br(\nu_h \to \gamma \nu) > 0.9 $.
Direct searches for  heavy neutrino decays were performed by many experiments \cite{pdg}. However, none of these experiments 
 has reported  a bound on the mixing strength $|U_{\mu h}|^2$ or on the product $|U_{\mu h}|^2 \mu_{\rm tr}$,
 for the radiative decays of heavy neutrino in the mass region 40-80 MeV.
The mixing $\mix$ would result in a contribution to the 
effective $\nu_\mu$ magnetic moment,
$\mu_{\nu_\mu}^{eff} \simeq \mix \mu_{tr} \simeq (0.4 -4.0)\times 10^{-10}\mu_B$, 
 due to the nonzero $\nu_h$ magnetic moment.
 This contribution  is  below the best direct LSND experimental 
limit derived from the muon neutrino-electron scattering 
$\mu_{\nu_\mu}^{eff} < 6.8\times  10^{-10} \mu_B$ \cite{lsndmag}. However, 
 in this particular case the LSND limit is not directly applicable to 
the  $\nu_h$ magnetic moment  as the limit was obtained for the DAR $\overline{\nu}_\mu$,
which, as discussed in Sec. II,   cannot produce $\nu_h$ in the LSND 
experiment in  $\nu_\mu$ scattering, due to its heavy mass.

Consider now again bounds from LEP experiments \cite{pdg}.
For the mass region around 50 MeV, 
the model-independent limit from  the  searches for 
the $Z \to \nu \nu_h$ decay is $|U_{\mu h}|^2 \lesssim 10^{-2}$, 
(see e.g. \cite{delphi}) which is compatible with Eq.(\ref{lsndmix}). 
Consider the constraint  \eqref{aleph} from direct searches 
for radiative decays of an excited 
neutrino $\nu^* \to \gamma \nu$ produced in $Z\to \nu^* \nu$ decays 
\cite{aleph}.
 The number of expected $\nuh$
events in ALEPH is proportional to $Br(Z\to \nu \nu_h) Br(\nuh) [1-\exp (-\frac{l m_{\nu_h}}{p_{\nu_h} \tau_{\nu_h}})]$, with $l\simeq 1$ m and $p_{\nu_h}\simeq 45$ GeV. 
Taking into account  
$\frac{Br(Z\to \nu \nu_h)}{Br(Z \to \nu \nu )} \simeq  |U_{\mu h}|^2$
and  using Eq.(\ref{ratemagmom}), we find
\begin{equation}
|U_{\mu h}|^2\times \Bigl(\frac{\mu_{tr}}{\mu_B}\Bigr)^2 < 1.9 \times 10^{-16}. 
\end{equation}
Using Eq.(\ref{lsndmix}) 
results in $\mu_{tr} \lesssim (2.6-1.4)\times 10^{-7} \mu_B $, which is 
consistent with Eq.(\ref{magmom}).

The limit on the $\mu_{tr}$ between the $\nu_h$ and the $\nu_\mu$ 
has been obtained in Ref.\cite{gk1}, based on the idea of the 
Primakoff conversion $\nu_\mu Z \to \nu_h Z$   
of the muon neutrino into a heavy neutrino in the external Coulomb field 
of a nucleus $Z$, with the subsequent $\nuh$ decay.  
By  using the results  from the NOMAD experiment \cite{nomad2}, a 
model-independent bound $\mu_{tr}^{\mu h} \lesssim 4.2\times 10^{-8} \mu_B$ was 
set for the $\nu_h$ masses around 50 MeV (see Table 1 and Fig.2 in 
Ref.\cite{gk1}), which is also consistent with  Eq.(\ref{magmom}).  
Values of $\mu_{tr}$ larger than $10^{-8} \mu_B$ for the $m_{\nu_h}> 40$ MeV
 could be obtained e.g. in the framework of the Zee model \cite{moh}.

\section{Proposed  searches for heavy neutrino}

In this section we propose experimental searches for  heavy neutrinos 
 in the  $K_{\mu 2}$ decay and muon neutrino interactions. The sensitivity of the
 proposed experiments is expected to cover the region of the LSND-MiniBooNE 
 parameter space shown in Fig.\ref{plot}.
  A discussion of 
the possible search for heavy neutrino in muon decays will be reported elsewhere.

\subsection{Search for the $\nu_h$ in $K$  decays}

As discussed in Sec. VI, the existence of heavy neutrinos with 
masses  $\lesssim$ 400 MeV 
can be effectively probed by searching for a peak from the $\nu_h$ 
in two body $K^+ \to \mu \nu_\mu$ decays  of charged kaons \cite{shrock1}.
Depending on the experimental method, one could also search for  
 a  peak in the missing mass distribution
 corresponding to the mass of the heavy neutrino. 
The number of $K^+ \to \mu \nu_h$ events in the peak is defined by the 
mixing $|U_{\mu h}|^2$ and by the phase space and helicity  factors which 
depend on the $\nu_h$ mass \cite{shrock1}. 
For the  mass interval $m_{\nu_h}\simeq 40-80$~MeV 
the chirality-flip is mostly due to the sterile neutrino mass  
which results in 
\begin{equation}
\Gamma(K \to \mu \nu_h) \approx 
\Gamma(K \to \mu \nu_\mu)|U_{\mu h}|^2
\Bigl(\frac{m_{\nu_h}}{m_\mu}\Bigr)^2\;.
\label{rate3}
\end{equation} 
Using Eq.~\eqref{lsndmix}, 
we find that the branching fraction of $K \to \mu \nu_h$ is in 
the experimentally accessible range:
\begin{equation}
Br(K\to \mu \nu_h)\approx 10^{-4}- 10^{-3}
\label{br1}
\end{equation}  
for heavy neutrino masses in the range 40 - 80 MeV. 
 There are two  advantages to searching for the $\nu_h$
in the  $K$ decay peak experiments. First, an observation of
a  peak in the muon and/or  the missing mass spectra in $K$ decays  
gives unambiguous evidence for the existence of heavy neutrinos.
Second, the expected number of 
signal  events  occurring in a detector, and hence the sensitivity of the search, 
 is $\propto |U_{\mu h}|^2$, as it follows from \eqref{rate3}.
In neutrino scattering experiments the $\nu_h$  decay 
signal rate is either  proportional to $|U_{\mu h}|^2\cdot |U_{\mu h}|^2$, 
or, if the dominant decay is $\nuh$, proportional to 
$|U_{\mu h}|^2\cdot \mu_{tr}^2$ 
and thus, is  more suppressed. Here, the first term $|U_{\mu h}|^2$ 
appears through the heavy neutrino production in a target,  
and the second term, $|U_{\mu h}|^2$ or $\mu_{tr}^2$, 
 through the heavy neutrino  decay in the detector.  Note that 
in our particular case the sensitivity of the LSND and MiniBooNE experiments is $\propto |U_{\mu h}|^2$ because 
the $\nu_h$ is a short-lived particle due to the large value of $\mu_{tr}$, 
which decays in the detector target volume 
with the probability $\simeq 1$. 

As discussed in Sec.VI, the major physics   background to the
experiments searching for the $\nu_h$ peak in $K_{\mu 2}$ decays at rest  is 
 the radiative kaon decay 
$K^+ \to \mu \nu \gamma$ which has a branching fraction of about 1.5 \% for photon
energy above 10 MeV \cite{pdg}. This background  
 results in an admixture of a 
continuous spectrum  to the muon 
momentum distribution below the main peak and 
essentially constrains 
the sensitivity of the search for the $\nu_h$ mass range  $\lesssim 100$ MeV. 
To improve the sensitivity, this background decay mode has to be suppressed by increasing 
the detection efficiency of the decay photons. Experimentally, improvement of the photon 
efficiency for the searches with the $K$ decays at rest is difficult due to the limitation 
factors discussed in Sec.VI. 

\begin{figure}[h]
\begin{center}
    \resizebox{8cm}{!}{\includegraphics{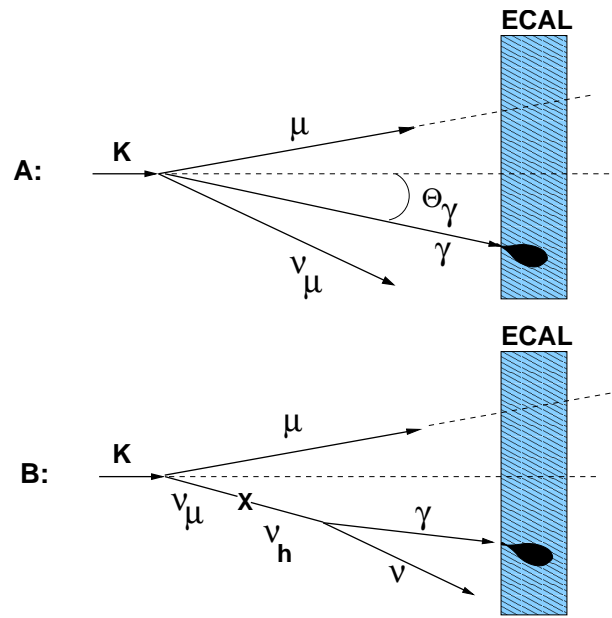}}
     \caption{ Schematic illustration of an
      experiment to search for radiative neutrino decay in $K_{\mu 2}$ decays in flight at 
     high energies: A) the main background decay  $K\to \mu \nu \gamma$
is suppressed because of the high detection efficiency of decay photons
 in the electromagnetic calorimeter (ECAL)  due to the Lorentz bust; B)
if the $\nu_h$ is a short-lived particle, a part of   
 the photons from the $\nuh$ decay is also detected. See text.}
\label{kaon}
\end{center}
\end{figure}

Here, we propose the use of $K$ decays in flight to improve the sensitivity 
against this background source.
A  substantial increase in  the detection 
efficiency of  radiative photons could be obtained   by  using the
$K$ decays in flight at high energies.
In this case, the vast majority of decay photons would be 
within the geometrical acceptance of the detector because they are  
 distributed 
within a narrow cone with 
 the maximal photon angle of the order 
$\Theta_\gamma \simeq m_K/E_K \simeq 7$ mrad
 for a kaon energy of $E_K = 70$ GeV, as  
schematically illustrated in Fig. \ref{kaon}A.
Thus,  the detection efficiency of high  energy decay photons 
from the decay  $K^+\to \mu^+ \nu \gamma $ in an electromagnetic calorimeter is
 expected to be almost 100\%.
If the $\nu_h$ is a relatively long-lived particle, with a lifetime $\gtrsim 10^{-10}$ s, then   
 it would rarely decay in the experiment, as the $\nu_h$ average decay length of $\gtrsim 300$ m is 
 much bigger than the typical length of a decay volume $\simeq$ 100 m. The 
detection of a muon and a photon in the final state would unambiguously signal 
the   detection of the radiative $K$ decay, as shown in Fig. \ref{kaon}A,
or another background decay mode. This would reduce the background significantly and allow 
the measurement of the muon energy distribution with a higher sensitivity.
 If the $\nu_h$ is a short-lived particle, with a lifetime $\simeq 10^{-11}$ [see Eq.(\ref{leptau})] and a 
 corresponding  decay length of about 30 m, 
   then the detection of a muon and a photon in the final state would mean either the
detection of a background process or the detection of the signal from the decay $\nuh$, as shown in Fig. \ref{kaon}B. In this case, one could still  suppress the background by rejection of the $\mu \gamma$ observed events
 at the cost of the signal efficiency loss. To avoid this reduction, one  could try to 
 identify signal events by using the  fact that 
the observed photon is originated from a secondary vertex, which is  displaced from the
 primary one by a large distance, provided  that precise measurements of 
the photon directionality can be done. 

A good example of an experiment where the proposed search could be 
performed is the  NA-62 at CERN \cite{na62}. 
The experiment is  running at a kaon energy of 74 GeV. 
The detector is well equipped to identify and measure the momenta/energy
  and directions of the charged particles. 
The photons are precisely  measured with  a LXe electromagnetic calorimeter.
To evade the KEK  limit
 for the region  
$m_{\nu_h} \lesssim 80 $ MeV the $\nu_h$ lifetime  is required to be 
in a slightly more restricted range 
 $\tau_{\nu_h}\lesssim 5\times  10^{-11}$ s. For example,
 for $\tau_{\nu_h}\simeq 3\times 10^{-11}$ s the most of the $\nu_h$'s  
 would decay in the $K$ decay volume \cite{na62}, thus producing a veto signal 
 in the LXe calorimeter. 
Other experiments capable of searching  for the $\nu_h\to \gamma \nu $ decay with their existing data are 
the E787 and its upgrade, the E949, at BNL \cite{bnl},
or ISTAR+  at IHEP \cite{istra}.
The former is  equipped with an almost $4\pi$ veto electromagnetic calorimeter, allowing 
good  rejection of photons from background  $K^+\to \mu^+ \pi^0 \nu$ and $K^+\to \mu^+ \nu \gamma $ decays. 

\subsection{Search for the decay $\nuh$ in $NC$ neutrino interactions}

\begin{figure}[h]
\hspace{-2.cm}{
    \resizebox{9cm}{!}{\includegraphics{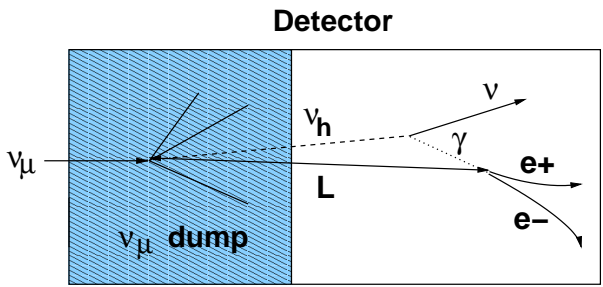}}}
     \caption{ Schematic illustration of the proposed
      neutrino experiment to search for radiative neutrino decay in $\nu_\mu NC$ interactions. 
      The electromagnetic and hadronic secondaries from 
the  $\nu_\mu NC$ event are absorbed in the
initial part of a neutrino detector serving as a dump. Heavy neutrino produced through 
muonic  mixing (see  Fig.\ref{diag}) penetrates the dump and decay into a photon 
and a light neutrino in  the downstream decay region of the detector. The experimental signature 
of the $\nuh$ decay  is the appearance of a single high energy  $e^+ e^-$ pair from the conversion 
of the decay photon at a  distance $L$ from the
primary vertex  significantly larger than the detector nuclear interaction length $L\gg \lambda_0$.
Precise  identification of the electromagnetic nature of the excess events is crucial for this 
experiment.  }
\label{icarus}
\end{figure}

As discussed above, in order to search for an excess of single converted photons  
from the radiative neutrino decay $\nuh$ in high energy $NC$ neutrino interactions the  
 background, mainly from the decays of $\pi^0$'s and bremsstrahlung photons produced either in the primary vertex or in the  secondary particles interactions has to be eliminated.    
To suppress the  background and to detect a clean and convincing sample of converted decay photons one can perform a neutrino "beam dump" experiment the main idea of which  is illustrated in Fig. \ref{icarus}.
A neutrino  detector is subdivided into two parts. The first part is an active  absorber part,  and the second one is the decay region for the detection of converted photons from the decay $\nuh$.
The secondary particles  from   $\nu_\mu \rm{NC}$ interactions in the detector are absorbed in its
first part. Heavy neutrinos produced through 
the muonic  mixing penetrate the dump and decay into a photon 
and a light neutrino in  the second downstream part of the detector,
 with a subsequent photon conversion. The experimental 
signature of the decay $\nuh$ is the appearance of a single $\pair$ pair originating 
from a secondary vertex displaced from the primary one at a distance $L$  significantly larger 
than the detector nuclear interaction length,  $L\gg \lambda_0$. 
The main  background sources for this setup are expected  from the secondary 
neutrons and/or  $K_L^0$'s  penetrating the dump and producing
 $\pi^0$'s either in hadronic  secondary interactions 
or in decays in flight  in the target. The  decays of these $\pi^0$'s  could be misidentified as a single decay photon event.
The suppression of these backgrounds can be achived by increasing  the $n,K_L^0$ absorption  in the first part of the detector simply by increasing its length or by selecting events with larger $L$. Obviously, the  precise  
identification of the electromagnetic nature of the signal event
 is of great importance  for this search. Interestingly, if the
 event excess is indeed originated from the 
 converted photons,   the proposed search could also distinguish whether the 
excess events are produced by photons from the $\nuh$ decays or by ones 
emitted  from the primary vertex due to  
 anomaly-mediated neutrino-photon coupling, as  discussed  in \cite{hill}.
The (almost) ideal detector to search for the decay $\nuh$ is a detector similar to the NOMAD one \cite{nomad1}.
The NOMAD  is equipped with a forward calorimeter (FCAL), where the 
secondaries from high energy neutrino interactions in FCAL could be dumped.
In addition, it has the excellent capability of identifying and reconstructing   
converted photons  due to the low mass target located in a magnetic field.
An example of the reconstruction of two conversion $\pair$ pairs  from the decay $\pi^0 \to 2 \gamma$ 
 can be found in Ref.\cite{nomadpi0}. One disadvantage of the detector is the
  short length of the tracking part. The overall detection efficiency  of the  $\nu_h$ production, 
  decay in flight with the subsequent 
conversion of the decay photons into an $\pair$ pair and the reconstruction of the conversion pairs is expected to be  low. The  advantage of NOMAD  is its great capability to  measure 
  the $\pair$ pair directionality with a precision of $(1-cos\Theta_{\pair}) \lesssim 10^{-5}$ 
\cite{nomadtaunu}. This will allow an effective suppression of converted photons originating 
from the primary vertex; see the discussion in Ref.\cite{nomadtaunu} on $\pi^0$ reconstruction.  

Another experiment capable of searching for the decay $\nuh$  is the  ICARUS T600, which is   
 currently taking data at the CERN-Gran Sasso neutrino beam \cite{icarus}. The detector is
 composed of two identical adjacent T300  half-modules filled  with liquid argon (LAr). A detailed 
 description of the apparatus can be found in \cite{icarus}. Each T300 half-module has the following 
internal dimensions: $3.6  \times 3.9 \times  19.9$(length) m$^3$. 
LAr has 
a radiation length of $X_0=14$ cm and a nuclear interaction length of $\lambda_0 =83.6$ cm,  and therefore  provide 
 good electromagnetic and hadronic secondary absorption  and detection
capabilities for the proposed search, assuming that the length of the 
decay region is $L\gtrsim 10$ m$\gg X_0,\lambda_0 $ .

The number of $\nuh$ events in ICARUS can be estimated as follows
\begin{equation}
\Delta N_{\nuh} \simeq  N_{NC}|U_{\mu h}|^2 P_{dec} P_{abs} \epsilon  
\end{equation}
where $N_{NC} \simeq 10^3$ is the number of the 
 detected neutral-current events, 
and $P_{dec}(\simeq 0.4), P_{abs} (\simeq 1), and
\epsilon (\simeq 0.7) $ are 
the probabilities for the $\nu_h$ decay  in the detector fiducial volume and  decay
photon conversion,  and the overall detection efficiency of the $\pair$ pair, respectively. 
In this estimate  the  average $\nu_h$ momentum is $<p_{\nu_h}>\simeq 10$ GeV,  
$\tau_{\nu_h}\lesssim  10^{-9}$, and the  length of the decay region is $L=12$ m. Finally, we find  
  that the number of expected $\nuh$ signal events 
 in ICARUS  is  
\begin{equation}
\Delta N_{\nuh} \simeq 6 \times10^2\times |U_{\mu h}|^2
\end{equation}
For the allowed mixing (see  Fig. \ref{plot}), this results in  
$\Delta  N_{\nuh} \simeq 1 - 3 $  events. If no candidates are seen above the expected background level,  ICARUS 
could set  a limit on the mixing strength  of the order 
$|U_{\mu h}|^2 \lesssim 10^{-3}$, which  is competitive for the mass range 40-80 MeV
with the bounds obtained from the TWIST experiment, possibly  allowing us 
to rule out  the LSND-MiniBooNE parameter region.
Note that the search for an excess of the $\nu_h$ decay events  
can also be performed in the  recently proposed ICARUS-like experiment at CERN PS \cite{icarusps}, or 
at FNAL with neutrino detectors such as MicroBooNE \cite{micro}, 
 HiResM$\nu$ \cite{hires}, and BooNE (a MiniBooNE near detector) \cite{boone}.

\section{Summary}

In summary, we reexamined neutrino oscillation results from the 
accelerator experiments LSND, KARMEN and MiniBooNE. We showed that the LSND evidence for $\overline{\nu}_\mu \to \overline{\nu}_e$ oscillations, 
its long-standing disagreement with the results from KARMEN, and  the 
 anomalous event excess observed by MiniBooNE in $\nu_\mu$ and $\overline{\nu}_\mu$ data,  can all be explained by the   production and decay of a heavy neutral lepton.
The shape of the excess events in several 
kinematic variables in the LSND and MiniBooNE $\nu_\mu$ and $\numub$ data 
is found to be consistent with the distributions obtained within this interpretation, 
 assuming that the  $\nu_h$'s are created
 by mixing in $\nu_\mu$ neutral-current interactions and decay radiatively  into $\gamma \nu$. Therefore,  our main 
prediction is that the  excess of events observed in
the LSND and MiniBooNE experiments originates from the Compton scattering or $\pair$  conversion of the decay photons
 in these detectors. In this context, the confirmation of 
the photon origin of the excess events by measurements with a detector able to distinguish 
electrons and photons becomes a crucial test for this scenario.

A combined analysis of the energy and angular 
distributions of the excess events observed in the LSND and MiniBooNE experiments 
suggests that  the $\nu_h$  mass is in the range from 40 to  80   MeV,  
the mixing strength is  $|U_{\mu h}|^2 \simeq 10^{-3}-10^{-2}$ and
 the lifetime  $\tau_{\nu_h} \lesssim 10^{-9}$ s.
 Surprisingly, this LSND-MiniBooNE  favorable parameters window  is found to be unconstrained 
by the results  from the most sensitive  $K_{\mu 2}$, neutrino scattering, and LEP experiments.
Because of  the short $\nu_h$ lifetime, 
the constraints coming from cosmological and astrophysical considerations, as well as 
the bounds from the atmospheric neutrino measurements, are also evaded.
 We set 
new limits on the mixing $|U_{\mu h}|^2$ for heavy neutrino masses in the range 40 to 80 MeV 
by using results on precision measurements of the Michel spectrum by the TWIST experiment.   
We also discuss the most natural model for the $\nuh$ decay through the  
transition magnetic moment between the $\nu_h$ and the  light neutrino and show that the
obtained values $|U_{\mu h}|^2 \simeq 10^{-3}-10^{-2}$ and $\mu_{tr}\gtrsim 10^{-8} \mu_B$ 
do not violate  bounds  from previous experiments. 

The results obtained provide a strong motivation for a 
sensitive search for the $\nu_h$ in a near future $ K$ decay or neutrino experiments.
We propose such experiments with the expected sensitivity to cover the region of the 
LSND-MiniBooNE parameter space and notice they fit well in the existing and planned experimental programs at CERN or FNAL.
 The radiative heavy neutrino decay 
could be present in various extensions of the standard model and, thus, could enhance the reported 
motivations  to search for this process. 
We note that an analysis of the excess of events due to 
the $\nuh$ decay may also  be possible with existing neutrino data; e.g. 
new results could be obtained from NOMAD \cite{nomad1}.

The reported analysis gives the estimated 
values of the parameters  $m_{\nu_h}$, $|U_{\mu h}|^2$, and $\tau_{\mu h}$
 and may be improved by more accurate
and detailed simulations of the LSND and MiniBooNE detectors. 
It would  also be interesting and important to have general analysis 
of the production of heavy neutrinos of Dirac or Majorana type,
 e.g. in $\nu_\mu NC$ interactions, for arbitrary weak couplings, 
including the leptonic mixing and helicity effects.

\begin{acknowledgments}
 I would like to thank  L. Camilleri, D.S. Gorbunov,  N.V. Krasnikov, 
S.A. Kulagin, L. Di Lella, V.A. Matveev, V.A. Rubakov, and A. Rubbia
  for discussions. I am grateful to W.C. Louis for comments and clarifications,
 and to D. Sillou for discussions and help in manuscript preparation.

\end{acknowledgments}

\end{document}